\documentclass{aastex63}
\newif\ifdraft \drafttrue
\usepackage[utf8]{inputenc}
\ifdraft \else \newcommand{\submitjournal}[1]{\relax} 
               \newcommand{\correspondingauthor}[1]{\relax} 
\fi
\submitjournal{ApJS}
\usepackage{lipsum}
\usepackage{graphicx}
\usepackage{xcolor}
\usepackage{upgreek}
\usepackage{amsmath}	
\ifdraft
   \usepackage{hyperref}
   \hypersetup{
               colorlinks=true,
               linkcolor=black,
               filecolor=black,
               citecolor=black,
               urlcolor=blue
              }
     \newcommand{\web}[1]{\Blb{\url{#1}}}
\else
     \newcommand{\web}[1]{#1}
\fi

\newcommand{\PIMA}{$\cal P\hspace{-0.067em}I\hspace{-0.067em}M\hspace{-0.067em}A$ }

\newcommand{\ntab}[2]{ \multicolumn{1}{#1}{#2} }
\newcommand{\nntab}[2]{ \multicolumn{2}{#1}{#2} }

\definecolor{Dred}{rgb}{0.312,0.070,0.070}
\definecolor{Dblue}{rgb}{0.070,0.070,0.312}
\definecolor{Dgreen}{rgb}{0.070,0.312,0.070}
\definecolor{Ugreen}{rgb}{0.070,0.660,0.070}
\definecolor{Db}{rgb}    {0.050,0.0,0.320}

\newcommand{\Blb}[1]{\textcolor{Dblue}{\bf #1}}

\newcommand{\Rdb}[1]{\textcolor{Dred}{\bf{#1}}}

\newcommand{\Fermi}{\textit{Fermi} }
\newcommand{\Gaia}{\textit{Gaia} }
\newcommand{\beq}{ \begin{eqnarray} }
\newcommand{\eeq}[1]{\label{#1}\end{eqnarray}}
\newcommand{\eeqn}{ \nonumber \end{eqnarray} }
\newcommand{\Frac}[2]{\frac{\displaystyle\strut #1}{\displaystyle\strut #2} }
\newcommand{\vex}{\vspace{1ex}}
\newcommand{\dss}{\displaystyle}
\renewcommand{\tau}{\uptau}
\renewcommand{\vec}[1]{{\mathbf #1}}
\newcommand{\vc}[1]{{\bf \vec{#1}}}

\newcommand{\lp}{ \left(  }
\newcommand{\rp}{ \right) }

\newcommand{\flo}[2]{\mbox{#1} \cdot 10^{#2}}

\newcommand{\der}[2] {\frac{ \partial #1 }{ \partial #2 } }

\newcommand{\mat}[1]{\widehat {\mathstrut \cal #1}}

\newcommand{\TEC}{\rm TEC}
\newcommand{\tss}{\textstyle}

\newcommand{\dintinf}{\int\limits_{\enskip-\infty}^{\enskip+\infty}\hspace{-0.75em}\int}

\newcommand{\Sp} { {\rm Sp}\hspace{0.2em} }
\newcommand{\Exp}{{\cal E}}
\newcommand{\tra}[1]{ {#1}^{^{\bf \top}}\! }
\newcommand{\ttra}[1]{ {#1}^{\bf \!\top} }

\newcommand{\tg}{\rm tg}

\newcounter{note}
\let\oldmarginpar\marginpar
\renewcommand\marginpar[1]{\-\oldmarginpar[\raggedleft\footnotesize #1]%
{\raggedright\footnotesize #1}}
\newcommand{\Note}[1]{\textcolor{Ugreen}{#1}}
\renewcommand{\Note}[1]{#1} 

\newcounter{myitemm}
\newcommand{\itemm}{\addtocounter{myitemm}{1}\themyitemm~}
\newcommand{\Number}[1]{\ifnum#1<10\relax0\number#1\else\number#1\fi}
\newcommand{\isodate}{
\count151=\time
\divide\count151 by 60
\count151=\count151
\multiply\count151 by 60
\count152=\time
\advance\count152 by -\count151
\divide\count151 by 60
\count152=\count151
\multiply\count151 by 60
\count153=\time
\advance\count153 by -\count151
\Number{\year}.\Number{\month}.\Number{\day}--\Number{\count152}:\Number{\count153}
}

\shorttitle{The Radio Fundamental Catalogue. I.}
\shortauthors{Petrov \& Kovalev}

\begin{document}
\title{The Radio Fundamental Catalogue. I. Astrometry}

\author[0000-0001-9737-9667]{L. Y. Petrov}
        \affil{NASA Goddard Space Flight Center,
        Code 61A, 8800 Greenbelt Rd, Greenbelt, 20771 MD, USA}

\author[0000-0001-9303-3263]{Y. Y. Kovalev}
\affiliation{Max-Planck-Institut f\"ur Radioastronomie, Auf dem H\"ugel 69, 53121 Bonn, Germany}

\received{August 12, 2024}
\revised{October 12, 2024}
\accepted{October 14, 2024}
\published{Janury 16, 2025}


\correspondingauthor{Leonid Petrov} \email{Leonid.Petrov@nasa.gov}

\begin{abstract}

We present an all-sky catalog of absolute positions and estimates of 
correlated flux density of 21,942 compact radio sources determined 
from processing interferometric visibility data of virtually all very 
long baseline interferometry (VLBI) observing sessions at 2--23 GHz 
from 72 programs suitable for absolute astrometry collected for 30 years. 
We used a novel technique of generation of a data set of fused observables 
that allowed us to incorporate all available data in our analysis. 
The catalog is the most complete and most precise to date. It forms the 
foundation and reference for positional astronomy, space geodesy, space 
navigation, and population analysis of active galactic nuclei (AGNs), 
and provides calibrators for phase referencing for differential astrometry 
and VLBI astrophysical observations. Its accuracy was evaluated through 
a detailed accounting of systematic errors, rigorous decimation tests, 
comparison of different data sets, and comparison with other catalogs. 
The catalog preferentially samples AGNs with strong contemporary 
parsec-scale synchrotron emission. Its milliarcsecond-level positional
accuracy allows association of these AGNs with detections in a wide 
range of the electromagnetic spectrum from low-frequency radio to 
$\gamma$-rays and high-energy neutrinos. We describe the innovative data 
processing and calibration technique in full detail, report the in depth 
analysis of random and systematic positional errors, and provide a list 
of associations with large surveys at different wavelengths.

\end{abstract}


\section{Introduction}

  The method of very long baseline interferometry (VLBI), first proposed 
by \citet{r:mat65} allows us to determine positions of compact radio
sources with a nanoradian level of accuracy (1 nrad $\approx$ 0.2~mas). 
The very first VLBI catalogue contained coordinates of 35~objects 
\citep{r:first-vlbi-cat}, which are active galactic nuclei (AGNs). Since then,
VLBI observations became routine for geodesy, astronomy, astrophysics, 
and space navigation. It was realized in 1980s that the list of sources 
with precisely known positions and their images at different frequencies
needed to be expanded. 

  To achieve the goals of space geodesy --- a millimeter level of accuracy
in ground station position determination --- a list of 100--300 bright 
extragalactic sources uniformly distributed over the sky with positions 
known at subnanoradian accuracy is necessary. To achieve 
the goals of space navigation --- a nanoradian level of accuracy for 
spacecraft tracking --- a list of 1000--2000 sources within $\pm 7^\circ$ of 
the ecliptic plane with positions known at a subnanoradian level is highly 
desirable. To achieve goals of astronomical VLBI observations with the use 
of the phase referencing technique, a larger pool of extragalactic sources 
with positional accuracy at the same level and with known distributions of 
flux density is required. When observations are preformed in the phase 
referencing mode, the radio telescopes of an array quickly switch from 
a target source to a calibrator within several degrees, which allows one to 
extend the integration time beyond the coherence limit set by the atmosphere 
and detect weaker targets and determine an offset of the target with respect
to a calibrator. We performed a Monte Carlo simulation and found that 
it is sufficient to have 6710 calibrator sources uniformly 
distributed over the celestial sphere in order to find a calibrator within 
$3\degr$ of a given direction with the probability of 99\% assuming the 
sources are distributed uniformly. This number raises to 15,100 if 
a calibrator within $2\degr$ of a given direction has to be found with that 
confidence level. The closer the calibrator to the target, the better the 
atmospheric propagation errors are compensated. \citet{r:mv10a} has 
characterized quantitatively the impact of residual errors on the quality 
of results as a function of target-to-calibrator separation. In general, 
a target-to-calibration separation less than $2\degr$ is desirable, 
a separation of 2--$5\degr$ is favorable, and a separation greater than 
$5\degr$ should be avoided. 

  Phase calibration is also used for differential astrometry. Accuracy 
at tens of microarcsecond can be achieved for the {\it positional offsets} 
from observations in a phase referencing mode 
\citep{r:honma_redi14,r:reid17}. A similar technique is also applicable 
to observations of spacecrafts \citep{r:jones20}. We should note that 
although differential phase referencing VLBI allows one to evaluate position 
offsets with respect to a calibrator with very high precision, this does 
not necessarily mean that a source position can be derived with that level 
of accuracy. The uncertainty of a target position is the sum in quadrature 
of the uncertainty of the positional offset and the calibrator positional
uncertainty. Therefore, positional accuracy of a target source cannot be 
greater than the positional accuracy of a calibrator, which may be orders 
of magnitude worse.

  A good calibrator should be strong and compact in order to be detectable 
with a relatively short integration time at all VLBI baselines, have a simple 
structure in order to reduce the errors of the fringe phase model, and have 
a precise position in the range of 0.2 to 10~mas depending on an application. 
Considering that not all calibrators fulfill these criteria and considering 
the nonuniformity of the distribution of AGNs over the sky, a list of 
$\sim\! 20,000$~calibrators with precise astrometry and photometry 
are required.

  A list of sources with precisely determined coordinates forms the 
reference to which positions of other objects are referred to and it 
serves as the basis for celestial object localization.  Therefore, such 
lists are often called ``reference frames.'' Since currently VLBI is one 
the two most precise astrometry techniques, a list of these sources 
provides the foundation to the positional astronomy.

  Detection of a large number of sources allows us to study a population 
of compact objects that are almost exclusively AGNs. Not all AGNs are 
compact enough to be detected with VLBI at baselines longer than 100~km. 
Correlating the milliarcsecond-scale morphology of detected sources with the 
arcsecond-scale morphologies, spectral indices, and variability, provides 
rich information about AGNs and is a key for understanding their nature. 
Lists of sources that form complete samples according to certain criteria 
are especially valuable because sample statistics derived from their 
analysis can be generalized to the entire population. Since AGNs are 
distributed uniformly over the sky, analysis of the dependence of their
apparent size versus frequency allows us to study not only their intrinsic 
properties, but also the scattering properties of the interstellar medium
in our Galaxy \citep[e.g.,][]{r:puskov15,r:korukova23}.

  Bright compact radio sources are relatively rare objects. The majority 
of them are blazars, a class of AGNs with jets pointing towards the observer 
with a viewing angle less than $10^\circ$. Knowing their statistics, 
the probability to find a compact AGN with the certain flux density within 
a specified search area from a given direction by chance can be evaluated. 
This probability if often small enough can be used to draw an important 
conclusion that the sources found in that specified search area are the same 
object with a high confidence level. This approach was successfully used for 
association of $\gamma$-ray sources detected by the \Fermi LAT space telescope 
with AGNs \citep{r:Fermi_VCS,r:aofus1,r:aofus2,r:aofus3,r:4fgl_agn}. Selection 
of high-energy neutrino associations, as well as estimates of 
an associated p-value can be achieved using a catalogue of compact radio 
sources, provided it covers all sky and it is complete at a certain flux 
density limit, as it has been demonstrated in 
\citet{r:Plavin_neutrino1,r:Plavin_neutrino2,r:Plavin_neutrino3,r:bel23,r:IC23}.
  
  Recognizing the values of a list of compact radio sources
detectable with VLBI with very precise positions, a number of observing 
programs for absolute astrometry were launched since 1980s. The criteria for 
specific programs varied, but the overarching goal was to produce an all-sky 
complete list of $\sim\! 20,000$ AGNs that are detectable at 2--20~GHz 
at baselines 1,000--10,000~km, determine their positions with 
a milliarcsecond level of accuracy, and get their images.

  It took over four decades to achieve the goals of this endeavor. Although
observing programs are still continuing and are expected to be continued,
we think we reached a turning point after which a further expansion of 
the VLBI absolute astrometry catalogue is expected to approach to a stall 
just because of a planet-wide resource limit, and the situation will
not change before a new generation of sensitive instruments, Square 
Kilometer Array (SKA) and next-geneeration Very Large Array (ngVLA),
will become fully operational in the 2030s.

  Here we present the Radio Fundamental Catalogue (RFC) which is the result 
of our analysis of virtually all publicly available VLBI data that are 
suitable for the goal, including a large number of observing programs that 
we have initiated. To date, these analysis efforts are the most extensive 
and counting the number of visibilities processed exceed by an order of 
magnitude previous efforts. The Radio Fundamental Catalogue contains precise 
positions determined with a method of absolute astrometry and estimates of 
correlated flux densities at three ranges of projected baseline lengths. 
We present the RFC in its current state on 2024.10.05 as data release 
rfc\_2024c. We will continue to provide online releases on a quarterly basis.

  We split the material in Paper I (here) and future Paper II focusing the 
first paper on description of the observations, their analysis and astrometry 
results. The scope of Paper II is a study of the source counts and sky 
distribution, sample completeness, parsec-scale observational characteristics 
and astrophysical properties of extragalactic objects with strong contemporary 
compact radio emission. In this Paper~I we make an overview of the observing 
programs in section~\ref{s:obs}, present the 
VLBI data analysis technique in section~\ref{s:data_anl},
describe in detail the analysis of positional error in 
section~\ref{s:error_anl}, outline the imaging results in 
section~\ref{s:image_res}, describe the catalogue in
section~\ref{s:the_cat}, and provide a discussion of produced
results in section~\ref{s:disc}. The summary and concluding 
remarks are presented in section~\ref{s:sum} followed by the 
Appendix that describes the machine-readable tables. The
Radio Fundamental Catalogue is accompanied with the Astrogeo 
VLBI FITS image database that cannot be put in the attachment
because of it size (212~GB) and it accessible at 
\dataset[\doi{10.25966/kyy8-yp57}]{\doi{10.25966/kyy8-yp57}}.

\section{Observing Campaigns}
\label{s:obs}

  VLBI observations are organized in campaigns that can contain one
or more segments also called experiments. Antennas of a given array
slew to a given source and collect voltage data during a specified
period of time, from 10 to 600~s. These data collections are called
scans. Some stations of an array may not participate in a given
scan either by design or are dropoed for technical reasons. 
Cross-correlated data from a given scan and a given pair of 
participating stations are called observations. When $N$ stations 
participated in a given scan, there are $N \times (N+1)/2$ observations.

\subsection{Dedicated astronomical experiments}

  A target source or several target sources, as well as a number of 
calibrator sources, are observed for 1--12 hours in a typical astronomical 
VLBI experiment. Such sources are  studied in detail at full sensitivity 
that is achieved owing to a long integration time. This allows us to 
reconstruct high fidelity images and/or determine highly accurate 
source positions using differential VLBI. In contrast, tens to hundreds 
os sources are observed in a single survey experiment, and a VLBI survey 
campaign may involve observations of up to several thousand sources. 
The goal of survey experiments is to study a population of sources. 
Inevitably, shorter integration times are used in survey experiments. 
That results in poorer images and worse positional accuracy than in 
dedicated experiments, but a much larger number of objects is observed
in a single experiment.

  Most of the surveys fall into three categories: pathfinder surveys,
follow-up surveys, and high-frequency extensions. The goal of a pathfinder 
survey is to detect target sources never observed with VLBI before, 
to determine their positions at a milliarcsecond level of accuracy, 
to measure their correlated flux density, and to synthesize their images 
from collected visibility data. Since VLBI has a small field of view, 
typically in the range of $10''$ to $5'$ at 2--24~GHz, blind surveys would 
be very inefficient, because the probability to find by chance a source 
with a VLBI flux density of 10--1000~mJy within such a narrow field of view 
is very low. Therefore, target sources in pathfinder surveys are selected 
among those that have been addlp{previously} detected in prior connected 
radio interferometers at resolutions of 1--$40''$ or in single dish 
observations at resolutions of $0.5'$--$5'$, and VLBI observations just 
follow up objects already detected at low resolutions. Only a fraction of 
target sources is detected with a given VLBI pathfinder survey. Depending on 
criteria used for source selection, which typically involve information 
about total flux density, sometimes supplemented by radio spectral indexes, 
the fraction of detected sources is in the range of 20 to 98\% 
\citep[e.g.,][]{r:vcs5,r:npcs}, with 59\% being the median fraction. Selecting 
targets with flat radio spectra significantly increases detectability since 
such samples strongly favour blazars with dominant Doppler-boosted opaque 
compact cores \citep{r:npcs}. However, this strategy imposes a selection 
bias. To overcome this bias, observing programs since 2015 gradually 
eliminated that criterion.

  The follow-up VLBI surveys target samples of sources previously 
detected in VLBI pathfinder programs with the goal to 
improve positional accuracy or get higher quality images. The radio telescope 
sensitivity is usually the highest in the range of 1--9~GHz, and source flux 
density is usually falling with frequency. Therefore, the chances to detect 
a source using given integration time are in general higher at lower 
frequencies. Sources detected at low frequencies are often followed up at 
higher frequencies in the third type of surveys called high-frequency 
extensions. The goal of these extensions is to get source images at 
higher frequencies that better characterize the core region close to the 
true jet base due to opacity conditions \citep[e.g.,][]{r:lob98}, 
evaluate the suitability of target sources as calibrators at high 
frequencies, and in some cases to improve their positional accuracy.

  In our work we collected data from {\it all} VLBI surveys above 2~GHz
for which visibility data are available from public archives, including 
surveys that we designed ourselves or participated as co-investigators. 
Additionally, we combed through VLBI data archives and examined observing 
campaigns in continuum  at frequencies above 4~GHz that observed 
16 or more target sources 
without the use of phase calibrators. We also included all geodetic VLBI data 
collected under observing programs for determination of station positions, 
station velocities, and the Earth orientation parameters (EOPs) since 
April 1980 as an auxiliary dataset. Although including these data in our 
solutions had only a marginal direct impact on source position estimates, 
their use significantly improved estimates of station positions and the EOPs
that are nuisance parameters in the context of this work, but are essential 
for reducing systematic errors related to the positional stability of the 
VLBI network and its motion with respect to the coordinate system origin.

  Radio wave propagation is described by differential equations against
source coordinates and other variables. Their solution requires three 
arbitrary initial conditions that define the orientation of the celestial 
coordinate system, as well as initial conditions that define the origin 
and orientation of the terrestrial coordinate system. If experiments 
have no common sources and common stations, i.e.\ are totally disjoint, 
source positions derived from each single experiment have arbitrary rotations
with respect to source positions derived from another experiments. Although 
it still possible to align positions derived from different experiments, 
the alignment procedure would introduce additional errors, which we would 
like to avoid. Therefore, a given VLBI survey experiment by design has 
a number of common sources with other experiments. 

  When each observing session has common stations and common sources, the 
whole dataset can be processed in a single least squares solution. Common
stations and common sources tie the dataset together. When the number of 
common stations or common sources is small, the impact of random errors
on positions of common stations and common sources on position estimates 
of other sources is nonnegligible. To avoid this, the number of common
stations and sources should be sufficiently large. In addition to avoiding 
degeneracies in estimation of coordinates, sources are observed in many 
campaigns for improving their positions and/or image quality by collecting 
more data, for examining their properties at different frequencies, or by 
mistake because a target source was not checked thoroughly whether it has 
been detected in previous campaigns. The share of unique sources that 
are detected only in a given campaign is in the range 
from 0 to 88\%.

Below we list all observational programs we used in our work. For those
observational programs for which we found no bibliographic reference
we used the principal investigator name instead. We present the experiment
name, acronym when available, the VLBI array, band, the experiment ID, 
duration, and the source selection criteria when known.

{
\renewcommand\labelenumi{\Roman{enumi}.}
\renewcommand\theenumi\labelenumi
\renewcommand\labelenumii{\arabic{enumii}.}
\renewcommand\theenumii\labelenumii
\begin{enumerate}
   \item Pathfinder surveys:

   \begin{enumerate}\setlength{\itemsep}{0.5ex}
      \item VLBA Calibrator survey 1 (VCS1), \citet{r:vcs1}; 
            VLBA BB023; 
            dual X/S bands; 
            11~segments; 
            since 1994.08.12 through 1997.08.27.
            Selection criteria:
            1)~declinations $>-30^\circ$ and
            2)~detected in the Jodrell Bank--VLA Astrometric Survey 
               JVAS \citep{r:jvas1,r:jvas2,r:jvas3}, an astrometric 
               snapshot survey of compact radio sources performed with 
               the NRAO Very Large Array during the period 1990--1993. 
   
      \item Phase-references superluminals, 
            PI: T. Beasley; 
            VLBA BB041;
            dual X/S bands; 
            2~segments;
            since 1995.06.25 through 1996.02.16.
   
      \item The VSOP Pre-launch VLBA Observations (VLBApls), \citet{r:vlbapls}; 
            VLBA BH019;
            C-band;
            1~segment;
            1996.06.05.
            Selection criteria:
            1)~total flux density at 5~GHz $>1.0$~Jy, and
            2)~spectral index flatter than $-0.5$\footnote{Spectral index
              $\alpha$ is defined as $f^{+\alpha}$ in this work, where $f$ 
              is frequency. We call sources with $\alpha>-0.5$ flat-spectrum
              objects.}, and
            3)~Galactic latitude $|b|>10^\circ$.
   
      \item A VLBA Survey of Flat-Spectrum FIRST Sources, \citet{r:ulv99};
            VLBA BU007;
            C-band; 
            1~segment; 
            1996.12.19.
            Selection criteria: 
            1)~total flux density greater than 50~mJy 
               at~4.85 GHz, from the GB6 survey \citep{r:gb6}, and
            2)~two-point spectral indices flatter 
              than $-0.5$ between 1.4~GHz FIRST \citep{r:first} and 
              4.85~GHz \citep[GB6,][]{r:gb6}, and 
            3)~right ascension range 
              6--$17^h$ and declination from $+29^\circ$ to $+34^\circ$.
   
      \item The Bologna Complete Sample of Nearby Radio Sources, \citet{r:bolsam}; 
            VLBA BG069, BG094, BG158;
            3~segments;
            since 1997.04.06 through 2000.01.22.
            Selection criteria: 
            1)~flux density $>0.25$~Jy at 408~MHz from the B2 Catalogue of 
               Radio Sources \citep{r:b2-4} and the Third Cambridge Revised 
               Catalogue \citep{r:3cr}, and
            2)~flux density at 178 MHz $>10$~Jy from the 3CR catalogue, and
            3)~declination $>10^\circ$, and
            4) galactic latitude $|b| > 15^\circ$, and
            5)~redshift $z < 0.1$.

      \item Caltech Jodrell Bank snapshot survey, \citet{r:bri07};
            VLBA BB119; 
            C-band;
            3~segments;
            since 1999.11.21 through 1999.11.26.
            Selection criteria: 
            1)~declination $>+35^\circ$, and
            2)~galactic latitude $|b| > 10^\circ$, and
            3)~flux density 5~GHz $>0.35$~Jy, and
            4)~spectral index flatter than $-0.5$ between 1.4 and 4.85~GHz 
               from analysis of NVSS \citep{r:nvss} and GB6 \citep{r:gb6} 
               surveys.
   
      \item Densification of the International Celestial Reference Frame, \citet{r:cha04}; 
            EVN EC013, EC017; 
            dual X/S bands;
            3~segments;
            since 2000.05.31 through 2003.10.17.
            Selection criterion: detected in JVAS.
   
      \item VLBA Calibrator survey 2 (VCS2), \citet{r:vcs2}; 
            VLBA BF071; 
            dual X/S bands; 
            2~segments; 
            since 2002.01.31 through 2002.05.14.
            Selection criteria: 
            1)~declination zone $-45^\circ < -30^\circ$, or
            2)~galactic latitude $|b| < 10^\circ$, and
            3)~not observed in VCS1.

      \item VLBA observations of compact 9C sources, \citet{r:bol06b}; 
            VLBA BB177; 
            C band; 
            1~segment; 
            2004.02.06;
            Selection criteria: sources with 15~GHz variability.

      \item VLBA Calibrator survey 3 (VCS3), \citet{r:vcs3}; 
            VLBA BP110; 
            dual X/S bands; 
            3~segments; 
            since 2004.04.30 through 2004.05.27.
            Selection criteria: 
            1)~declination $>-45^\circ$, and
            2)~total flux density $>100$~mJy at both at 2.3 and 8.6~GHz, and
            3)~spectral index flatter than -0.5, and
            4)~have no known calibrator within $3.9^\circ$.

      \item VLBA Calibrator survey 4 (VCS4), \citet{r:vcs4};
            VLBA BP118; 
            dual X/S bands; 
            3~segments;
            since 2005.05.12 through 2005.06.30.
            Selection criteria: multiple criteria with the primary goal to
            observe sources in the areas where no prior VLBI calibrator
            objects within $4^\circ$ radius are known.
   
      \item Observations of compact sources selected at 15 GHz, \citet{r:bol06}; 
            VLBA BC151; 
            X/C band; 
            4~segments; 
            since 2005.06.16 through 2005.08.04.
            Selection criteria: sources with 15~GHz variability.

      \item VLBA Calibrator survey 5 (VCS5), \citet{r:vcs5}, 
            VLBA BK124; 
            dual X/S bands; 
            3~segments; 
            since 2005.07.08 through 2005.07.20.
            Selection criteria: 
            1)~declination $>-30^\circ$, and
            2)~spectral index flatter than $-0.5$, and
            3)~flux density interpolated at 8.6~GHz using
               data from multiple radio astronomy catalogues 
               $>$ 150~mJy.
   
      \item VLBA Imaging and Polarimetry Survey at 5 GHz (VIPS), \citet{r:vips,r:astro_vips};
            VLBA BT085; 
            C-band;
            16~segments;
            since 2006.01.03 through 2006.08.12.
%
            Selection criteria: 
            1)~declination $>+15^\circ$ and $<+65^\circ$, and
            2)~flux density $>0.085$~Jy, and
            3)~present in the Cosmic Lens All-Sky Survey (CLASS) 
              \citep{r:class}, and
            4)~present in the Sloan Digital Sky Survey 
               SDSS \citep{r:sloan} footprint.

      \item The VLBA Galactic Plane Survey (VGaPS), \citet{r:vgaps}; 
            VLBA BP125; 
            K-band;
            3~segments;
            since 2006.02.04 through 2006.10.20.
            Selection criteria: 
            1)~all detected sources from the
               VERA 22 GHz Fringe Search Survey \citep{r:vera22GHz}, and
            2)~flux density interpolated at 22~GHz using
               data from multiple radio astronomy catalogues $>$ 0.2~Jy, and
            3)~spectral index flatter than $-0.5$, and
            4)~galactic latitude $|b|<10^\circ$.
  
      \item Northern Polar Cup Survey, \citet{r:npcs}; 
            VLBA BK130; 
            dual X/S bands;
            3~segments;
            since 2006.02.14 through 2006.02.23.
            Selection criteria: 
            1)~declination $>+75^\circ$ and
            2)~flux density $>0.2$~Jy from NVSS.
   
      \item Compactness of weak radio sources at high 
            frequencies, \citet{r:majid09};
            VLBA BM252; 
            X-band;
            2~segments;
            since 2006.11.06 through 2006.11.13.
            Selection criteria: 
            1)~flux density at 31~GHz $>10$~mJy and
            2)~two right ascension fields near $2^h$ and $20^h$.
   
      \item VLBA Calibrator survey 6 (VCS6), \citet{r:vcs6}, 
            VLBA BP133; 
            dual X/S bands; 
            3~segments; 
            since 2006.12.18 through 2007.01.11.
            Selection criteria: 
            1)~declination $>-30^\circ$, and
            2)~spectral index flatter than $-0.5$, and
            3)~flux density interpolated at 8.6~GHz using 
               data from multiple radio astronomy catalogues $>0.2$~Jy , or
            4)~intra-day variable sources observed in the framework of the 
               MASIV survey \citep{r:masiv} and flux density $>130$~mJy.
   
      \item VERA Galactic Plane Survey, PI: L.~Petrov; 
            VERA R07030A, R07100A; 
            K-band;
            2~segments;
            since 2007.01.30 through 2007.03.21.
            Selection criteria: 
            1)~either within $6^\circ$ of the Galactic plane, or
            2)~within $11^\circ$ of the Galactic center, or
            3)~within $2^\circ$ of a known maser source, and
            4)~detected in the VERA Fringe Search Survey \citep{r:vera22GHz}.
   
      \item LBA Calibrator Survey-1 (LCS--1), \citet{r:lcs1}; 
            LBA V230R, V254, V271AR, V271BR, V271CR; 
            X-band; 
            5~segments
            since 2008.02.05 through 2009.12.12.
            Selection criteria: 
            1)~declination $<-45^\circ$, and
            2)~flux density at 8.3~GHz $>$~150~mJy from 
               20~GHz AT20G \citep{r:at20g}, and
            3)~spectral index flatter than $-0.6$.
   
      \item Searching for candidate radio sources for the GAIA 
            astrometric link (OBRS-1), \citep{r:obrs1}; 
            VLBA+EVN GC030;
            1~segment;
            2008.03.07.
            Selection criteria: 
            1) cross-match of NVSS and the catalogue of quasars and active 
               nuclei \citep{r:vv13}, and
            2)~Bmag $< 18$, and
            3) $\delta>-40^\circ$.
   
      \item The EVN Galactic Plane Survey (EGaPS), \citep{r:egaps}; 
            EVN EP066; 
            K band;
            1~segment;
            2009.10.27.
            Selection criteria: 
            1)~galactic latitude $|b|<6^\circ$, and
            2)~declination $>-20\degr$, and
            3)~flux densities extrapolated to 22~GHz $>$ 80mJy using 
               at least two measurements above 2~GHz, and
            4)~spectral indices flatter than $-0.5$.

      \item Bessel Calibrator Search (BeSSel), \citep{r:bessel}; 
            VLBA BR145; 
            X-band;
            34~segments;
            since 2009.11.16 through 2010.08.29.
            Selection criteria: 
            1)~point-like sources from the NVSS (sizes $< 20''$), and 
            2)~present in CORNISH catalogs \citet{r:cornish} with flux 
               densities above 30 mJy, and
            3)~within circles with a radius of $1.5^\circ$ around the 
               109 target maser sources formed of the BeSSeL 
               program.\footnote{
            \web{https://www3.mpifr-bonn.mpg.de/staff/abrunthaler/BeSSeL/index.shtml}}

      \item Low Luminosity gamma-ray blazars, \citep{r:linford12};
            VLBA S2078, BT110; 
            C-band;
            7~segments;
            since 2009.11.22 through 2010.07.30.
            Selection criteria: 
            1)~present in the $\gamma$-ray Fermi Large Area Telescope 
               First Source Catalog 1FGL \citep{r:1fgl} and
            2)~brighter than 30~mJy at 8~GHz from CRATES catalogue 
               \citep{r:crates}, or 
            3)~detected with VIPS survey \citep{r:vips}.
   
      \item LBA Calibrator Survey--2 (LCS--2), \citep{r:lcs2}; 
            LBA V271DR, V271ER, V271F, V271G, V271H, V271I, V271J, V271K, V271L, V271M, V271N, V271O, V441, V493;
            X band;
            14~segments; 
            since 2010.03.11 through 2016.06.28.
            Selection criteria: 
            1)~declination $<-40^\circ$, and
            2)~spectral index flatter than -0.5, and
            3)~present in he Parkes quarter-Jansky survey 
               \citet{r:qjy} with flux density interpolated to 8~GHz $>0.2$~Jy, or 
            4)~present in ATC20G catalogue \citep{r:at20g} with flux density 
m              interpolated to 8~GHz $>0.15$ Jy, or 
            5)~present in the Parkes-MIT-NRAO (PMN) catalogue 
              \citep{r:pmn1,r:pmn2,r:pmn3,r:pmn4,r:pmn5,r:pmn6,r:pmn7} 
               with flux density interpolated to 8~GHz $>0.18$~Jy, or 
            6)~present in the Australia Telescope Parkes-MIT-NRAO ATPMN 
               \citep{r:atpmn} catalogue with flux density interpolated 
               to 8~GHz $>0.17$~Jy.

      \item Searching for candidate radio sources for the Gaia astrometric link and
            Global VLBI observations of weak sources (OBRS-2), \citep{r:obrs2}; 
            VLBA+EVN GC034,GB073; 
            7~segments;
            since 2010.03.23 through 2012.05.27.
            Selection criteria: 
            1)~declination $\delta>-40^\circ$, and
            2)~present either in NVSS, and 
            3)~present in the catalogue of quasars and active nuclei \citep{r:vv13}, and
            4)~Bmag $< 18$.
   
      \item A systematic search for inspiraling, binary, and recoiling black holes 
            in nearby galaxies (V2M), \citep{r:v2m};
            VLBA BC191, BC196, BC201; 
            X band;
            94~segments;
            since 2010.07.15 through 2012.06.05.
            Selection criteria: 
            1)~declination $>-40^\circ$, and
            2)~identified as a galaxy with K20fe $<$ 12.25~mag 
               from 2MASS catalogue \citep{r:2mass}, and
            3)~NVSS flux density $> 50$~mJy.
   
      \item 1FGL Active Galactic Nuclei at parsec scales, PI: Y. Kovalev; 
            VLBA S3111; 
            X-band;
            3~segments;
            since 2010.12.05 through 2011.01.09.
            Selection criteria: 
            1)~declination $>-40^\circ$ and
            2)~detected $\gamma$-ray emission with Fermi in 1FGL 
               \citep{r:1fgl} and associated with a radio source not observed
               before with VLBI.
   
      \item Bessel Calibrator Search follow-on, PI: M.~Reid; 
            VLBA BR149, BM317; 
            X-band;
            14~segments;
            since 2010.02.06 through 2013.08.04.
            Selection criterion: sources within $3^\circ$ of the 
            Galactic plane in a close distance to target sources
            of the BeSSeL program.

      \item VLBA Calibrator Densification 7 (VCS7), \citep{r:wfcs};
            VLBA BP171; 
            dual X/C bands; 
            17~segments; 
            since 2013.02.08 through 2013.08.01.
            Selection criteria: 
            1)~declinations $>-45^\circ$, and
            2)~flux densities extrapolated at 8~GHz $>0.1$~Jy, and
            3)~spectral index $> -0.55$, and
            4)~no planetary nebulae or HII region within $2'$.

      \item 2FGL Active Galactic Nuclei at Parsec Scales, PI: Y. Kovalev; 
            VLBA S4195; 
            X-band;
            3~segments;
            since 2013.05.07 through 2013.06.22.
            Selection criteria: 
            1)~declination $>-40^\circ$ and
            2)~detected $\gamma$-ray emission with Fermi in 2FGL 
               \citep{r:2fgl} and associated with a radio source.
   
      \item VLBI follow-up of Fermi sources, \citep{r:aofus2}; 
            VLBA S5272; 
            X-band, 
            4~segments; 
            since 2013.08.06 through 2013.12.05.
            Selection criteria: 
            1)~declination $-45^\circ$, and
            2)~flux density $>10$~mJy at 4.5 or 8.4 GHz, and
            3)~detected with Very Long Array (VLA) or Australia 
               Telescope Compact Array (ATCA) within the 95\% 
               localization error ellipse of $\gamma$-ray sources
               reported in 2FGL Fermi catalogue.
   
      \item VLBA Calibrator Densification 8 (VCS8), \citep{r:wfcs}; 
            VLBA BP177; 
            dual X/C bands;
            10~segments;
            since 2014.01.07 through 2014.02.23.
            Selection criteria: 
            1)~declinations $>-45^\circ$, and
            2)~flux densities extrapolated at 8~GHz and greater 150~mJy, and
            3)~spectral index $> -0.55$, and
            4)~no planetary nebulae or HII region with $2'$.
   
      \item VLBI Ecliptic band survey with the CVN (VEPS-1), \citep{r:veps1}; 
            CVN VEPS; 
            X band;
            17~segments;
            since 2015.02.13 through 2017.12.14.
            Selection criteria: 
            1)~ecliptic latitude $|\beta|<7.5^\circ$ and
            2)~present in GB6 and PMN catalogues with flux density 
               at 5~GHz $>-0.05$~Jy.
   
      \item 2FGL AGNs at parsec scales, 2nd survey, \citep{r:aofus2}; 
            VLBA BS241;
            X-band;
            7~segments;
            since 2015.02.16 through  2015.07.01.
            Selection criteria: 
            1)~declination $>-40^\circ$ and
            2)~detected $\gamma$-ray emission with Fermi in 2FGL 
               \citep{r:2fgl} and associated with a radio source.
   
      \item VLBA Calibrator Densification 9 (VCS9), \citep{r:wfcs}; 
            VLBA BP192;
            dual X/C bands;
            99~segments; 
            since 2015.08.07 through 2016.09.07.
            Selection criteria: 
            1)~declinations $>-40^\circ$ and
            2)~flux density at 4.8~GHz $>0.07$~Jy from GB6 or PMN
               catalogues.
   
      \item 3FGL at parsec scales, \citep{r:aofus3}; 
            VLBA S7104; 
            X-band;
            9~segments;
            since 2016.06.27 through 2016.07.26.
            Selection criteria: 
            1)~declinations $>-40^\circ$ and
            2)~detected with Very Long Array (VLA) or Australia 
               Telescope Compact Array (ATCA) within the 95\% 
               localization error ellipse of $\gamma$-ray sources 
               reported in 3FGL Fermi catalogue \citep{r:3fgl}, and
            3)~flux density at 5 or 9 GHz $>10$~mJy.
   
      \item Search for SOuthern Fermi Unassociated sources (SOFUS), PI: L.~Petrov;
            LBA SOFUS, V592, WARK1; 
            X-band;
            4~segments;
            since 2017.04.07 through 2021.05.08.
            Selection criteria: 
            1)~declinations $<-40^\circ$ and
            2)~detected with Australia Telescope Compact Array (ATCA) 
               within the 95\% localization error ellipse of $\gamma$-ray sources
               reported in 3FGL Fermi catalogue \citep{r:3fgl}
            3)~Flux density at 5 or 9 GHz $>10$~mJy.
   
      \item VLBA Survey of unassociated gamma-ray objects in the 7-year 
            Fermi/LAT catalog.
            PI: F. Schinzel, 
            VLBA BS262; 
            dual X/C bands;
            21~segments;
            since 2018.04.08 through 2018.07.24.
            Selection criteria: 
            1)~declinations $>-40^\circ$ and
            2)~detected with Very Long Array (VLA) within the 95\% localization
               error ellipse a $\gamma$-ray source reported in 
               4FGL Fermi catalogue \citep{r:fermi-4fgl}. Flux density at 
               5 or 9 GHz $>10$~mJy.
   
      \item VLBA Survey of unassociated gamma-ray objects in the 7-year Fermi/LAT catalog, 
            2nd survey.
            PI: F. Schinzel;
            VLBA SB072; 
            dual X/C bands;
            31~segments;
            since 2018.08.25 through 2019.02.17.
            Selection criteria: 
            1)~declinations $>-40^\circ$ and
            2)~detected with Very Long Array (VLA) within the 95\% localization
               error ellipse a $\gamma$-ray source reported in 
               4FGL Fermi catalogue \citep{r:fermi-4fgl}, and
            3)~flux density at 5 or 9 GHz $>10$~mJy.
   
      \item Study of the population of steep-spectrum compact radio sources, XC part (VCS10);
            VLBA BP242, BP245; 
            dual X/C bands;
            19~segments;
            since 2019.07.24 through 2020.02.11.
            Selection criteria: 
            1)~declinations $>-40^\circ$ and $<0^\circ$, and
            2)~all sources from AT20G not observed before with VLBI, regardless
               of their spectral index, or
            3)~flux density from GB6 and PMN catalogues $>0.07$~mJy and ecliptic 
               latitude $|\beta|<7.5^\circ$.
   
      \item A search for high-frequency calibrators within 10 degrees of 
            the Galactic center, PI: L.~Petrov;
            KVN N20LP01; 
            K and Q bands;
            14~segments;
            since 2020.03.05 through 2020.06.16.
            Selection criteria: 
            1)~angular distance to the Galactic center less than $10^\circ$, and
            2)~known objects detected with VLBI at 2--8~GHz but never 
               observed at 22/43 GHz, and
            3)~sources from VLASS \citep{r:vlass} with peak flux density 
               $>30$~mJy and the ratio total/peak flux density less than 
               1.5, and 
            4)~have never been observed with VLBI. 
   
      \item VLBA flux-limited Surveys of VLASS Fields --- Pilot Observations;
            PI: A. Beasley
            VLBA  BB409;
            C band;
            4~segments; 
            since 2020.05.20 through 2020.07.20.
            Selection criterion: detected with VLASS in three fields.
   
      \item Completion of Surveys for a Gravitational Lens Search to Explore Dark Matter 
            (VCS11), PI: T. Readhead; 
            VLBA BR235;
            18~segments; 
            since 2020.09.11 through 2021.02.16.
            Selection criteria: 
            1)~declinations $>-40^\circ$ and
            2)~sources from CLASS and CRATES catalogues with flux density 
               $>49.4$~mJy.

      \item Reaching completeness of the VLBI-selected AGN sample North of -40 deg (VCS12)
            PI: L.~Petrov
            VLBA BP252;
            dual X/C bands;
            53~segments; 
            since 2021.09.21 through 2022.12.02.
            Selection criteria: 
            1)~declinations $>-40^\circ$, and
            2)~flux density $>0.1$~Jy at 5~GHz from  GB6, or
            3)~flux density $>0.1$~Jy at 5~GHz from  PMN, or
            4)~flux density $>0.1$~Jy at 3~GHz from  VLASS.

\end{enumerate}
   \item Astrometric follow-ups:
   \begin{enumerate}\setcounter{enumii}{45}\setlength{\itemsep}{0.0ex}
      
      \item Regular geodesy with VLBA (RDV), \citep{r:rdv}; 
            VLBA RV, RDV, BE010, BF012, BF025, BF090, BP138, BR005, BR025, BW008, BW025, CN18, CN19, RDGEO, RDS, RDV, RDWAPS, RDWPS, TC001, BR, TC, BW, RDG, WAP, CN18, CN19; 
            dual X/S bands; 
            207~segments, 
            since 1994.07.08 through 2023.04.25.
   
      \item Dual X/S Astrometry Program, \citep{r:bf025};
            VLBA BF025; 
            dual X/S bands;
            2~segments;
            since 1997.01.10 through 1997.01.11.
   
      \item Investigation of residual systematic errors in dual-band linear 
            combinations of delays caused by the ionosphere, \citep{r:wfcs};
            VLBA BP175; 
            dual X/C-band;
            10~segments;
            since 2013.10.26 through 2013.12.26.
            Selection criteria: 
            1)~declination $>-40^\circ$ and
            2)~median correlated flux density $>0.2$~Jy at 8.4~GHz at baseline
               projection lengths longer than 5000~km.

      \item The second epoch VLBA Calibrator survey (VCS-II), \citep{r:vcs-ii};
            VLBA BG219;
            dual X/S bands; 
            9~segments; 
            since 2014.01.04 through 2015.03.17.
            Selection criterion: re-observations of the sources detected in 
            VCS1, VCS2, VCS3, VCS4, VCS5.
   
      \item VLBA Ecliptic Plane Survey (VEPS-V1), \citep{r:veps1}; 
            VLBA BS250; 
            dual X/S bands;
            4~segments;
            since 2016.03.22 through 2016.05.19.
            Selection criterion: re-observations of the sources detected in 
            the prior VLBI Ecliptic Band Survey with the CVN.
   
      \item The third epoch VLBA Calibrator survey (VCS-III), \citep{r:vcs-eyes};
            VLBA UF001;
            dual X/S bands;
            20~segments;
            since 2017.01.16 through 2017.10.21.
            Selection criterion: re-observations of the sources detected in 
            VCS1, VCS2, VCS3, VCS4, VCS5.
   
      \item Revealing milliarcsecond optical structure through VLBI observations 
            of Gaia detected AGNs at Southern Hemisphere, PI: L.~Petrov; 
            LBA V561; 
            dual X/S bands;
            2~segments;
            since 2017.06.16 through 2018.03.14.
            Selection criteria: 
            1)~detected in LCS--1 and LCS--2,  and
            2)~declinations $<-45^\circ$, and
            3)~correlated flux density at 8~GHz within the range 
               of [0.07, 0.3]~Jy, and
            4)~have a \Gaia counterpart, and
            4)~no prior X/S VLBI observations.
   
      \item SOuthern Astrometry Program (SOAP), PI: L.~Petrov; 
            LBA AUA, V515; 
            dual X/S bands;
            26~segments;
            since 2017.06.18 through  2019.12.04.
            Selection criteria: 
            1)~declinations $<-45^\circ$, and
            2)~correlated flux density at 8.4~GHz from prior VLBI 
               observations $>0.25$~Jy, and
            3)~no prior X/S VLBI observations.
   
      \item The fourth epoch VLBA Calibrator survey (VCS-IV), \citep{r:vcs-eyes};
            VLBA UG002; 
            dual X/S bands;
            24~segments;
            since 2018.01.18 through 2019.01.21.
            Selection criterion: re-observations of the sources detected in 
            VCS1, VCS2, VCS3, VCS4, VCS5.
   
      \item VLBA Ecliptic Plane Survey 2  (VEPS-3), PI: L.~Petrov; 
            CVN EPA; 
            dual X/S bands;
            2~segments;
            since 2018.01.24 through 2018.02.10.
            Selection criterion: re-observations of the sources detected in 
            the prior VLBI Ecliptic Band Survey with the CVN.

      \item VLBA Ecliptic Plane Survey 2  (VEPS-2), PI: F.~Shu; 
            VLBA BS264; 
            dual X/S bands;
            6~segments;
            since 2018.03.21 through 2018.06.15.
            Selection criterion: re-observations of the sources detected in 
            the prior VLBI Ecliptic Band Survey with the CVN.

      \item Probing milliarcsecond optical structure through VLBI 
            observations of Gaia detected AGNs, PI: L.~Petrov; 
            VLBA BP222, BP236;
            dual X/S bands;
            38~segments; 
            since 2018.05.15 through 2020.04.19.
            Selection criterion: re-observations of the sources with large
            offsets between VLBI and Gaia positions and with low 
            quality of their VLBI images.
   
      \item The Asian VLBI Galactic Plane Survey, PI: L.~Petrov; 
            EAVN AP001A; 
            K band;
            4~segments;
            since 2018.10.09 through 2019.01.28.
            Selection criteria: 
            1)~declinations $>-40^\circ$, and
            2)~Galactic plane defined as the region with galactic longitude
               $|l|< 15^\circ$ and galactic latitude $|b| < 12^\circ$ or 
               $|l| > 15^\circ$, and
            3)~VLBI positional accuracy worse than 0.5~mas, and
            4)~correlated flux density at any band within 4 to 24~GHz, and
            5)~detected with Gaia, or 
            6)~ecliptic latitude $\beta|<7.5^\circ$.

      \item The fifth epoch VLBA Calibrator survey (VCS-V), \citep{r:vcs-eyes};
            VLBA UG003; 
            dual X/S bands;
            26~segments;
            since 2019.01.27 through 2020.08.09.
            Selection criterion: re-observations of the sources detected in 
            VCS1, VCS2, VCS3, VCS4, VCS5.

      \item The sixth epoch VLBA Calibrator survey (VCS-VI), \citep{r:vcs-eyes};
            VLBA UH007; 
            dual X/S bands;
            28~segments;
            since 2020.09.18 through 2022.12.12.
            Selection criterion: re-observations of the sources detected in 
            VCS1, VCS2, VCS3, VCS4, VCS5.
   
      \item Study of the population of steep-spectrum compact radio sources, XS part (VCS10);
            VLBA BP245; 
            dual X/S bands;
            6~segments;
            since 2020.03.02 through 2020.03.23.
            Selection criteria: 
            1)~declination $>+75^\circ$, and
            2)~flux density $>0.2$~Jy from NVSS, and
            3)~no detection or weak detection 
               in the prior Norther Polar Cup Survey campaign. 
           
\end{enumerate}
   \item High frequency extensions:
   \begin{enumerate}\setcounter{enumii}{61}\setlength{\itemsep}{0.0ex}

      \item K/Q survey, \citep{r:kq_astro,r:kq_image};
            VLBA BR079, BL115, BL122, BL151, BL166; 
            X/K/Q bands;
            14~segments;
            since 2002.05.15 through 2011.02.05.
            Selection criterion: sources with correlated flux density
            at 8~GHz brighter than 0.3~Jy.
   
      \item K-band KVN calibrator survey, \citep{r:lee2017,r:lee2023};
            KVN N13JL01, S14TJ05, S14JL01; 
            K-band;
            7~segments;
            since 2013.09.04 through 2014.12.24.
            Selection criterion: sources with flux density $>0.2$~Jy 
            at 22~GHz and not previously detected with VLBI.
   
      \item Improving the K-band Celestial Reference Frame in 
            the North, \citep{r:witt23}; 
            VLBA BJ083; 
            K band;
            5~segments;
            since 2015.07.21 through 2016.06.20
            Selection criterion: sources that have been detected in prior
            22~GHz VLBI surveys;
   
      \item K-band EVN observations for geodesy and astrometry \citep{r:gomez23};
            EVN EL054, EC076;
            K band;
            2~segments;
            since 2016.06.15 through 2020.10.23
   
      \item Improving the K-band Celestial Reference Frame in the North, \citep{r:witt23}; 
            VLBA UD001; 
            K band;
            24~segments;
            since 2017.01.08 through 2018.07.22.
            Selection criterion: sources that have been detected in prior
            22~GHz VLBI surveys.
   
      \item Improving the K-band Celestial Reference Frame in the 
            North (the 2nd campaign), \citep{r:witt23};
            VLBA UD009; 
            K band;
            35~segments;
            since 2018.09.09 through 2021.06.12.
            Selection criterion: sources that have been detected in prior
            22~GHz VLBI surveys.
   
      \item Detection of the background position noise due to non-stationary 
            of the Galactic gravitational field,
            PI: L.~Petrov, 
            KVN GAJI; 
            K/Q bands;
            5~segments;
            since 2018.09.25 through 2018.12.29.
            Selection criteria:
            1)~declinations $>-40^\circ$, and
            2)~galactic latitude $|b| < 1.5^\circ$, and
            3)~galactic longitude $|l| < 20^\circ$.
   
      \item Asian K-band observations for geodesy and astrometry;
            PI: S. Xu;
            EAVN S20TJ, A20, A21, A22, A23;
            K band;
            9~segments;
            since 2020.11.05 through 2013.06.08.
   
      \item K- and Q-band VLBI Calibrators near the Galactic Center, PI: Y. Pihlstrom;
            VLBA BP251;
            K/Q bands;
            2~segments;
            since 2021.03.19 through 2021.04.15.
            Selection criteria: 
            1)~angular distance to the Galactic center less than $10^\circ$ and
            2)~detected in prior K/Q observations with KVN.

      \item Improving the K-band Celestial Reference Frame in the 
            North (the 3rd campaign);
            PI: A. de Witt; 
            VLBA UD015; 
            K band;
            18~segments;
            since 2021.07.26 through 2023.01.06.
            Selection criterion: sources that have been detected in prior
            22~GHz VLBI surveys.
   
      \item Further improving the K-band Celestial Reference Frame 
            in the North (the 4th campaign);
            PI: A. de Witt; 
            VLBA UD018; 
            K band;
            4~segments;
            since 2023.07.03 through 2023.07.24.
            Selection criterion: sources that have been detected in prior
            22~GHz VLBI surveys.
   
   \end{enumerate}
\end{enumerate}
}

Starting at 2013, pathfinder surveys switched to the upgraded wide
C-band receiver at VLBA because of a low level of radio interference 
at the time and its high sensitivity. It covers 4--8~GHz, and we call 
such observations dual-band X/C. The dual-band X/S and X/C data were 
used in a joint dual-band solution in this work.

   Table \ref{t:camp} presented in the Appendix shows the list of 
72 observing campaigns that we processed. Most of the observations 
were made at the Very Long Baseline Array (VLBA), which covers the 
declinations $>-40\degr$. Sources at declinations $<-40\degr$ were 
observed with the Australian Long Baseline Array (LBA). We used 
also data, mainly at 22 and 43~GHz, from the European VLBI 
Network (EVN), the East Asian VLBI Network (EAVN), the Korean 
VLBI Network (KVN), and the VLBI Exploration in Radio Astronomy 
array (VERA). 

\subsection{Geodetic VLBI experiments}

  The International VLBI Service for geodesy and astrometry  
\citep[IVS,][]{r:ivs} coordinates observing programs dedicated to geodesy. 
These programs are similar to astrometric programs. The main differences 
are schedule optimization, antenna sensitivity, and source selection.
There are two flavors of geodetic programs: 24~hr experiments that
typically involve 8--10 stations \citep{r:r1r4} and 1~hr experiments
dedicated for determination of the Earth orientation parameter UT1
that typically run at a single baseline 
\citep{r:ut1_int_1st,r:ut1_int_2nd}. There are $\sim\!$ 250 regular 24~hr 
geodetic sessions per year, i.e.\ on average, a geodetic experiment runs
every second day. The regular 1~hr observing sessions started on 
April 01, 1985. Their number gradually increased from 236 in 1988 to 
860 in 2023, i.e.\ on average, more than two such experiments per day 
run since the 2020s.

  Observing schedules of geodetic experiments are optimized to 
a determination of site positions and the Earth orientation parameters, 
which is almost orthogonal to optimization of astrometric programs. 
Antennas used in geodetic programs are less sensitive than those used in 
astronomy programs. The source list with rare exceptions is limited to 
50--100 frequently observed bright objects. Positions of these sources are 
determined so well in astrometric experiments that additional observations 
have virtually no impact. There were attempts to include additional sources
of interest to astrometry into schedules of geodetic experiments.
Although it was demonstrated that it is possible to determine positions
of several dozens of sources with a nanoradian level of accuracy
\citep{r:bail16}, these observations are not competitive with respect to
dedicated astrometric campaigns.

  We used all publicly available geodetic experiments in our work in order
to improve estimates of station positions and Earth orientation parameters.
The high density of geodetic observations helps to stabilize the global 
VLBI solution and allows us to make it fully self-consistent without the 
use of any external geodetic or astrometric information.

\subsection{Scheduling Observations}

  A VLBI schedule consists of a table with entries called scans that
for each station defines the start time, slewing to a program source, 
the start time for recording baseband data that are the digitized voltage 
samples from a receiver, and the scan end time. Upon completion of one scan, 
an antenna executes another scan. A campaign design sets a goal to observe 
sources from a given list in given number of scans at at least the minimum
number of stations with a given integration time per scan. If the number 
of scans per source is greater than one, additional 
requirements are set, such as the minimum time interval between 
observations of a given source or observing a source in the given minimum 
number of scans at the specified number of ranges at hour angle of the 
array reference antenna. Observing at different azimuths, elevations, and 
hour angles reduces systematic errors in estimates of source coordinates 
and improves the $uv$-coverage of program sources, which makes imaging more 
robust. Only a fraction of target sources in pathfinder surveys is 
detected at some baselines, and even a smaller fraction is detected at 
more than one half of the baselines. Therefore, in order to minimize losses 
of antenna time for observing sources that we cannot detect, pathfinder 
surveys observe target sources in one or two scans only.

A sequence of scans is generated with a specialized software.
It consecutively computes for each program source 
(i)~the number of antennas that see it above the physical horizon mask, 
(ii)~slewing time, 
(iii)~the likelihood that a given source can be visible at a given minimum 
number of stations in the future either during the current  observing session 
or during the entire campaign, and 
(iv)~the score that depends on all these factors. A scan with the 
highest score is selected for the schedule and the process is repeated. 
The algorithm for computing the final score is adjusted in such a way 
that the maximum number of sources is included into the schedule that 
satisfy the campaign design criteria, and the overall slewing time 
is close to the minimum.

  In addition to program sources, a schedule includes observations of
known strong sources that are considered calibrators. A common 
practice is to include every hour observations of blocks of four strong 
sources selected in such a way that at each station at least one of them 
is observed at low elevations, for instance $10\degr$--$30\degr$, and one 
source is observed at high elevations, say $45\degr$--$90\degr$. 
The purpose of including calibrators in survey observations is fourfold:
(1)~these sources are used as fringe-finders for initialization of the 
correlation process; (2)~these sources are used for computation of the 
complex bandpass calibration; (3)~these sources are used for improving 
separation of variables when estimating residual atmospheric path delay; 
and (4)~these sources are frequently observed in many other programs and 
therefore, provide a connection of coordinate estimates of the program
sources with the core sources that define the orientation of the coordinate 
system. In general, 10 to 25\% observing time is spent for observing 
calibrators. In addition, some telescope, like the Green Bank Telescope 
(GBT), require observing every 2--4 hours so-called pointing calibrators 
that are used for adjusting the pointing model. Phased VLA or ATCA require 
observing so-called phasing calibrators for adjusting phases of individual 
telescopes of the array in such a way that the phased arrays can be used 
for further processing as if it is a single telescope. High frequency 
surveys may require observations of planets for flux density calibration.

  Optimization of the observing schedule takes into account campaign design 
goals, placement of calibrators, and other constraints. It is a fairly 
complicated task that is performed by a specialized software 
\citep[see for more details]{r:wfcs,r:sch20,r:sch21}. A campaign consists 
of segments that are scheduled separately and run at different days. 
The scheduling procedure keeps records which sources were observed in prior 
surveys. For pathfinder surveys that are designed to have one observation 
per source, a source is removed from the list after putting it into a 
schedule. In order to facilitate optimization, the input source list has 
more sources than a campaign can observe. The oversubscription rate is 
a modest, 2--30\%, for follow-up surveys and large, a factor of 1.5 to 4, 
for pathfinder surveys. Because of that, the number of combinations of 
admissible observations that can fit a given time slot is very large. 
The scheduling process selects that combination of admissible scans that 
maximizes the metric of a given campaign, for instance, the total number 
of observed sources with a given minimum number of scans per source. 
A chance of a given source to be observed can be altered by assigning 
a weight to such a source that impacts the score calculation. This 
mechanism is used for fine-tuning the source selection process: the 
sources are split into several categories and their weights are assigned 
according to categories they belong.

  For some survey campaigns segment durations are fixed, and observing 
schedules are prepared in advance, while most of pathfinder campaigns after 
2010 with VLBA were scheduled dynamically. That means the array operator 
launches the schedule generation process by using a web form when the array 
has a gap between high priority programs that are more demanding to weather 
conditions and/or the required range of local sidereal 
time. The principal investigator of observing campaigns scheduled that way 
does not have direct control when and even whether a given source will be 
observed. But following that approach, more observing time can be allotted 
because otherwise, the array would have stayed idle.

\section{Data Analysis}
\label{s:data_anl}

  Radio telescopes synchronously track a sequence of radio sources.
VLBI hardware records voltage from receivers sampled at several 
intermediate frequencies (IFs) at rates from 8 to 256~Mbps at 2~bits 
per IF. Each recording block with a typical size of 5--32~kB is called 
a frame, and it has metadata that includes time stamps from a local 
hydrogen maser. The first stage of data analysis is performed by 
a correlator that computes time series of cross- and auto-correlation 
spectra with a resolution in the range from 15.6~KHz to 2~MHz 
averaged over the correlator accumulation time that is in the range from 
0.1 to 4~sec. The original raw VLBI data from radio telescopes are purged 
upon the initial quality control after correlation, since 
currently it is still not feasible to keep them because of their large 
volume. All astrometric voltage data reported here amount to 66~PB. Time 
averaging at the correlator reduces the output data volume to 146~TB. 
These data are kept at the data archives listed in the acknowledgment 
section indefinitely as a legacy of observing facilities. Three other 
stages of data processing are a)~visibility analysis that takes the 
correlator output as an input, and computes phase, group delays and 
visibilities averaged over frequency and time; b)~astrometric analysis 
that takes group delays as an input and adjusts for source coordinates 
and other nuisance parameters and flags for outliers; and c)~imaging data 
analysis that uses the time and frequency-averaged visibilities, as well 
as flags determined at the previous step and reconstructs source images. 
These steps are, in general, interdependent and at least one iteration 
is required between visibility analysis, astrometric analysis, and imaging.

\subsection{Correlation}

  Data acquisition terminals at stations split the input radio frequency
signal from radio telescope receivers into a number of subband
channels called here intermediate frequencies (IFs) and write them into disks 
or tapes with time tags from H-masers. The original records of voltage are 
played back at an array operating centers, shifted according to the a~priori 
model of path delay, correlated, and time averaged with either special 
hardware complex or software Mark-III, Mark-IV, S2, K3, K4, K5, VLBA, SFXC, 
Mitaka, KJCC, and DiFX \citep{r:difx1,r:difx2}. The correlator produces time 
series of auto- and cross- correlation of input data streams. These time 
series, augmented with auxiliary information, form a Level~1, or 
visibility, VLBI dataset.

  Time averaging and frequency resolution affect the field of view of the 
interferometer. The higher spectral resolution within an IF and finer time 
averaging, the wider the field of view. Once the correlator setup is made
for a given experiment, data are correlated, and the recording media 
with raw data is released, that choice is final, and it is not feasible to 
fix the correlator settings. Early hardware correlators were inflexible in 
setting spectral and time resolutions due to their architecture and they 
limited the output record rate. A common correlator setup provided a field of 
view of 5--$20''$. If an a~priori positional error was greater, time or 
frequency smearing would result in a reduction of interferometric signal.
Newer software correlators do not have that limitation, and imaging 
of the entire prime beam of telescopes became feasible, although that 
feature is not frequently used.

\subsection{Analysis of visibilities}

  Using visibility data, we evaluate residual phase and group delay, as well 
as their time derivatives. A given visibility, i.e.\ a constituent of the 
cross-spectrum averaged over a given interval of time called an accumulation
period and a given spectral range can be presented as a complex number.

  We treated observations that were used for absolute astrometry and for 
geodesy differently. We reprocessed all astrometric data at the visibility
level using our software \PIMA\! \citep{r:vgaps} that is a part of 
the Space Geodesy Data Analysis Software 
Suite\footnote{\web{See https://astrogeo.smce.nasa.gov/sgdass}} (SGDASS), 
but only a portion of geodetic data. For the remaining geodetic datasets 
we used in our solution group delays computed by the IVS correlation 
centers. 

  Source positions were determined using not the complex visibility data 
$v^\mathrm{o}_{ij}$ but derived quantities, group delays $\tau_\mathrm{g}$:
\beq
   v^\mathrm{o}_{ij} = g_i \; S_\mathrm{c} \; e^{-2\pi i
                    (  f_0 (\tau^\mathrm{o}_p - \tau^\mathrm{a}_p) \: + \:
                      (f_i - f_0) (\tau^\mathrm{o}_g - \tau^\mathrm{a}_g) \: + \:
                       f_0 (\dot{\tau}^\mathrm{o}_p - \dot{\tau}^\mathrm{a}_p) (t_j - t_0) \: + \:
                      (f_i - f_0) (\dot{\tau^\mathrm{o}_g} - \dot{\tau^\mathrm{a}_g}) (t_j - t_0) 
                    )},
\eeq{e:e1}
where $g_i$ is antenna gain, 
$S_\mathrm{c}$ is correlated flux density, $\tau_p$ is
phase delay, $f_i$ is frequency of the $i$-th spectral channel, $f_0$ is
the reference frequency, $t_j$ is time of $j$-th accumulation period, and
$t_0$ is the fringe reference time. Superscripts `o' and `a' denote
observed and a~priori delays and their rates, respectively. Quantities 
$\tau_p$, $\tau_g$, as well as their time derivatives, are evaluated in 
the fringe fitting procedure. The procedure of fringe fitting used in 
our work is described in detail in \citet{r:vgaps} and \citet{r:wfcs}. 
We only outline it here while making an emphasis on computation of group 
delay uncertainties.

  Since fringe visibility depends on path delay strongly nonlinearly, 
estimation of these quantities is performed in two steps. First, the sum of 
visibilities over all IFs, all spectral channels, and all accumulation periods 
is computed on a 2D grid of trial $\tau_\mathrm{g}$ and $\dot{\tau}_p$, and the 
maximum is sought. The location of the maximum and its magnitude is found by 
a parabolic fit using the element at the 2D grid that provides the maximum and
four adjacent elements. The group path delay and phase delay rate that 
correspond to the maximum are considered coarse estimates. 

  Then we compute the mean amplitude of $\sum\sum v^\mathrm{o}_{ij}$ when
no signal is present. We select randomly $N$ elements at the 2D grid used
for the coarse fringe search and put the sums of visibilities in an array.
$N$ is the minimum of 32768 and 1/4 of the total number of visibilities 
of a given observation. We sort this array in the ascending order, compute
the mean and root mean square (rms), and run an iterative outlier 
elimination procedure. We remove an element with the maximum amplitude, 
recompute the average and rms, and repeat the procedure till the maximum 
element is less than 3.5 times of the rms. This procedure cleans the noise 
array from a possible contamination with a signal from observed sources.  
The ratio of the amplitude of the maximum to the mean amplitude of noise 
we call a signal to noise ratio (SNR)\footnote{One may define SNR as the 
ratio of the maximum to the rms amplitude of the noise. The SNR defined 
that way is $\sqrt{2/\pi}$ of the SNR according to our definition.}.

  In order to get optimal estimate of group delay and phase delay rate,
as well as their uncertainties, the visibility data were transformed to
a form suitable for least square adjustment. The 2D array of visibilities
of a given observation is split into segments with an SNR at each segment
equal to approximately 1. We applied coarse estimates of group delays and 
phase delay rates denoted with superscript `c' to segmented visibilities,
i.e.\ we counter-rotated phases of visibilities averaged over a certain
range of time and frequency that we call a segment:
\beq
v^s_{lm} = \sum_{i=a}^{i=b} \sum_{j=c}^{j=d} \; v^\mathrm{o}_{ij} \;
           e^{2\pi i( (f_i - f_0) (\tau^\mathrm{c}_\mathrm{g} - \tau^\mathrm{a}_\mathrm{g}) \: + \:
           f_0 (\dot{\tau}^\mathrm{o}_p - \dot{\tau}^\mathrm{a}_p) (t_j - t_0)
           )}.
\eeq{e:e2}

  The phases of segmented visibilities were used for consecutive data
analysis. Based on the abovementioned estimates of the signal to noise 
ratios in segmented visibilities and assuming the real and image parts
of the noise in the visibilities are independent {\it within the 
segment} and normally distributed, we computed uncertainties of 
visibility phases. These uncertainties are the base of the ladder of 
error propagation to source positions.

  Then we used weighted least squares to find 
$\tau_\mathrm{g}$, $\dot{\tau}_p$, as well as their rates, using phases 
of segmented visibilities as observables and reciprocal phase 
uncertainties as initial weights. Visibility phases are considered 
uncorrelated. In general, systematic errors in the signal chain and in 
the propagation media affect visibility phases to a greater extent than
visibility phases and therefore, uncertainties in segmented visibility 
phases computed on the basis of SNR in amplitudes are underestimated. 
In order to account for the contribution of these factors that increase 
the phase scatter within a scan, we adjust initial weights by adding in 
quadrature an extra variance, which is constant within a scan. That makes 
the ratio of the weighted sum of residuals to their mathematical expectation 
close to unity. This procedure to a greater extent affects observations with 
high SNR, say $>30$, and makes uncertainties less unrealistic. At the same 
time, the additive reweighting technique cannot fully account for the 
impact of the systematic errors affecting group delay estimates since 
visibility phases are still considered uncorrelated.

\subsection{Analysis of group delays: dual-band, single-band, fused}

   Astrometric analysis of group delays involves two major steps: 
a)~computation of theoretical path delays and forming small differences 
between the observed and theoretical delays called o-c, as well as 
computation of partial derivatives of o-c over parameters, 
b)~preprocessing; and c)~parameter estimation using o-c as a right-hand 
side. We used our software VTD for computation of path delay and software
pSolve for astrometric analysis. They are components of SGDASS
package.

   Computation of group delay in general followed the so-called IERS 
Conventions \citep{r:iers2010}, with a number of improvements. We mention
here six of them. First, we applied site displacements for the atmospheric 
pressure loading, land water storage loading, tidal ocean loading, and 
non-tidal ocean loading on the observation level. The loading time series of 
3D displacements were taken from the International Mass Loading
Service \citep{r:malo15}. Second, we computed displacements caused by solid 
Earth tides using rigorous equations \citep[see][for details]{r:harpos}.
Third, we applied in data reduction a~priori slant path 
delays computed by a direct integration of equations of wave propagation 
through the heterogeneous atmosphere \citep{r:padel} using the output of 
NASA numerical weather model GEOS-FPIT \citep{r:geos18}. Fourth, we 
modeled the contribution of Galactic aberration rate to path delay.
We used the distance to the Galactic center and the velocity of the Sun 
with respect to the Galactic center 8.34~kpc and 255.2~km/s respectively 
according to \citet{r:reid14}. This gives us the acceleration toward the 
Galactic center $\flo{2.531}{-10}\ \rm{m/s}^2$, which corresponds to the 
$5.49~\mu$as/yr annual change of the Galactic aberration. Fifth, we applied 
the data reduction for parallax for known radio stars that have parallaxes 
determined with \Gaia and published in the Early Data Release~3 
\citep{r:gaia_edr3} and for Sgr\,A${}^\star$ determined from 
dedicated differential VLBI observations \citep{r:oyama24}. Sixth, we 
included in our data reduction the ionospheric contribution computed from 
the GNSS global ionospheric model CODE \citep{r:schaer99} with important 
modifications: elevation for the ionospheric mapping function was scaled by 
0.9782, the nominal height of the ionosphere was increased by 56.7~km, and 
the total electron content (TEC) was scaled by 0.85. A thorough discussion 
of the impact of these modifications is given in \citet{r:sba}. 

  The contribution of the ionospheric path delay can be expanded into series
of frequency. For accounting for path delay in the ionosphere, it is 
sufficient to retain only one term that is reciprocal to the square of the 
effective frequency. As it was shown by \citet{r:iono2nd}, the impact of 
a higher order of expansion on the group delay, namely proportional to $f^{-3}$, 
does not exceed several picoseconds and is not detectable. The impact of 
the ionosphere on group delay is almost entirely eliminated if to observe 
simultaneously at two or more widely separated frequency bands. The following 
linear combination of two group delays at the upper and lower bands, 
$\tau_{u}$ and $\tau_{l}$, respectively, is ionosphere free:
\beq
  \tau_{\rm if} = \Frac{f_\mathrm{u}^2}{f_\mathrm{u}^2 - f_\mathrm{l}^2} \, \tau_\mathrm{u} -
                  \Frac{f_\mathrm{l}^2}{f_\mathrm{u}^2 - f_\mathrm{l}^2} \, \tau_\mathrm{l}.
\eeq{e:e3}
  Here $f_\mathrm{u}$ and $f_\mathrm{l}$ are effective ionospheric 
frequencies at the upper and lower bands, respectively. We do not apply 
the contribution of the ionosphere to group delay from the GNSS global 
ionospheric model when we process dual-band observations.

  The downside of this approach is that the uncertainty of an ionosphere-free 
group delay is greater than uncertainties of $\tau_\mathrm{u}$ and 
$\tau_\mathrm{l}$. For instance, when observations are made at 2.2 and 8.4~GHz 
and uncertainties of group delays at both bands are the same, the uncertainty 
of $\tau_{\rm if}$ is increased by a factor of 1.08. When observations are 
made at 4.3/7.6~GHz, the uncertainty of the ionosphere free path delay is 
increased by a factor of 1.56. However, there are two cases when the ionosphere 
free path delays cannot be used. 

  First, an experiment can use only one band by design. In that case we
apply the a~priori contribution to group delay from the GNSS global 
ionospheric model. However, the model accounts only for a part of the 
contribution. In \citet{r:sba} we performed a detailed study and evaluated 
the residual errors of the ionospheric contribution. We found that the rms of 
the residual errors $\sigma_{\rm rr}$ can be represented as the 
following regression through the scatter of total ionospheric path 
delay at a given baseline:
\beq
  \sigma_{\rm rr}(f,e) = \biggl(\Frac{f_{\rm 8GHz}}{f}\biggr)^{\!2} \; 
                         \sum_{k=-2}^{k=n-1} c_k \, B^3_k(\sigma(\tau_{\rm gt})) \, 
                         \sqrt{M^2(e_1) + M^2(e_2)}.
\eeq{e:e4}

  Here $f$ is teh frequency in GHz, $e$ is the elevation, $M(e)$ is the 
ionospheric mapping function that describes the elevation dependence of 
the ionospheric model, $B^3_k(x)$ is the B-spline function of the 3rd 
degree with the pivotal knot $k$, and $\sigma(\tau_{\rm gt})$ is the rms 
of the total ionospheric path delay. For computation of 
$\sigma(\tau_{\rm gt})$ we calculated the coordinates of 16384 points 
uniformly distributed over the celestial sphere using a random number 
generator. Then for each baseline and each time epoch of a given 
VLBI experiment, azimuth and elevation angles of those points, $A_i$, 
and $e_i$, are computed at both stations of the baseline. If elevations above 
the horizon are greater than $5^\circ$ at both stations, that point is 
selected for further computations. If not, the next point is drawn. Then the 
total ionospheric path delay $\tau_i(A_1,e_1,A_2,e_2)$ is computed using 
the GNSS TEC maps. We normalize it by dividing by the mean mapping function 
$\tilde{M} = (M(e_1) + M(e_2))/2$. The process is repeated for 1440 time 
epochs that cover the time interval of a given VLBI experiment under 
consideration with a step of 1~minute. Then for each baseline we computed 
$\sigma(\tau_{\rm gt})$ over this time series of 1440 normalized $\tau_i$ 
values. Validation of this regression model, as well as the values of 
numerical coefficients, can be found in \citet{r:sba}. The uncertainty of 
group delay $\sigma_{\rm rr}(f,e)$ is added in quadrature to the 
uncertainty of group delay determined by the fringe fitting procedure. 

  As it was shown in \citet{r:sba}, the use of GNSS global ionospheric model
to account for ionospheric path delay causes a declination-dependent 
bias in declination that can reach 0.4~mas at 8~GHz. The origin of this bias
is an oversimplification of the dependence of the total electron content
with height as a thin shell layer for computation of the GNSS ionospheric 
models. 
  
  Second, there are situations when fringes can be detected for a given 
observation only at one band in a dual-band VLBI experiment. Usually, 
only a small fraction of dual-band observations is affected: from 2 to 20\%. 
For the remaining observations we compute the ionospheric path delay 
$\tau^v_i$  from VLBI group delay at the upper and lower frequency bands. 
We represent this ionospheric delay at stations $j$ and $k$ as:
\beq
      \tau^v_i(t) = b_j(t) - b_k(t) \: + \:
                   \Frac{e^2}{ 8\, \pi^2 \, c \, m_e \,  \epsilon_o }
                   \Frac{1}{f_u^2} 
                   \biggl( \Bigl(\TEC_j(\phi_j,\lambda_j,t) + a_j(t) \Bigr) M(e_j) \: - \: 
                           \Bigl(\TEC_k(\phi_k,\lambda_k,t) + a_k(t) \Bigr) M(e_k) \biggr),
\eeq{e:e5}
   where TEC is the total electron content from the GNSS global ionosphere
model, $b_j(t) = \dss\sum_{i=1}^{i=n} b_{ij} \, B^0_i(t) $ is a delay bias 
expanded over the B-spline basis of the 0th degree, 
$a_j(t) = \sum_{i=-2}^{i=n-1} a_{ij} \, B^3_i(t) $ is the TEC bias 
expanded over the B-spline basis of the 3rd degree, $\phi, \lambda$ are 
coordinates of the ionosphere piercing point, which depend on positions of 
observing stations, as well as on azimuths and elevations of observed 
sources, $e$ is the charge of the electron, $m_e$ is the mass of the
electron, $\epsilon_o$ is the permittivity of free space, and $c$ is the 
velocity of light in vacuum. The clock bias has jumps at epochs of clock 
discontinuities. If clock had no jumps in a given experiment, $b_j$ is 
a constant. The value $a_j(t)$ describes the time variable bias in the 
total electron content from the GNSS global ionospheric map.

  We estimated coefficients $a_{ij}$ and $b_{ij}$ in a single weighted least 
squares run for each dual-band astrometric VLBI experiment. Weights were chosen
to be reciprocal to the uncertainty of the ionospheric contribution from VLBI
dual-band group delays with a floor 12~ps added in quadrature. The model in
equation~\ref{e:e5} is an approximation, and its use causes systematic errors.
That floor was added to accommodate systematic errors of the model and avoid 
the dominance of few observations with small uncertainties in the solution. 
The B-spline knots for modeling $a_{ij}$ were selected with spans equal 
to 900~s. Constraints on $a_{ij}$, its first and second time derivatives with 
reciprocal weights $\flo{5}{-10}$~s, $\flo{4}{-14}$, and $\flo{2}{-18} s^{-1}$
respectively, were imposed in order to stabilize the solution when there are 
too few observations at some intervals and to enforce the continuity of TEC 
bias evolution with time. Using estimates of $a_{ij}$ and $b_{ij}$ 
coefficients, we computed $\tau^m_i(t)$ following equation~\ref{e:e5}. Invoking 
the law of error propagation, we computed the full covariance matrix of 
$a_{ij}$ and $b_{ij}$, and then using that full covariance matrix we computed
the uncertainty of $\tau^m_i(t)$, scaled it by the empirical fudge factor of
0.889, and added in quadrature that uncertainty to the of group delay 
uncertainty. The empirical scaling factor of 0.889 was found by comparison of 
modeled $\tau^m_i(t)$ with observed $\tau^m_i(t)$ using a dataset of 4 million 
observations.  The validation procedure for computation of $\tau^m_i(t)$ 
is described in full detail in \citet{r:sba}.

  This approach allows us to treat dual-band and single-band data uniformly.
When processing single-band delay experiments, we computed ionospheric path
delays using the GNSS global ionospheric model, as well as their 
uncertainties, from the regression expression~\ref{e:e4}. When processing 
dual-band delay experiments, we used ionosphere-free linear combinations 
of group delays at lower and upper bands when observables at both bands 
were available. If observables only at one band were available, we computed 
the ionospheric contribution $\tau^m_i$ from coefficients $a_{ij}$ and 
$b_{ij}$ evaluated by processing dual-band data of that experiment, and 
computed the uncertainty of $\tau^m_i$ using the full covariance matrix.
We call a dataset with a mixture of dual-band and single-band data of a 
given experiment ``fused.'' The summary of fused delay computation 
is presented in Table~\label{t:fused_summery}. This is a new technique. 
The validity of this approach hinges on the accuracy of $\tau^m_i$ and its 
uncertainty and on a lack of significant biases based on comparison of 
results from processing fused data when some dual-band data were 
artificially treated as single-band data against the reference solution. 
In our prior papers \citep{r:sba,r:radf} we presented an in depth 
investigation of this approach and its validation.

\begin{table}[h]
    \Note{
    \caption{Summary of generation of fused observables and their 
             uncertainties. }
    \newcommand{\LP}{\bigl(}
    \newcommand{\RP}{\bigr)}
    \hspace{-3em}
    \begin{tabular}{l l l l}
        \hline
        \nntab{c}{Available band} & & \\
        upper & lower & \ntab{c}{$\tau_{\rm fused}$}           & 
                        \ntab{c}{$\sigma^2(\tau_{\rm fused})$} \\
        \hline
           yes  &  yes & $\kappa_1 \tau_u - \kappa_2 \tau_l$ & 
                         $\kappa^2_1 \sigma^2(\tau_u) + \kappa^2_2 \sigma^2(\tau_l)$  \\
           yes  &  no  & $\tau_u + \kappa_3/f^2_u \LP(TEC_1 + a_1) \, M(e_1) - 
                                                     (TEC_2 + a_2) \, M(e_2)\RP $ & 
                         $\sigma^2\tau_u + \kappa^2_3/f^4_u \LP\sigma^2 a_1  \, M^2(e_1) + 
                                                               \sigma^2 a_2  \, M^2(e_2)\RP$ \\
           no   &  yes & $\tau_l + \kappa_3/f^2_l \LP(TEC_1 + a_1) \, M(e_1) - 
                                                     (TEC_2 + a_2) \, M(e_2)\RP$ & 
                         $\sigma^2\tau_l + \kappa^2_3/f^4_l \LP\sigma^2 a_1  \, M^2(e_1) + 
                                                               \sigma^2 a_2  \, M^2(e_2)\RP$ \\
           no   &  no   & not used & not applicable \vex\vex \\
           only & ---  & $ \tau_u + \kappa_3/f^2_u \LP TEC_1 \, M(e_1) - 
                                                       TEC_2 \, M(e_2)\RP $ & 
                         $ \sigma^2\tau_u + \sigma^2_{rr}(f_u) $ \\
        \hline
    \end{tabular}
    \tablecomments{ The first four rows describe a case when fused 
             data are generated from a dual-band experiment depending on 
             which band data are available for a given observation. 
             The fifth row describes a case of a single-band experiment.
             $\kappa_1 = \Frac{f_\mathrm{u}^2}{f_\mathrm{u}^2 - f_\mathrm{l}^2}, \enskip
             \kappa_2 = \Frac{f_\mathrm{l}^2}{f_\mathrm{u}^2 - f_\mathrm{l}^2}, \enskip 
             \kappa_3 = \Frac{e^2}{ 8\, \pi^2 \, c \, m_e \,  \epsilon_o }$
             }
    }
    \label{t:fused_summery}
\end{table}

\subsection{Parameter estimation}
\label{s:pe}

  We estimated parameters in a single least squares run using a given 
dataset. The total number of estimated  parameters exceeds 5 million
in the fused solution. Inversion of a normal matrix of this size 
is possible through partitioning. We used three partitioning classes: 
global parameters that were estimated using the entire dataset, local 
parameters that were estimated for each observing session, and segmented 
parameters that were estimated for each station for an interval of time 
that is shorter than an observing session. 

  The parametric model included estimation of the following segmented
parameters: 

\par\vspace{-1ex}\par
\begin{itemize}\setlength{\itemsep}{-0.5ex}
  \item clock function, except for the reference station;
  \item residual atmospheric path delay in zenith direction; 
  \item the tilt of the symmetry axis of the refractivity field, 
        also known as atmospheric gradients. 
\end{itemize}

  These parameters were modeled as an expansion over the B-spline basis of 
the 1st degree. The span between knots was 60~minutes for clock function, 
20~minutes for the atmospheric path delay in zenith direction, and 6~hours 
for tilt angles. 

  The parametric model included estimation of the following local
parameters: 

\par\vspace{-1ex}\par
\begin{itemize}\setlength{\itemsep}{-0.1ex}
  \item baseline-dependent clock;
  \item UT1, polar motion, and their time derivatives for experiments
        prior January 01, 1990.
\end{itemize}

  Weak stabilizing constraints were imposed on UT1, polar motion, and their
rates: $\flo{2.18}{-7}$~rad and $\flo{2.53}{-12}$ rad/s. These 
constraints allowed to process experiments that use a single baseline or
experiments that had too few observations at some stations, or experiments
that had only short baselines.

  The parametric model included estimation of the following global
parameters:.
\par\vspace{-1ex}\par
\begin{itemize}\setlength{\itemsep}{-0.1ex}
   \item Positions of all the stations at the reference epoch 2000.01.01.

   \item Linear velocities of all the stations.

   \item Antenna axis offsets of 104 stations.

   \item Sine and cosine components of harmonic site position variations
         at diurnal, semi-diurnal, annual, and semi-annual frequencies
         of 69 stations with a long history of observations. This technique
         is described in \citet{r:harpos} in detail. Estimation of harmonic
         variations allows to mitigate remaining systematic errors, for
         instance, the impact of thermal variations.

   \item B-spline coefficients that model the nonlinear motion of 
         29 stations. The nonlinear motion includes sudden co-seismic 
         position changes at {\sc fortords, gilcreek, kashim11, kashim34, 
         koganei, mk-vlba, miura, mojave12, presidio, sintotu3, sourdogh, 
         tateyama, tigoconc, tsukub32, usuda64, veramzsw, wark12m, whthorse, 
         yakataga}, smooth post-seismic relaxation at {\sc gilcreek, 
         kashim11, kashim34, tigoconc, tsukub32, sintotu3, veramzsw}, 
         nonlinear change of the antenna tilt at {\sc pietown}, nonlinear 
         uplift due to glaciers melting at {\sc nyales20}, nonlinear local 
         motion of {\sc hras\_085}, and discontinuities due to station repair 
         at {\sc dss15, dss65, eflsberg, ggao7108, medicina, sintotu3, 
         tsukub32, urumqi}, and {\sc yebes40m}. The degree and placements 
         of B-spline knots varied. Some knots were multiple to describe 
         discontinuities in positions. The optimal placements of B-spline 
         knots was determined by a trial. In order to assess the validity 
         of the estimation model, we performed a residual solution where we 
         used the estimates of the B-spline coefficients, as well as 
         estimates of sine and cosine components of harmonic position 
         variations as a~priori. We estimated station positions for each 
         experiment independently in the residual solution and examined 
         time series of baseline lengths for the presence of residual 
         discontinuities and non-liner motions.

   \item Coefficients of the empirical model of the perturbational Earth 
         rotation vector $\vec{q}_e(t)$ with respect to the a~priori model. 
         In the framework of this formalism, a station vector in the 
         co-rotating terrestrial coordinate system $\vec{r}_{{}_T}$ is 
         related to a vector in the inertial celestial coordinate system 
         $\vec{r}_{{}_C}$ from the a~priori Earth rotation matrix 
         ${\mathstrut\cal M}_a(t)$ and a small vector of the perturbational 
         rotation $\vec{q}_e(t)$ as
         \beq
           \vec{r}_{{}_C} = \widehat{\mathstrut\cal M}_a(t) \, \vec{r}_{{}_T} +
                            \vec{q}_e(t) \times \vec{r}_{{}_T}.
         \eeq{e:e6}

         We model vector $\vec{q}_e(t)$ in the terrestrial coordinate system 
         as a sum of the coefficients of a B-spline that describe the slow 
         constituents in the Earth rotation, coefficients of harmonic variations 
         in the Earth rotation at periods 32 hours and shorter, and a
         cross-term $t\times\sin(t)$, $t\times\cos(t)$ for one harmonic:
         \beq
            \vec{q}_e(t) = \left(
              \begin{array}{ l@{\;} l@{\;} l l@{\;} l@{\;} l}
                \displaystyle\sum_{k=-2}^{n-1} e_{1k} \, B_k^3(t) \:&  + &
                \displaystyle\sum_{j}^{N} \left( P^c_{j} \cos \omega_m \, t \: + \:
                                                 P^s_{j} \sin \omega_j \, t \right)   
                & + & t \, \left( S^c \cos \! - \Omega_n \, t \: + \:
                                  S^s \sin \! - \Omega_n \, t \right) 
                \vspace{0.5ex} \vspace{2ex} \\
                \displaystyle\sum_{k=-2}^{n-1} e_{2k} \, B_k^3(t) \: & + & 
                \displaystyle\sum_{j=1}^{N} \left( P^c_{j} \sin \omega_j \, t \: - \:
                                      P^s_{j} \cos \omega_j \, t \right)
                & + & t \, \left( S^c \sin \! - \Omega_n \, t \: - \:
                                  S^s \cos \! - \Omega_n \, t \right)
                \vspace{0.5ex} \vspace{2ex} \\
                \displaystyle\sum_{k=-2}^{n-1} e_{3k} \, B_k^3(t) \: & + &
                \displaystyle\sum_{j=1}^{N} \left( E^c_{j} \cos \omega_j \, t + 
                                      E^s_{j} \sin \omega_j \, t \right) 
                \\
              \end{array}
              \right),
         \eeq{e:7}
         where $B_k^m(t)$ is the B-spline function of degree $m$ determined 
         at a sequence of knots $ t_{1-m}, \, t_{2-m}, \, \ldots, \, t_0, 
         \, t_1, \, \ldots \, t_k$; $\omega_j$ are the frequencies of external 
         forces; the coefficients $e_{ik}, P^c_j, P^s_j, S^c, S^s, E^c_{j}, 
         E^s_{j}$; are the parameters of the expansion; and $\Omega_n$ is the 
         nominal frequency of the Earth's rotation. Here $n$ is the dimension of 
         the B-spline basis and $N$ is the dimension of the Fourier basis. 

         We estimated harmonic variations at 877 frequencies. That included:
         \par\vspace{-1ex}\par
         \begin{enumerate}
             \item all the frequencies with nutation amplitude exceeding 
                   10~prad from the REN2000 rigid Earth nutation series 
                   \citep{r:ren2000a,r:ren2000b,r:ren2000c} to model 
                   nutation ($q_1$, and $q_2$ only);

             \item all harmonics of the tide generating potential 
                   \citep{r:hw95} with amplitudes greater than 0.002 of the 
                   $M_2$ tide amplitude to model all tidal variations in 
                   polar motion and UT1, except zonal tides;

             \item 31 frequencies in the range of $\flo{-7.31149}{-5}$ to
                   $\flo{-7.29622}{-5}$ rad/s with a step of 
                   $f_s = 2\pi/\Delta t = \flo{4.9273}{-9}$ rad/s to model 
                   the retrograde free core nutation ($q_1$, and $q_2$ only),
                   where $\Delta t$ is the interval of time for estimation
                   of harmonic variations in the Earth orientation parameters,
                   40.7 years;

             \item 19 frequencies in the range of $\flo{-7.28370}{-5}$ to 
                   $\flo{-7.27385}{-5}$ rad/s with a step of 
                   $f_s$ rad/s to model the prograde
                   free inner core nutation ($q_1$, and $q_2$ only);

             \item 47 frequencies in the range of $\flo{-2.188774}{-5}$ to 
                   $\flo{-2.188774}{-5}$ rad/s, as well as 
                   $\flo{2.188774}{-5}$ to $\flo{2.188774}{-5}$ rad/s
                   with a step of $f_s$ to model 
                   prograde and retrograde variations within the 
                   ter-diurnal frequency band with a broadened spectrum
                   due to seasonal modulations;

             \item 47 frequencies in the range of $\flo{-2.915706}{-5}$ to 
                   $\flo{-2.917986}{-5}$ rad/s, as well as 
                   $\flo{2.915706}{-5}$ to $\flo{2.917986}{-5}$ 
                   rad/s with a step of $f_s$ to model 
                   prograde and retrograde variations within the 
                   quad-diurnal frequency band with a broadened spectrum 
                   due to seasonal modulations.

         \end{enumerate}

         The knot sequence of the B-spline basis used for modeling
         $\vec{q}_e(t)$ spanned time interval January 01, 1990 through
         September 05, 2024 with a step of 2 days. This technique is described
         in full detail in \citet{r:erm}.

   \item Positions of all the sources with at least three usable 
         observations.

   \item Proper motions of Galactic sources that had a lest two epochs:
         radio stars and Sgr\,A${}^\star$. The reference epoch for source
         positions of these sources was 2016.01.01.
\end{itemize}

  This parameter estimation model allowed us to adjust all parameters
in a single least squares run. We used all VLBI data: 24~hr astrometric
experiments, 3--8~hr survey style astrometric experiments, 24~hr
geodetic experiments, and 1-hr so-called intensive experiments dedicated
to estimation of UT1. These experiments were optimized for different
goals. Many astrometric experiments, especially surveys, are not well 
suitable for estimation of the Earth orientation parameters and station 
positions. Treating these parameters as local, i.e.\ estimating them in 
each experiment, as it was often made in the past \citep[see, for 
example,][] {r:dia23}, causes solution instabilities due to the 
cross-talk  between radio source positions and the Earth orientation 
parameters. This adds a jitter in source position estimates. 

  Using the a~priori nonlinear motion and the Earth orientation parameter 
series from external results that are regarded reputable in the geodetic 
community, opens a door for propagation of errors from these results to 
our solution. For instance, the ITRF2020 solution endorsed by the 
International Union of Geodesy and Geophysics that recommends its use 
as a standard, does not model the nonlinear motion of VLBA radio telescope 
{\sc pietown} that was known for decades \citep{r:rdv}. Omission of 
peculiar motion of {\sc pietown} causes noticeable errors in source 
positions. Our careful analysis reveals noticeable errors in 
IERS time series of the Earth orientation parameters in 1997--2000. 
In addition to errors in the external solutions that can be identified
during quality control, the use of external solutions introduces biases 
due to model inconsistencies. For instance, the IERS time series is the 
smoothed weighted mean of solutions from different analysis centers that 
either use different mass loading models or do not use them at all.

\begin{figure}
   \centerline{\includegraphics[width=0.61\textwidth]{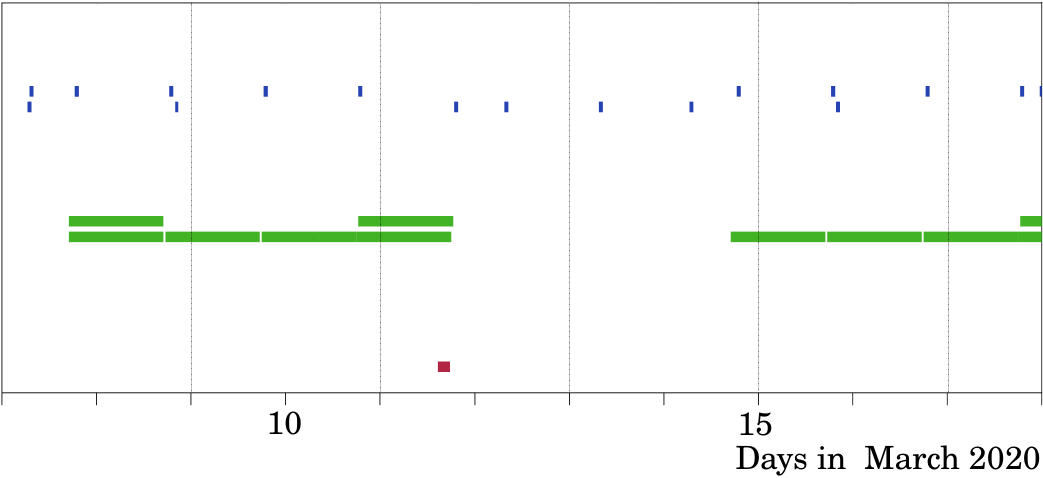}}
   \caption{Time allocation for VLBI experiments for 10 days in March 2020.
            Upper (blue): 1-hr IVS Intensive experiments. Middle (green):
            24-hr IVS geodetic experiments. Bottom (red): an astrometric 
            experiment. Thin vertical lines correspond to epochs of 
            B-spline knots.
           }
   \label{f:dens_eop}
\end{figure}

  Estimating B-spline coefficients directly using all the 
observations, we have implemented an assimilation scheme. The B-spline of 
the 3rd degree is sensitive to data within three span intervals before and 
after a given epoch, i.e.\ within an interval of 14 days. All data, 
astrometric, 24~hr geodetic, and 1~hr geodetic contribute to estimates of 
the Earth rotation. As we can see in Figure~\ref{f:dens_eop}, the density
of geodetic data is high, and there is a number of overlaps. The density 
of geodetic data was low in the 1980s, which is not always sufficient 
for the use of the assimilation scheme. Therefore, we started B-spline in 
1990.01.01 and estimated the Earth orientation parameters in the
old-fashion geodetic style as local parameters prior that date. Data 
prior to 1990 have a negligible direct impact on source position estimates. 
We included data since 1984.01.04, when geodetic VLBI observations 
became regular to make estimates of station positions more stable.

  Estimating harmonic variations in the Earth orientation parameters,
we eliminated the need to adjust nutation daily offsets 
introduced by \citet{r:her86a} as a temporary measure for processing 
datasets that were shorter than the 18.6~year principle nutation term.
The retrograde free core nutation and putative prograde 
inner core free nutation are not harmonic processes, but we assume
they are band-limited. To account for their contribution, we just
sampled them within their bands at the Nyquist frequency $2\pi/T$, 
where $T$ is the total interval of observations, 40.7 years.
Thus, instead of estimating the time series of free core nutation,
we directly estimated their spectrum, together with known spectral
constituents of forced nutations, variations in the Earth rotation
caused by ocean tides and other ad hoc harmonic processes.

\subsection{Constraints used in the solution}

  We used constraints of three types in our solutions: identifying 
constraints that eliminate a degeneracy in observation equations,
decorrelation constraints that substantially reduce correlations
between parameters, and weak constraints that stabilize estimation 
of some parameters when there are not enough data to provide a reliable
estimate. Constraints are equations that do not originate from 
observations: they augment them. The choice of constraints and their 
weights is up to some degree subjective. Since results depend on 
constraints, we disclose below all constraints that we have imposed 
in full.

\subsubsection{Identifying constraints}

  The equation for path delay $\tau$ of a two element interferometer
with stations $\vec{r}_1$ and $\vec{r}_2$ in the terrestrial coordinate
system observing a source with a unit vector $\vec{s}$ in the inertial 
celestial coordinate system can be simplified to a from:
\beq
   \tau = \Frac{1}{c} \mat{E} (\vec{r}_1 - \vec{r}_2) \cdot \vec{s} + O(c^{-2}),
\eeq{e:e8}
   where $\mat{E}$ is the Earth rotation matrix. We can see immediately
that equation~\ref{e:e8} is invariant with respect to a translation of
the coordinate system, as well as to its time derivatives. Rotating 
the terrestrial coordinate system by matrix $\mat{P}_t$ is equivalent
to a replacement of the Earth rotation matrix $\mat{E}$ with 
$\mat{E} \, \ttra{\mat{P}_t}$. Therefore, path delay is invariant with 
respect to a rotation of the terrestrial coordinate system and its time 
derivative. Rotating the celestial coordinate system by matrix 
$\mat{P}_s$ is equivalent to a replacement of the Earth rotation matrix 
$\mat{E}$ with $\ttra{\mat{P}_s} \, \mat{E}$. Therefore, path delay is 
invariant with respect to a rotation of the celestial coordinate system. 
When we estimate positions of all the stations, velocities of all the 
station, and coordinates of all the sources, there are five invariant 
conditions with 15 degrees of freedom. Observations themselves does not 
provide enough data and cannot provide them in principle to determine 
a unique solution. In order to invert the normal matrix, we have to 
augment the system of equations with 15 constraint equations that 
identify the solutions. These constraints make the rank of the system of 
equations equal to the number of parameters. Using the same logic, we 
conclude that the specific choice of identifying constraints that makes 
the system of equations invertible is not important, since any choice 
satisfies observations in exactly the same way.

  We formulated identifying constraint by requiring that the net translation 
of station positions, net translation of station velocities, the mean of 
the Earth orientation parameters, their rate of change, and the net rotation 
of source positions be equal to some right-hand side vector that is not 
necessarily zero.

  Constraint equations in a general form can be written as 
\beq
   C \, (x_e - x_r) = R,
\eeq{e:e9}
   where $C$ is a matrix of constraint equations, $x_e$ is the vector of 
estimates of constrained parameters, $x_r$ is the vector of reference values 
of those parameters, and $R$ is the vector of the rand-hand side constraint 
equations.

  We specified the translational constraints of a coordinate system in a 
form that requires the unweighted net translation in positions and net 
translation of velocities of 27 stations with respect to the reference 
catalogue ITRF2000 \citep{r:itrf2000} be zero. General equation~\ref{e:e9} 
is reduced to:
\beq
   \sum_i^n \Delta x^{j}_{ei} = \sum_i^n x^{j}_{ri} - \Delta x^{j}_{ai},
\eeq{e:e10}
  where $\Delta x_{e} = x_a - x_e$, index $i$ runs over stations,
superscript $j$ runs over 3 components of station positions and 
three components of station velocities, and indices $a$, $r$, and $e$ 
denote vectors of a~priori, reference, and estimates, respectively. 
The reciprocal constraint weights are 0.1~mm for station positions and 
0.1~mm/yr for station velocities.

  We specified the rotational constraints of the coordinate system in a 
form that requires the unweighted differences in estimates of the Euler 
angles describing the Earth rotation $E_e(t)$ with respect to the time 
series IERS~C04 $E_e$ to have no bias and no linear trend over 
[2000.0, 2024.0] time interval. The general equation~\ref{e:e9} 
is reduced to:
\beq
  \begin{array}{l @{\enskip}c@{\enskip} l}
     \dss\sum^N_i  E^j_{ei} - E^j_{ri}           & = & 0  \vex \\
     \dss\sum^N_i (E^j_{ei} - E^j_{ri})(t - t_0) & = & 0, \\
  \end{array}
\eeq{e:e11}
  where $E^j_{ei}$ is the estimate of the $j$th component of the Euler 
angle at $i$th epoch and $E^j_{ri}$ is that Euler angle from the reference 
time series. Summation is performed over $N$ epochs within the interval
[2000.0, 2024.0], which is narrower than the total interval of the EOP 
estimation with the B-splines. Then these equations are transformed to:
\beq
  \begin{array}{l @{\enskip}c@{\enskip} l}
     \dss\sum^N_i  \Delta E^j_{ea} & = & \Phi^j                 \vex \\
     \dss\sum^N_i  \Delta E^j_{ea} & = & \dot{\Phi}^j (t-t_0),       \\
  \end{array}
\eeq{e:e12}
  where $\Delta E^j_{ea} = E^j_{ei} - E^j_{ai}$, $\Phi^j$ and $\dot{\Phi}^j$
are the mean value and linear trend in differences $E^j_{ri} - E^j_{ai}$.
These equations are transformed to 
\beq
    \begin{array}{l @{\enskip}c@{\enskip} l @{\enskip}c@{\enskip} l}
        \dss\sum_{k=1}^{k=N} \sum_{i=1-m}^{i=n-1} e_{ji} \, B^m_i(t_k)              & = & 
        \Phi^j \, N                                                                 & + & 
        \dot{\Phi}^j \dss\sum_{k=1}^{k=N} (t_k - t_0) 
\vex \\
        \dss\sum_{k=1}^{k=N} \sum_{i=1-m}^{i=n-1} e_{ji} \, B^m_i(t_k) \, (t_k-t_0) & = &
        \Phi_i \dss\sum_{k=1}^{k=N} (t_k - t_0)                                     & + & 
        \dot{\Phi}^j \dss\sum_{k=1}^{k=N} (t_k - t_0)^2,
    \end{array}
\eeq{e:e13}
   where $e_{ji}$ are B-spline coefficients for the $j$th component
of the Euler angle and $i$th time epoch. 
The reciprocal constraint weights are $\flo{7}{-9}$~rad on the mean value
and $\flo{3}{-17}$~rad/s on the linear trend.

  We impose identifying constraints on source positions in a form that 
requires the net rotation of estimated positions $\vec{s_e}$ of the 
subset of 212 sources with respect to the ICRF1 catalogue \citep{r:icrf1} 
denoted as $\vec{s_r}$ be zero, i.e.\ $ \sum_i^N \vec{s_{ei}} \times 
\vec{s_{ri}} = 0$. These conditions expressed 
via $\Delta \vec{s} = \vec{s_e} - \vec{s_a}$ lead to 
\beq
    \sum_i^N \Delta \vec{s}_i \times \vec{s_{ri}} = \sum_i^N (\vec{s_{ri}} - \vec{s_{ai}}).
\eeq{e:e14}

  This equations can be easily expanded for components in right ascensions
and declination:

\beq
   \left\{
   \begin{array}{l @{\enskip}c@{\enskip} l}
       \dss\sum_i^N - \cos \alpha_{ri} \; \tg \delta_{ri}  \: \Delta \alpha_{ri} 
                    + \sin \alpha_{ri} \: \Delta \delta_{ri} & = & 
       \dss\sum_i^N - \cos \alpha_{ri} \; \tg \delta_{ri} \: (\alpha_{ai} - \alpha_{ri}) 
                    + \sin \alpha_{ri} \: (\delta_{ai} - \delta_{ri})  
   \vex \\
       \dss\sum_i^N- \sin \alpha_{ri} \; \tg \delta_{ri}  \: \Delta \alpha_{ri} 
                - \cos \alpha_{ri} \: \Delta \delta_{ri} & = & 
       \dss\sum_i^N - \sin \alpha_{ri} \; \tg \delta_{ri} \: (\alpha_{ai} - \alpha_{ri}) 
                - \cos \alpha_{ri} \: (\delta_{ai} - \delta_{ri})  
   \vex \\
       \dss\sum_i^N \Delta \: \alpha_{ri} & = & 
       \dss\sum_i^N (\alpha_{ai} - \alpha_{ri})  
   \end{array}
   \right..
\eeq{e:e15}
  The reciprocal constraint weights are $\flo{1}{-10}$~rad.

   These identifying conditions with the associated reference catalogues,
list of objects, and weights (unity in our case) unambiguously define the 
origin and orientation of the catalogues of station positions and source 
coordinates. We should stress that the choice of identifying conditions 
is a matter of convention in a similar way as the Greenwich meridian is 
used as a fiducial reference for longitude. We used relatively old 
catalogues ITRF2000 \citep{r:itrf2000} and ICRF1 \citep{r:icrf1}, despite 
the positions of individual objects are not the best because we think 
it is important to provide the continuity in the convention.

\subsubsection{Decorrelation constraints}

  Some combinations of estimated parameters are not exactly linearly
dependent, but close to that. Their estimation makes the system
of equations ill-conditioned and causes correlations between some 
groups of parameters be very close to $\pm 1$. To overcome these problem,
we impose decorrelation constraints. They include the following.

\begin{itemize}
    \item Constraints between the constituents of the harmonic expansion
          of the Earth orientation parameters with a frequency separation
          less than $0.8 \, f_s = \flo{3.941}{-9}$ rad/s. We require
          that the ratio of the complex amplitudes of estimated parameters
          $P$ be the same as the ratio of the a~priori complex 
          amplitudes $A$:
\beq
          \frac{\displaystyle\strut P^c_1 + i \, P^s_1}
          {\displaystyle\strut P^c_2 + i \, P^s_2} = 
          \frac{\displaystyle\strut A^c_1 + i \, A^s_1}
          {\displaystyle\strut A^c_2 + i \, A^s_2} \,,
\eeq{e:e16}
          where  index 1 denotes the main components of the spectra 
          that correspond to nutations and tidal variations and index 
          2 denotes the secondary close component. In this context the 
          constituents that have a frequency separation between each other
          $ >0.8 \, f_s$ are called primary. The reciprocal constraint 
          weights are $\flo{3}{-11}$~rad.

    \item Net-translation constraints on sine and cosine components of
          station position variations at each frequency:
\beq
          \begin{array}{l @{\enskip}c@{\enskip} l}
              \dss\sum x^c_{ei} & = & 0\,, \vex \\
              \dss\sum x^s_{ei} & = & 0\,, \\
          \end{array}
\eeq{e:e17}
          where the superscript runs over cosine and sine components and 
          the subscript $i$ runs over four frequencies. These parameters are
          considered nuisance in the context of this work. We should note
          that since we estimate harmonic position variations of only
          a subset of stations, this does not lead to the singularity of
          the normal matrix, but makes it ill-conditioned. The reciprocal 
          constraint weights are 0.3~mm.

    \item Net-rotation constraints on sine and cosine components of
          station position variations at each frequency:
\beq
          \begin{array}{l @{\enskip}c@{\enskip} l}
              \dss\sum X_a \times x^c_{ei} & = & 0\,, \vex \\
              \dss\sum X_a \times x^s_{ei} & = & 0\,, \\
          \end{array}
\eeq{e:e18}
          where $X_a$ is a vector of a~priori site positions. The reciprocal 
          constraint weights are 0.3~mm.

    \item Decorrelation between station position at the reference epoch
          and the B-spline model of the station position evolution.
          B-spline of the degree 0 and greater is linearly dependent
          with station position estimates. We require that the integral
          of the B-spline model over the interval of observations be zero:
\beq
          \displaystyle\int\limits_{t_0}^{t_n} \sum_{k=1-m}^{k=n-1} x_{ik} \, B^m_k(t) \, dt = 0\,,
\eeq{e:e19}
          where $x_{ik}$ is the B-spline coefficient at the $k$-th knot
          of the $i$th station position component, which is reduced to:
\beq
          \sum_{k=1-m}^{k=n-1} x_{ik} I^m_k(t) \, = 0\,,
\eeq{e:e20}
          where $I^m_k(t) = \displaystyle\int\limits_{0}^{t} B^m_k(t) \, dt$.
          The reciprocal constraint weights are 0.3~mm.

    \item Decorrelation between station velocity and the B-spline model of 
          station position evolution when the B-spline degree is $>0$.
          B-spline of the degree 1 and greater is linearly dependent
          with station velocity estimates. We require that the momentum
          integral of the B-spline model over the interval of observations 
          be zero:
\beq
          \displaystyle\int\limits_{t_0}^{t_n} \sum_{k=1-m}^{k=n-1} x_{ik} \, t \, B^m_k(t) \, dt = 0\,.
\eeq{e:e21}
          This is reduced to
\beq
          \sum_{k=1-m}^{k=n-1} x_{ik} K^m_k(t) \, = 0\,,
\eeq{e:e22}
          where $K^m_k(t) = \displaystyle\int\limits_{0}^{t} t \, B^m_k(t) \, dt$.
          We should note there exist recurrent relationships for computation of 
          functions $I^m_k(t)$ and $K^m_k(t)$ analogous to computation of 
          B-spline functions themselves. The reciprocal constraint weights 
          are 0.3~mm/yr.
          
    \item Tie velocities. We constraint the differences in velocity estimates
          at stations that are located within 0.03--2~km because they are 
          the subject of the same tectonic motions. Constraints on the $i$th
          component of velocities of stations $X$ and $Y$ are imposed in this 
          form:
\beq
          \dot{X}_i - \dot{Y}_i = 0.
\eeq{e:e23}
          The reciprocal constraint weights are 0.3~mm/yr.
\end{itemize}

\subsubsection{Weak constraints}

   There are situations when there are no sufficient data to get realistic 
estimates of some parameters. Imposing weak constraints allows us 
to stabilize the solution by expense of causing biases of estimated 
parameters towards the a~priori values. We imposed the following weak 
constraints.

\begin{itemize}
    \item Weak constraints on time derivative of clock function to
          process data with gaps and to make estimates smoother.
          Default reciprocal constraint weights: $\flo{2}{-14}$.
          The reciprocal constraint weights were increased in rare
          cases when clock function had large variations ($>$ 500~ps)
          due to hardware malfunctioning.

    \item Weak constraints on time derivative of atmospheric path 
          delay in the zenith direction. The reciprocal constraint 
          weights are $\flo{1.39}{-14}$.

    \item Weak constraints on time derivative of tilts of the refractivity
          field symmetry axis. The reciprocal constraint weights 
          are $\flo{7.72}{-17} \, {\rm s}^{-1}$.

    \item Weak constraint on the second derivative of B-spline that models
          nonlinear station position evolution. The reciprocal constraint 
          weights are 1~mm/yr${}^2$.

    \item Weak constraints on the Earth orientation parameters modeled 
          with B-spline, its first, and second time derivative. This
          allows to stabilize the Earth orientation parameter estimates
          in the periods of time that cover gaps in observations, usually 
          due to public holidays. Imposing constraints on the second time
          derivatives makes the time series smoother. The reciprocal 
          constraint weights are $\flo{2}{-6}$~rad, $\flo{3}{-14}$~rad/s, 
          and $\flo{1}{-19}$~rad/s${}^2$ for the value and the first and 
          second derivatives, respectively.

    \item Weak constraints on the Earth orientation parameters and their
          rate models as local parameters in the range of 1980--1990.
          Imposing weak constraints allows to process single-baseline 
          data and data at short baselines. The reciprocal constraint
          weights are $\flo{2.18}{-7}$ rad for Earth orientation parameters 
          and $\flo{2.53}{-12}$ on their rate of changes.

    \item Constraints on estimates of velocities of 49 stations. 
          This allows to process data with a short history of observations.
          The reciprocal weights are 0.1~mm/yr for the vertical components
          and 3.0~mm/yr for the horizontal component. Strictly speaking
          these constraint are not weak.
          
\end{itemize}

\subsection{Filtering data for outliers}
\label{s:filtering}

  A VLBI dataset of a given experiment may have from 5 to 80\% 
outliers, i.e.\ observations with group delay errors much greater than 
reported uncertainties. The most common reasons of such errors are
a)~failures in the fringe fitting procedure because of insufficient
baseline sensitivity; b)~failures in the fringe fitting procedure because 
of poor phase calibration; c)~failures in the fringe fitting procedure
due to radio interference; d)~scattering in the ionosphere or in rare
cases in the solar corona; e)~malfunctioning of either a Hydrogen maser
or a maser signal distributor. In addition, some observations may have
excessive residuals due to deficiencies of the model of radio wave
propagation in the atmosphere, especially when observing at elevations 
below $20^\circ$ or because of unaccounted source structure. 
It is essential that these outliers are identified and excluded from the 
final data analysis during the preprocessing part of the analysis.

  The fringe fitting process provides an estimate of group delay even
in the absence of the contribution of a signal from the observed source
in visibility data. However, when the fringe fitting procedure is applied
to the noise without a signal, the amplitude of the best fit is low. The 
distribution of the amplitudes weakly depends on the frequency and time 
resolution of visibility data, and therefore, varies within 10--30\% from 
a campaign to campaign. The distribution density of the SNR has two 
components: from the noise and from the signal of the target sources 
\citep[see Figure~9 in][]{r:wfcs}. The first component vanishes 
at the SNR $>5.5$--7.0. Fitting the first component, we can separate them and
compute the probability of false detection.

  At the beginning, we compute the distribution density of fringe amplitudes 
and determine the SNR when the probability of false detection is less than 
0.001, 0.01, and 0.1. We process a given experiment with suppressing 
observations with the SNR cutoff that corresponds to the probability of false 
detection 0.001. Then we run a preliminary solution estimating station clock 
function, atmospheric path delay in zenith direction, polar motion, and UT1,
as well as positions of some sources. Initially, we apply weights that are 
reciprocal to group delay uncertainty determined by fringe fitting. Then we 
run the procedure of outlier elimination and reweighing. 

  The presence of outliers distorts the solution and the residuals. 
To overcome this difficulty, we implemented the following iterative
outlier elimination procedure with five steps:

\begin{enumerate}\setlength{\itemsep}{0.0ex}
   \item Computation of postfit residuals and normalizing them by 
         multiplying by the weight.

   \item Sorting the normalized residuals.

   \item Flagging the observations with the highest by modulo residual.

   \item Update of the vector of estimate $\vec{x}$ for exclusion of the 
         flagged observation with index $k$. Invoking the lemma of the 
         inversion of an extended matrix, also known as 
         Sherman-Morrison lemma, this can be done on the basis
         of the prior least squares solution:
\beq
    \vec{x}_u = \vec{x} - \hat{V} \, \vec{a}_k \, w_k \,
                \Frac{ y_k - \vec{a}_k \cdot \vec{x}   }
                     {1 + w_k \, \ttra{\vec{a}}_k \, \hat{V} a_k }\,,
\eeq{e:e24}
    where $\vec{a}_k$ is the equation of observation, $w_k$ is weight, 
and $\hat{V}$ is the covariance matrix of the estimate vector $\vec{x}$.

   \item Update of the covariance matrix for exclusion of the flagged 
         observation
\beq
   \hat{V}_u = \hat{V}+ \Frac{\hat{V} \, \vec{a}_k \, w_k \, \ttra{\vec{a}}_k \, \hat{V} }
                         { 1 + w_k \, \ttra{\vec{a}}_k \, \hat{V} \, a_k }\,.
\eeq{e:e25}
\end{enumerate}
The procedure is repeated, untill the maximum by modulo of a normalized 
residual exceeds the specified limit.

  The procedure can be reversed and we can update the estimate vector 
and its covariance matrix from the least squares solution when we include 
observation $r$ that has been previously excluded by changing a sign after 
$\vec{x}$ in equation~\ref{e:e24} and after $V$ in equation~\ref{e:e25}. 
If we have more than one excluded observation, we find among them 
the observation with the smallest by modulo normalized residual considered. 
We call that reversed procedure data restoration. Some observations can be 
considered ineligible for restoration, for instance, because their SNRs are 
less than the threshold or phase calibration data were missing.

  Processing pathfinder astrometric experiments poses an additional 
complication. When these experiments include observations of sources never
before observed with VLBI, their positional errors may reach arcminutes.
We have to estimate their coordinates. Estimation of positions affects 
robustness of the outlier elimination process. According to our experience, 
the following situation happens with the probability of 5--20\%: only two 
observations of a given source remained at the end of the outlier 
elimination process, and only one of them is the outlier. When we estimate 
right ascension and declination using two observations, the residual is 
zero and the observations look good. We single out sources with estimated 
positions that have only two good observations and two or more 
outliers. We check for other combinations of flags for a given source in 
a given experiment using the brute force approach starting with short 
baselines first. This procedure has a chance of 20--30\% to end up with 
a combination of flags that restores three or more observations and 
provides the normalized residuals less than 4, and thus, fixes the failure 
of the outlier elimination procedure.

  Another complication emerges in processing double sources with 
a component separation greater than $\sim\! 100$~mas and with a ratio
of flux densities less than 3--5. The fringe fitting process may catch
different components of a source at different baselines. These sources
are singled out as objects with an excessive outlier rate among
observations with the probability of false detection $< 0.01$.
In that case we inverse the suppression flags for observations of 
these sources and repeat the outlier elimination procedure. This
procedure may have two outcomes: (1)~only two observations will remain and
(2)~three or more observations will remain. In the first case we 
discard results of this procedure. In the second case we further examine
results by running the imaging process if the difference in positions 
of that source is less than $1''$ or examine Very VLA images
from the archive, when available. If images confirm the presence 
of the second component close to the derived positions, we assign a new 
pointer to the visibility data and treat the dataset of these observations 
as having two or more sources in the field of view. We rerun the procedure 
of visibility data analysis from the very beginning. We fix positions of two 
or more sources in the field of view to the values determined in the 
previous round of data analysis during the first run of the outlier 
elimination, then estimate positions of these sources or source 
components, and then run one more iteration of restoration of observations 
and outlier elimination. In rare cases when we cannot confirm the second 
component in images, we keep questionable observations suppressed.

  We performed the outlier elimination for each fused, dual-band, and 
single-band observables independently and kept flags and reweighting 
parameters $q_b$ separately for each combination
of group delays. Upon completion, we re-run the fringe fitting for 
eligible outliers with the narrow group delay window, 0.7--2.0~ns
depending on band. Observations of sources that had fewer than three
usable observations or with failed phase calibration are considered 
ineligible. Then we repeat the procedure of outlier elimination using
the flags that were set in the prior analysis. This round restores 
a fraction of previously eliminated observations, from 10 to 80\%.

  Sources with two or fewer usable observations at a given pathfinder
session but with more than three or more usable observations among all the 
observing sessions were additionally checked. We ran a preliminary
global solution using all the sessions, estimated positions of these
sources, and examined residuals. Then we performed the outlier elimination
procedure for these sources, but using a whole dataset. All observations 
of those sources that had fewer than three unflagged observations in 
a global dataset were not used in the final solution. Three observations 
per source provide a minimum redundancy. A failure of the fringe fitting 
or the radio interference distorts group delays. Fitting right ascension 
and declination to one or two observations will result in zero residuals, 
regardless whether group delays were correct or not, while fitting to 
three or more observations will cause large residuals if one of the group 
delays was wrong, and examining residuals will allow us to detect an 
anomaly. Thus, this redundancy provides a safeguard against 
contamination of the output catalogue by spurious results.

  Finally, in order to mitigate subjectivity, the outlier elimination
procedure that uses prior results as an initial guess is executed in 
the fully automatic mode once again.

\subsection{Weight update}
\label{s:reweihging}

  In addition to outlier elimination, we ran the weight update procedure. 
First, we added in quadrature the elevation-dependent weights in the form
of $\beta * \sqrt{\tau_{w1}^2 + \tau_{w2}^2}$, where $\tau_{wi}$ is the wet
path delay in the direction of the observed source at the $i$-th station
and $\beta$ is a scaling factor. We used $\beta=0.1$ in our work.
As a trial, we ran a set of geodetic solutions when we estimated station
positions from each experiment individually. We got a time series of 
baseline lengths, fitted the linear model with discontinuities due to 
seismic events at some stations, and computed the rms of the residuals to 
that model, the so-called baseline length repeatability. We repeated these
trial solutions with different $\beta$ and found that $\beta=0.1$ resulted 
to the smallest repeatabilities for most of baselines. The use of this
reweighting scheme accounts for errors in modeling path delay as 
a sum of the a~priori path delay derived from numerical weather models,
adjusted corrections to the zenith path delays, and adjusted tilts 
of the refractivity field symmetry axis. Considering that the wet 
constituent of the zenith path delay averaged over all experiments and 
all stations is 371~ps, the added elevation-dependent noise was on 
average 52~ps when the source was in zenith at both stations and 260~ps 
when the source was at $10^\circ$ elevation.

  We compute the ratio of the sum of weighted residuals $R$ and its 
mathematical expectation $\Exp(R)$ using the following approximation:
\beq
   \Exp(R) = n - \lp m - \Sp(\hat{V} \, \tra{\hat{B}} \beta^{-1} \hat{B}) \rp,
\eeq{e:e26}
   where $n$ is the number of observations, $m$ is the number of equations,
$\hat{V}$ is a covariance matrix of estimates, $\hat{B}$ is a matrix
of constraints, and $\beta$ is a matrix of constraint weights, and $\Sp$
denotes a matrix trace. When no constraints are imposed, eq.~\ref{e:e26} is 
reduced to $n-m$, also known as the number of degrees of freedom (ndf). 
Analysis of statistics showed that adding the elevation-dependent noise 
is insufficient to make $\Exp(R)$ close to unity.

  If the used weights were reciprocal to the true uncertainties of 
observations and correlations between observations were zero, this ratio 
would have been one. Usually, this ratio is greater than one. Using an estimate 
of $R$, we can improve weights under two assumptions: a)~observations are not 
correlated and b)~there is another unknown independent source of errors 
in group delays with the zero mean and unknown baseline-dependent variance, 
i.e.\ the used weights are $1/w^2 = \sigma^2 - q_b^2$. Then after some algebra 
we arrive to the estimate of $q$ parameter for a given baseline $b$:
\beq
     q_b = \sqrt{ \Frac{ \Exp{(R_i)} -
          \lp n_b - \Sp( V \tra{A} W^{-1}_i A ) \rp }
         { \Sp(W^{-1}_b) \: - \: \Sp ( V \, \tra{A} W^{-2}_i A) } },
\eeq{e:26}
   where $W$ is the a~priori weight matrix and index $n_b$ is the number 
of equations used in the solution at a given baseline $b$.

  Cleaning the dataset involves several cycles. First, we discard 
observations with an SNR less than a quantity which corresponds to the 
probability of false detection of 0.001. We execute the sequence: 1)~outlier 
elimination at a given maximum by modulo normalized residual $N_\sigma$;
2)~weight update, and 3)~outlier elimination. Observations with an SNR 
less than the threshold are barred from restoration. We start with 
$N_\sigma$ = 8, then reduce it to 6, 5, and 4. Then we gradually reduced 
the threshold of the probability of false detection to 0.01, 0.1, and 
finally to 0.2. Those observations that we discarded in the previous cycle 
because of their SNR are automatically flagged as outliers, but become 
eligible for restoration.

  Although we raised the threshold of false association to 0.2 in the 
filter, that does not mean that the final dataset has 1/5 spurious group 
delays. The weight root mean square of residuals (wrms) of detected
observations is in range of 20 to 100~ps for astrometric experiments and
the residuals have a distribution that is close to Gaussian.
Nondetections have a uniform distribution within a search window that 
ranges from $\pm 1$ to $\pm 16$ ${\mu}s$. In a typical case when the wrms
of postfit residuals is 50~ps and the fringe search window is $\pm 2~\mu$s,
the probability of that a nondetection will pass $4 \sigma$ filter 
is $10^{-4}$. Therefore, the overall rate of presence of nondetections
in the final data set is $\flo{2}{-5}$, which is acceptable.

\subsection{Fused solution, single frequency, and  dual-frequency solutions}

   We performed six solutions that uses different observables. Of them, 
two are full solutions are four we residual solutions. 

   The main solution uses the fused group delay observables, i.e.\ dual-band
ionosphere-free combinations of group delay observables when usable group
delays were available at both bands, and single-band delays when usable
group delays were available only at one band. We call an observable 
good if it was retained in the solution after an iterative outlier
elimination procedure that keeps the normalized residuals greater than
$4.5\sigma$ at S~band, $4.0\sigma$ at other bands and in the fused dataset, 
and $3.5\sigma$ at dual-band ionosphere-free liner combinations. All global 
parameters were estimated in the fused and dual-band solutions, which is why 
we call them full solutions.

  The advantage of the fused data approach is that it takes the most from 
the existing observations and treats inhomogeneous data in 
the most consistent way. The disadvantage of this approach is that it blends 
genuine positional offsets between frequency bands. A small fraction of 
sources, 2 to 6\,\%, depending on the statistical criteria used, have 
different positions at different frequencies. The loss of information about 
these offsets is undesirable. To overcome this problem, we performed five 
auxiliary solutions. The first auxiliary solution used dual-band data 
only, i.e.\ X/S and X/C datasets. We mixed together X/S and X/C data because 
a dedicated pilot 48~hr campaign of observing 394 sources in a mode when 
each scan was observed two times in X/S and X/C showed no measurable biases 
between X/S and X/C observables \citep{r:wfcs} and between source position 
estimates. We can consider the fused solution as an extension of the 
dual-band solution by including single-band observations when dual-band 
linear combinations of observables are missing.

  We ran four  single-band solutions in a different mode. We applied 
estimates of station positions, including harmonic station position
variations and nonlinear station motions in a form of the B-spline, antenna 
axis offsets, coefficients of the harmonic variations in the Earth 
parameters, and the coefficients of the expansion of the Earth 
orientation parameters from the fused solution as a~priori in the 
single-band solutions. The only global estimated parameters in 
single-band solutions were source positions. We had to use this 
approach because we do not have enough data to derive positions of 
all the stations and the Earth orientation parameters from all the 
epochs from single-band solutions with an accuracy comparable to 
the accuracy of dual-band or fused solutions. 

  Strictly speaking, results of single-band solutions are not entirely 
independent from the dual-band solution because they implicitly depend 
on estimates of station positions and the Earth orientation derived from 
dual-band delays. We neglect this statistical dependence.
The reduced set of estimated parameters decreases estimates of position 
uncertainties with respect to the full solution. However, reciprocal weights
in single-band solutions were inflated to account for errors in the 
ionosphere path delay modeling, and this observation down-weighting affects
the position uncertainties to a greater extent than a reduction of the 
number of parameters.

  We represent source positions from single-band and dual-band solutions as 
offsets with respect to the positions from the fused solution that is 
considered the primary result. Positions from single-band and dual-band 
solutions are considered as auxiliary results.

\subsection{Datasets used in astrometric solutions}

  Tables~\ref{t:rfc1_sol} and \ref{t:rfc1_it} show the statistics of 
the datasets used in six RFC solutions. The fused solution used dual-band 
observables from the geodetic experiments, both 24~hr and 1~hr, fused 
observables from dual-band astrometric experiments, and single-band 
observables from single-band astrometric experiments.

\begin{table}[h!]
   \caption{Statistics of the RFC solutions.}
%
%
   \begin{center}
      \begin{tabular}{lrrrr}
           \hline
           \hline
           \ntab{c}{Solution type}   & \ntab{c}{\# src} & 
           \ntab{c}{\# exp} & \ntab{c}{\# obs} & \ntab{c}{postfit wrms} \\
           \hline
           fused     & 21,942 & 20,575 & 26,005,718 &  25.367 ps \\
           dual-band & 17,461 & 20,150 & 22,561,262 &  25.097 ps \\
           S-band    &   6636 &    466 &  6,124,272 & 270.133 ps \\
           C-band    & 16,379 &    318 &  1,186,786 &  68.418 ps \\
           X-band    & 19,742 &   1209 &  7,356,535 &  38.883 ps \\
           K-band    &   1872 &    179 &  1,723,933 &  23.462 ps \\
           \hline
      \end{tabular}
   \end{center}
   \tablecomments{dual-band means ionosphere-free combinations of either
                  X/S or X/C observables.}
   \label{t:rfc1_sol}
\end{table}

\begin{figure}[h!]
   \includegraphics[width=0.48\textwidth]{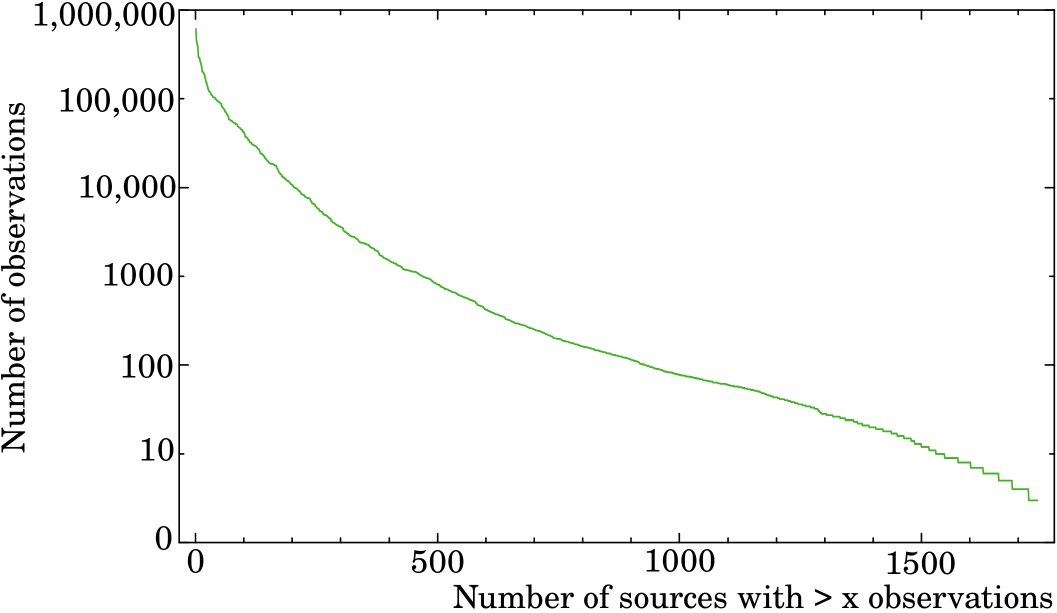}
   \hspace{0.039\textwidth}
   \includegraphics[width=0.48\textwidth]{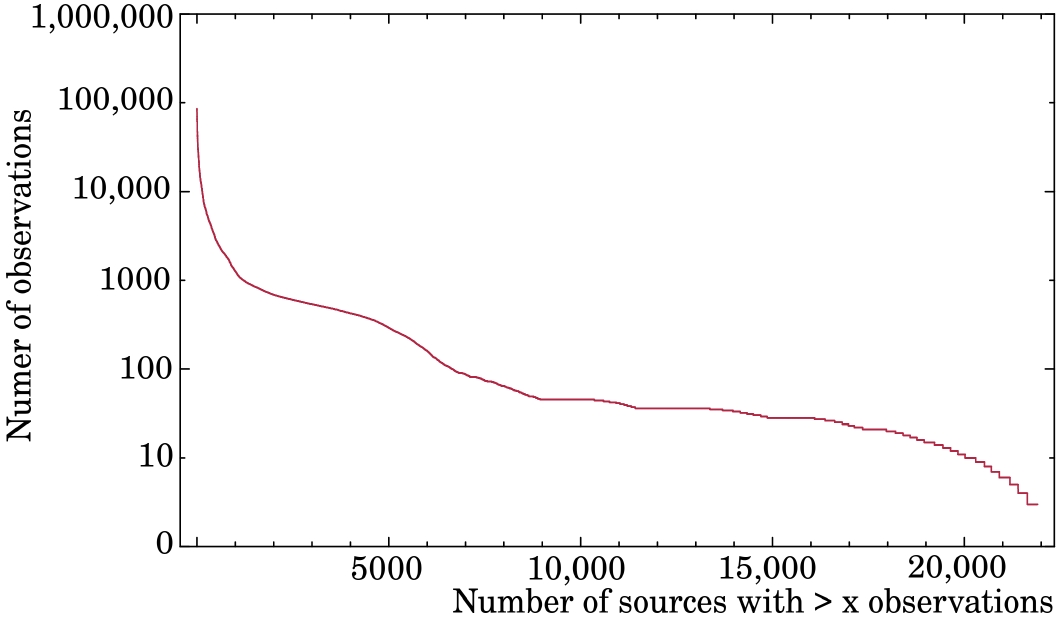}
   \caption{The distributions of sources by the logarithm of the 
            number of observations used in the solutions in geodetic 
            observing sessions (left) and astrometric sessions (right).
           }
   \label{f:num_obs}
\end{figure}

  In total, 1740 sources have been detected in three or more observations 
in geodetic experiments. However, the frequency of observations of sources
is substantially uneven. For instance, observations of 244 sources, or 14\%
of the total number, provided 95\% of geodetic data. Among 21,942 sources 
detected in three or more observations in astrometric experiments, 
observations of 7222 objects, or 33\%, provided 95\% astrometric data. 
The median number of observations of a given source in the astrometric
experiments is 41. See Figure~\ref{f:num_obs} for the distributions of
the number of sources with a given number observations that are used in 
solutions.

\begin{table}[h!]
   \caption{Types of data used in the RFC solutions.}
   \begin{center}
      \begin{tabular}{lrr}
           \hline
           \hline
           \ntab{c}{Data type} & \ntab{c}{\# sess} & \ntab{c}{Duration (hr)} \\
           \hline
           Geodetic 1~hr       & 11,848            &  12,118.4 \\
           Geodetic 24~hr      &  7,587            & 182,291.4 \\
           Astrometric         &  1,140            &  16,943.9 \\
           \hline
           Total               & 20,575            & 211,353.3 \\
           \hline
      \end{tabular}
   \end{center}
   \label{t:rfc1_it}
\end{table}

\begin{figure}
   \centerline{\includegraphics[width=0.61\textwidth]{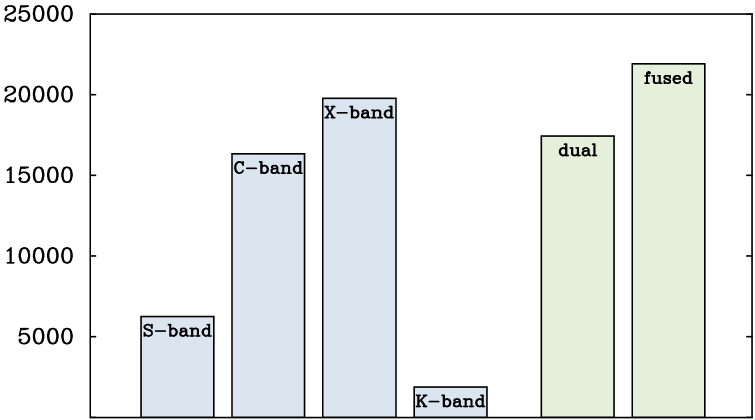}}
   \caption{A histogram of the number of detected sources per band.
            The dual-band box shows the number of sources that have been 
            detected in both bands simultaneously. A detected source has
            at least three observations that have passed the outlier detection
            tests.
           }
   \label{f:num_obs_band}
\end{figure}

   The histogram in Figure~\ref{f:num_obs_band} illustrates the statistics
of detected sources per band. This histogram demonstrates the choice of 
frequencies in survey programs and does not reflects a source detectability 
at given frequencies. 

\section{Error analysis}
\label{s:error_anl}

  It is common to call uncertainties of source positions derived from 
uncertainties of observable via the error propagation law formal. In the
case where the parametric model fully describes the observations and 
the stochastic model of the observables used in the data analysis correctly
describes the measurement noise, the formal uncertainties provide a 
realistic measure of errors of the parameter estimates. Violations of 
these conditions cause biases in source positional error estimates, 
usually as in a form of underestimation. We performed a number of 
statistical tests for an assessment of the reported errors.

\subsection{Decimations}

  A common technique to assess the realism of positional errors is to 
perform a decimation test: to split the dataset into several groups, 
process them separately, and then compare. If the errors are 
uncorrelated and Gaussian, the differences in positions divided by the 
sum of individual solution errors in quadrature will be Gaussian with 
a zero first moment and the second moment equal to 1. In the presence 
of systematic errors, the moments of the distribution will be different 
than 0 and 1, and the distribution itself may deviate from Gaussian.

  A dataset of $n$ points can be split into two subsets with $n!$ 
combinations. In the presence of the red noise the decimation results will 
depend on the way how a dataset is split. In this paper we call 
a noise with the power spectral density at lower frequency greater
than at high frequencies red. The longer the history of observations, 
the more prominent the impact of the red noise. In order to fully 
assess the impact of the red noise on source position estimates, we 
considered two extreme cases that we dubbed as local and global decimation. 
We sorted data of each source first in the chronological order and then 
in alphabetic order of baseline names. We labeled the sequence of 
observations as OEOEOEOEOEOE, where ``O'' marks an odd 
observation and ``E'' marks an even observation. The first dataset 
downweights even observations by a factor of 1000, and the second 
datasets downweights odd observations. We call this decimation local. 
Then we relabeled observations of each sources as FFFFFFLLLLLL, 
where ``F'' marks first $n/2$ observations of a given source and ``L'' 
marks last $n/2$ observations. We call this decimation global. 

  We have to limit the list of sources eligible for the decimation 
solution. We did not consider for our decimation tests 3205 sources that 
have fewer than 16 usable dual-band observations, i.e.\ 8 in each 
decimation subset, and 106 sources with more than 10,000 observations. 
Sources with too few observations provide unstable statistics. The 
frequently observed sources heavily contributed to geodetic experiments 
and their down-weighting would have caused a numerical instability and 
a degradation of estimates of the Earth orientation parameters. We would 
like to exclude the impact of distorted Earth orientation parameter on 
our results. We created four datasets of updated weights for 14150 
sources: their weights were divided by a factor of 1000 for one half 
of observations used in the solution. Weights of the remaining 3311 
sources were not modified. Two updated weight lists corresponded to 
odd and even subsets for the local decimations. Two other weight lists 
corresponded to first and last subsets of the global decimations. 
   
\begin{figure}[h!]
   \includegraphics[width=0.48\textwidth]{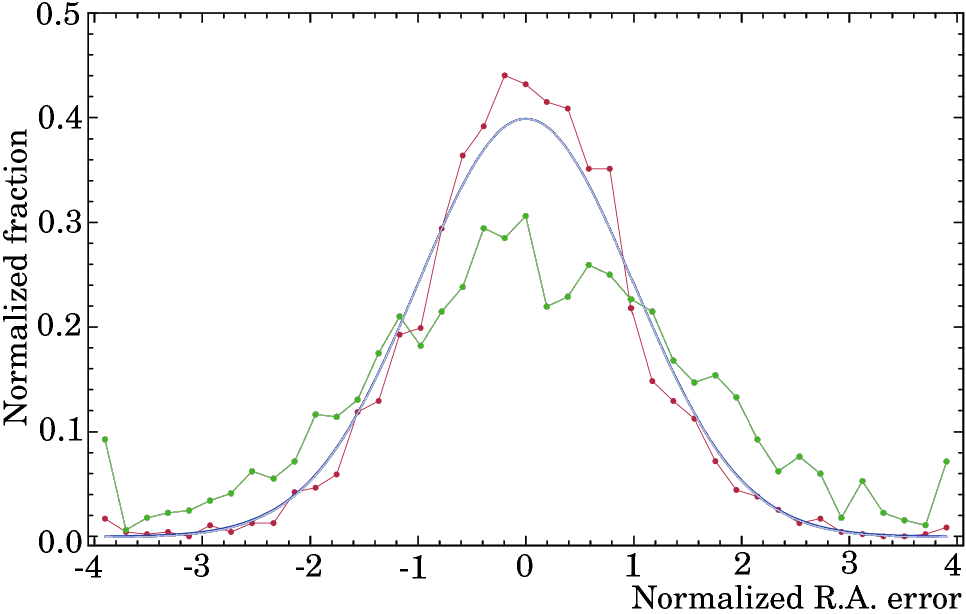}
   \hspace{0.039\textwidth}
   \includegraphics[width=0.48\textwidth]{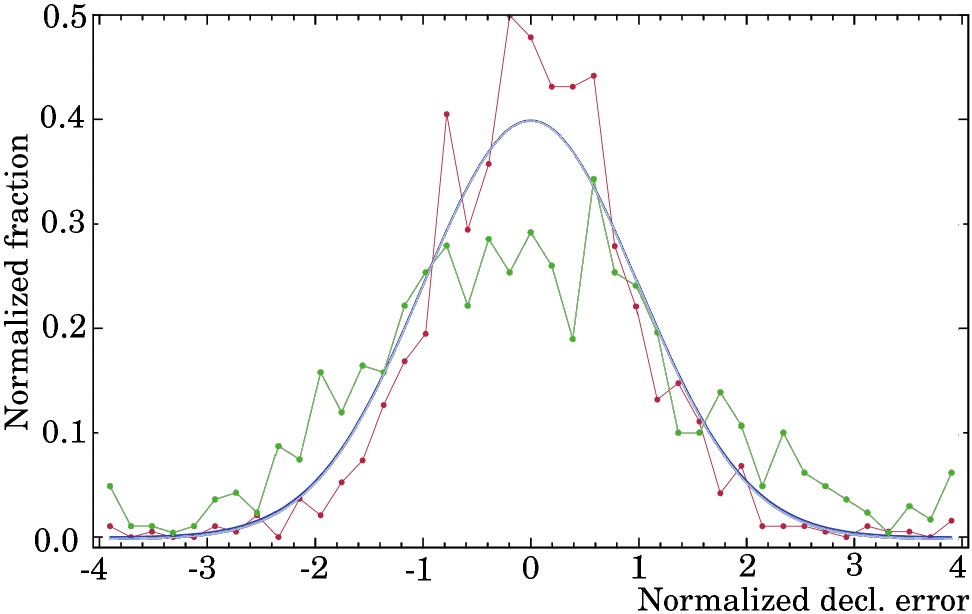}
   \caption{A histogram of normalized source positional errors
            for right ascensions scaled by the $\cos \delta$ factor 
            (left) and declination (right) derived from decimation
            solutions. The upper red curve corresponds to the 
            local decimation solutions. The lower green curve 
            corresponds to the global decimation solutions. Only 
            sources with positional errors $< 0.2$~mas are 
            considered. The central blue curve is the N(0,1) Gaussian 
            function shown as a reference.
           }
   \label{f:raw_err_hist}
\end{figure}

  Figure~\ref{f:raw_err_hist} shows the distribution of residuals
divided by $\sqrt{\sigma_1^2 + \sigma_2^2}$ of decimation solutions
1 and 2. The Gaussian distribution is shown for comparison. We indeed,
see significant differences in the statistics. In that figure we
limited the dataset to sources with positional errors $< 0.2$~mas.
These are mainly sources with a long history observation. In contrast,
most of the sources with uncertainties $> 2.0$~mas were observed in one 
scan, and therefore, both global and local decimations would pick 
up observations at the same epoch, but at different baselines.

  We see that results of global decimation (low green line) indicate
that the errors are underestimated, while results of local decimation
(red upper line) indicate the opposite. Let us recollect that we
added in quadrature baseline-dependent errors computed over all observations
of a given session to provide the ratios of the square of weighted 
residuals to their mathematical expectation close to unity. This added
variance accounts for the full impact of unmodeled errors for
{\it that experiment}. This added noise affects both ``O'' and ``E'' 
observables and it is partially canceled in observable differences. 
On the other hand, considerably fewer observations in ``F'' and ``L'' 
subset will be from the same observing session in the global decimation 
scheme. They would be affected with the red noise that is not accounted 
for by reweighting.

  As we mentioned above, global decimation provides a strong evidence that 
the errors in source positions are underestimated. We seek error re-scaling 
in the most simple empirical form: 
$\sigma_{\rm final} = \sqrt{(s  \, \sigma_{\rm original})^2 + F^2}$,
where $s$ is the scale that describes multiplicative errors and $F$ 
is the floor that describe errors that are independent on observations. 
Most likely, the realistic model is more complicated, but we do not have 
evidence to advocate for a more sophisticated model.

\begin{figure}[h!]
   \includegraphics[width=0.48\textwidth]{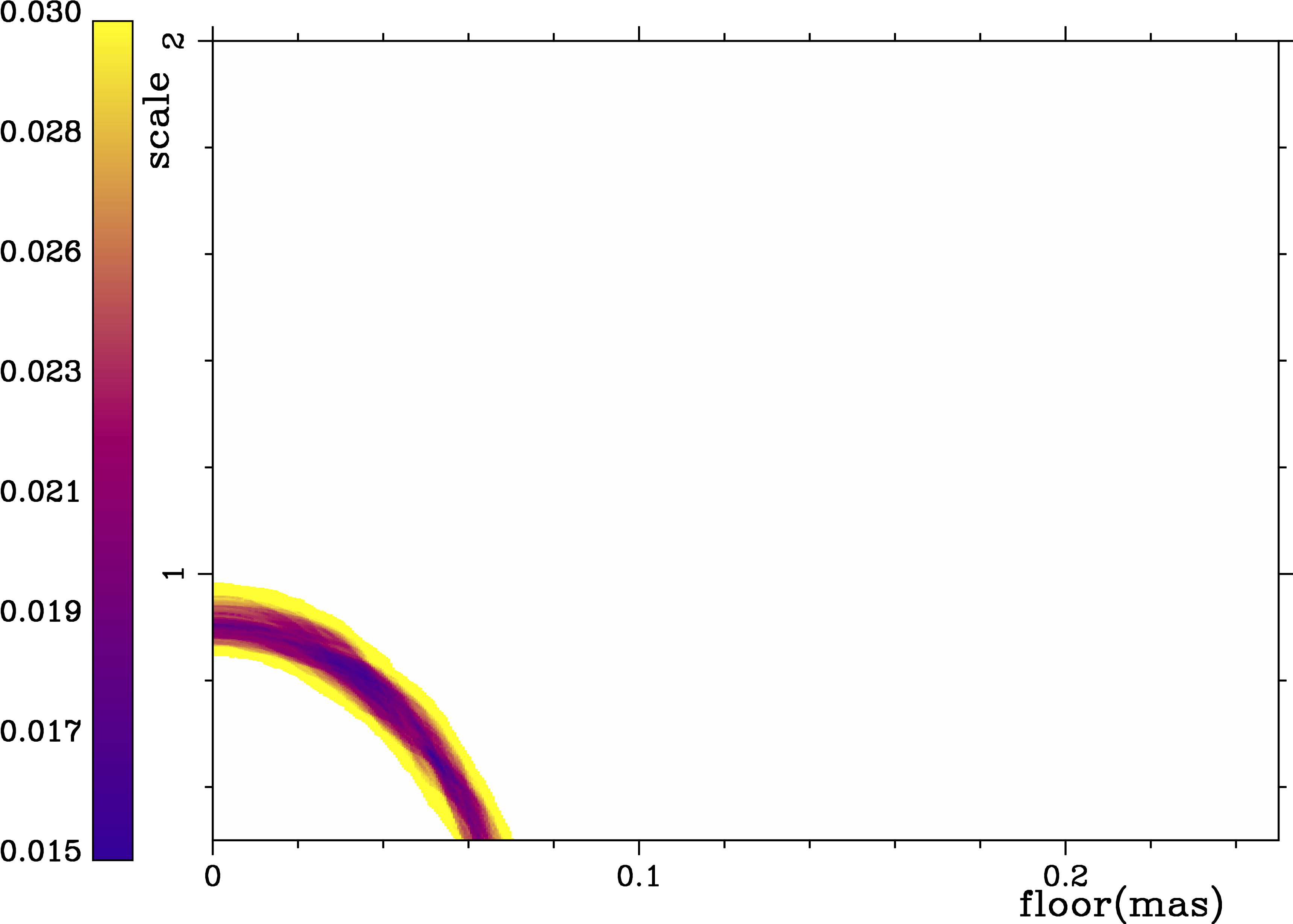}
   \hspace{0.039\textwidth}
   \includegraphics[width=0.48\textwidth]{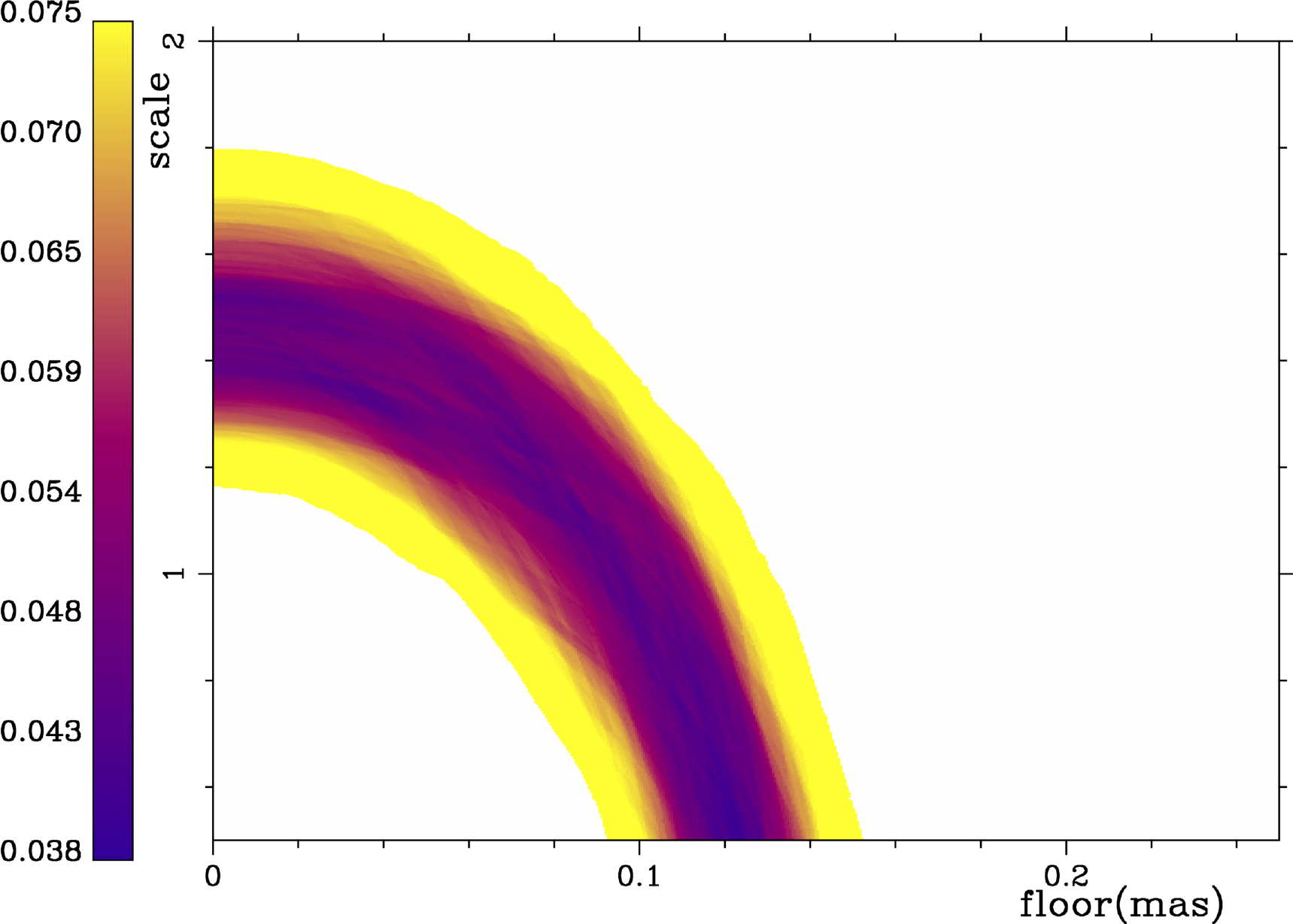}
   \caption{A floor-scale diagram for the right ascensions scaled
            by the $\cos \delta$ factor: local decimation (left) 
            and global decimation (right). The color corresponds to 
            the rms of the fit. Only the sources with positional errors 
            $< 0.4$~mas are considered.
           }
   \label{f:floor_scale_ra}
\end{figure}

\begin{figure}[h!]
   \includegraphics[width=0.48\textwidth]{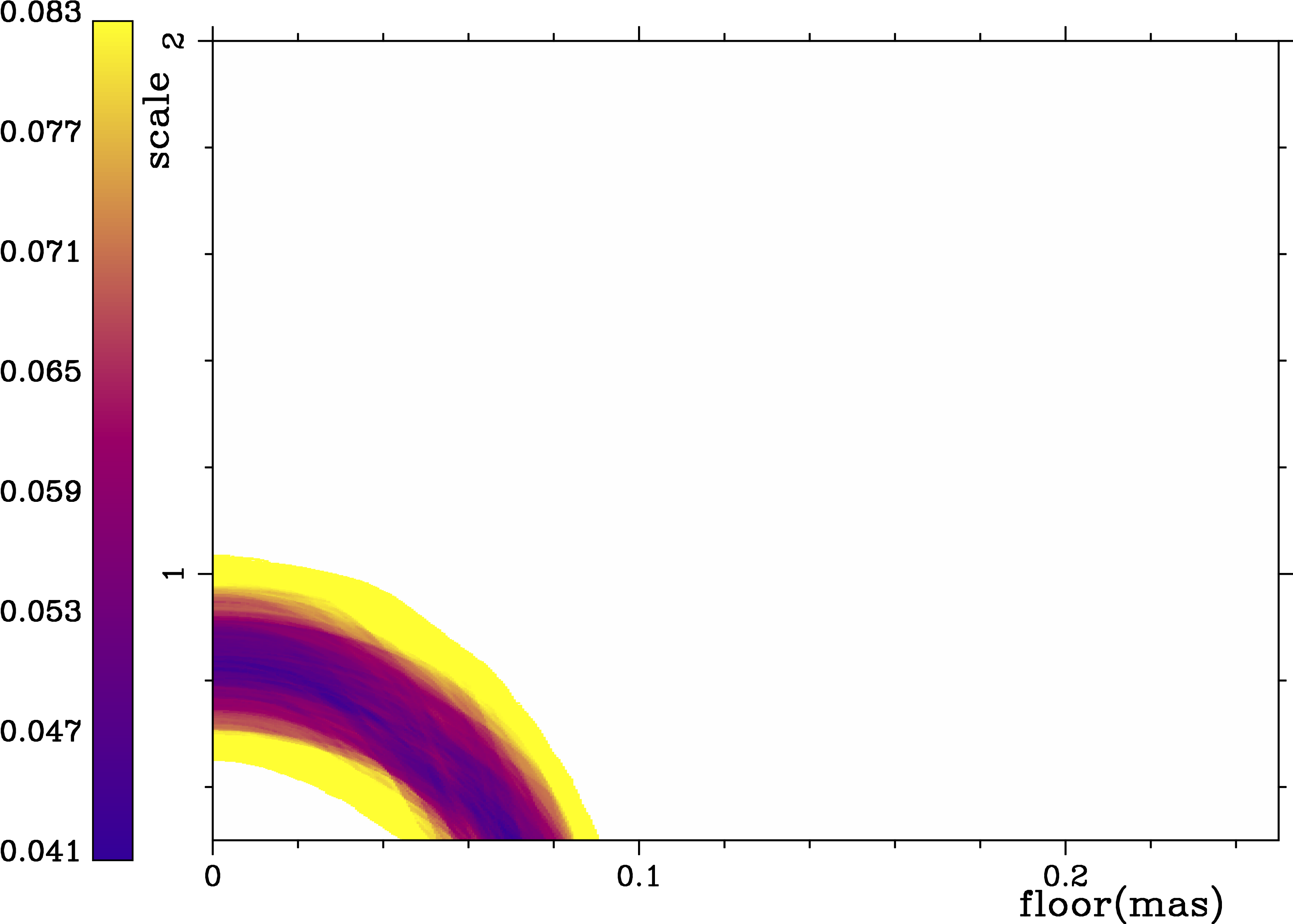}
   \hspace{0.039\textwidth}
   \includegraphics[width=0.48\textwidth]{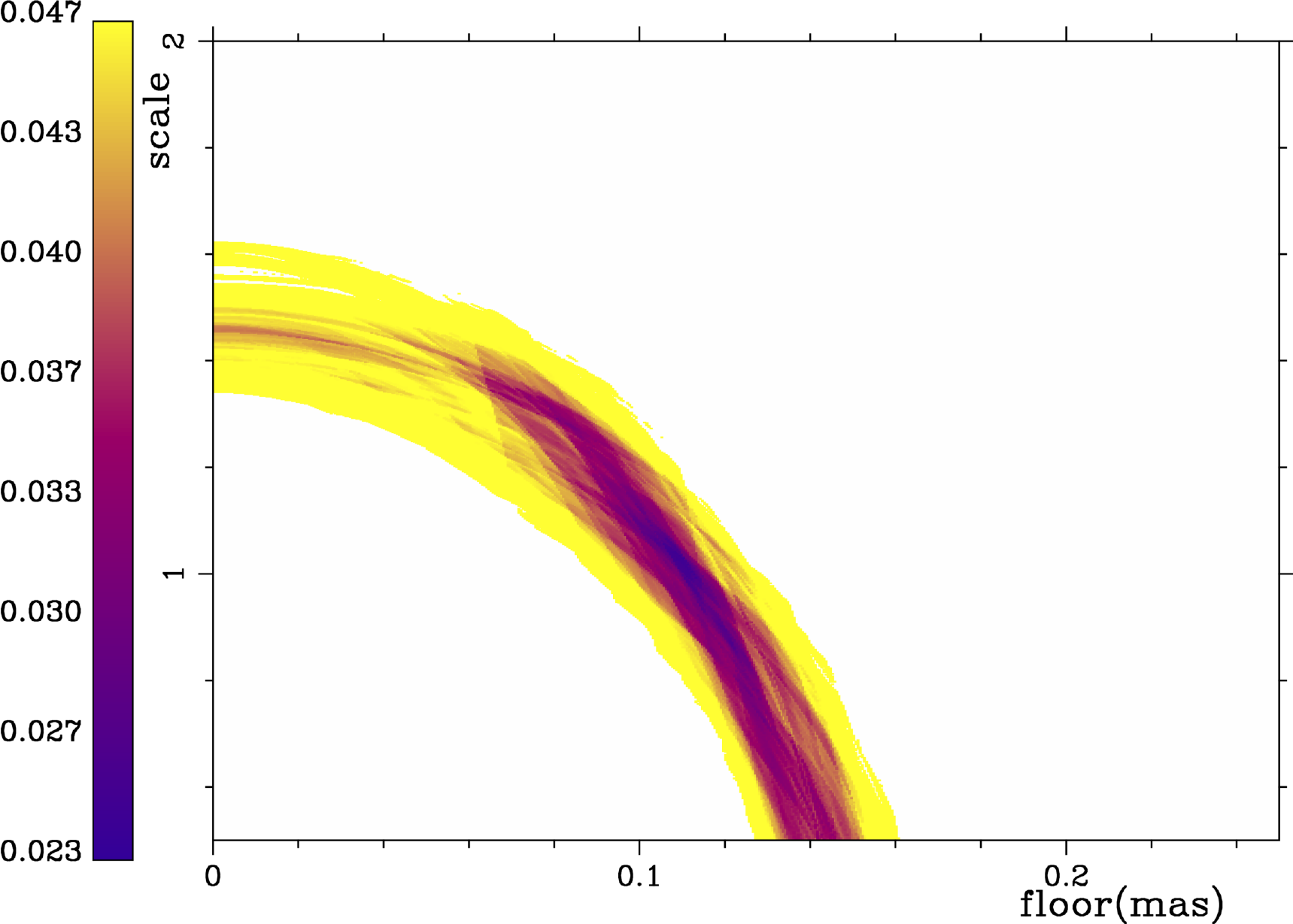}
   \caption{A floor-scale diagram for the declinations:
            local decimation (left) and global decimation (right). 
            The color corresponds to the rms of the fit. Only 
            the sources with positional errors $< 0.4$~mas are 
            considered.
           }
   \label{f:floor_scale_dec}
\end{figure}

  We fitted for the error scale $s$ and the floor $F$ to the histogram 
of normalized residuals and sought for the minimum in residuals. 
Obviously, if the source positional errors are large, the impact of 
a small error floor is small. Therefore, we limited out analysis to 
sources with position errors less than 0.4~mas to exercise a balance 
between the sample size and the sensitivity to the error floor. 
Figures~\ref{f:floor_scale_ra}--\ref{f:floor_scale_dec} show the 
residuals of the best fit as a function of the error floor and scale.
The residual are dimensionless. Dark colors correspond to a better
fit. For clarity, the dynamic range of these figures was restricted
to 1:2. The spread of these diagrams illustrates the poor separability
of the floor and scale parameters.

  These floor-scale diagrams are very different for local and global
decimations. The diagrams demonstrate rather clearly the upper limits
of the error floor. They are 0.06 and 0.075~mas for right ascension 
scaled by  $\cos\delta$ and declinations for the local decimation,
respectively, and 0.12~mas for the global decimation.

  The local decimation is appropriate in a situation where one investigates 
the potential of the observing technique to determine source positions 
by focusing on the short-term noise and discarding the long-term noise. 
The global decimation better characterizes the impact of the long-term 
noise. Since we focus this study to characterization of source positions 
averaged over the 30 year period of observations, we restrict further 
analysis to the global decimation.

\begin{figure}[h!]
   \centerline{\includegraphics[width=0.61\textwidth]{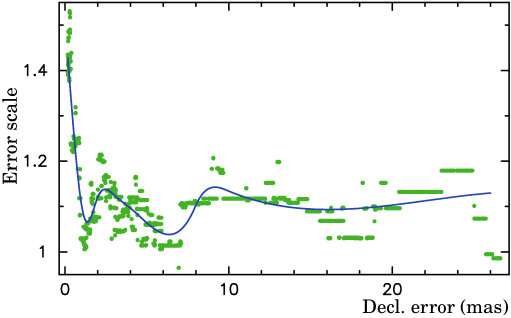}}
   \caption{Estimates of the scale $s$ for declination errors from the 
            global decimation solution with the error floor fixed to zero
            as a function of the mean declination error within a sliding
            window $[\sigma_l, \sigma_u]$. There were 1000 declination
            error estimates within each window. The blue line shows the 
            result of smoothing with B-spline.
           }
   \label{f:error_scale_decl}
\end{figure}

  Figures~\ref{f:floor_scale_ra} and \ref{f:floor_scale_dec} present 
evidence that the floor and scale estimates are correlated, which does 
not allow us to separate the floor and scale parameters reliably without 
additional assumptions. We sorted positional errors in a rising order and 
computed a set of 14159 histograms of a partial datasets with a sliding 
window of 1000 positional errors and estimated the scaling factor keeping 
the floor parameter fixed to zero. Figure~\ref{f:error_scale_decl} shows 
the scale estimates as a function of the average error in the sliding 
window. We see that for errors $>1$~mas, the error scale is around~1.0. 
The scale error wiggles from 1.0 to 1.5 for declination errors $<1$~mas. 
We interpret this as the contribution of the error floor, which is 
negligible for large declination errors. Analysis of these scale 
estimates prompted us to make an assumption that the scaling parameters 
affect all the sources regardless of their positional errors. We performed 
a similar analysis for right ascension errors scaled by $\cos\delta$ and 
using all the positional errors $>1$~mas, we derived the scaling factors
1.08 for right ascensions scaled by $\cos \delta$ and 1.16 for declinations.

\begin{figure}[h!]
   \includegraphics[width=0.48\textwidth]{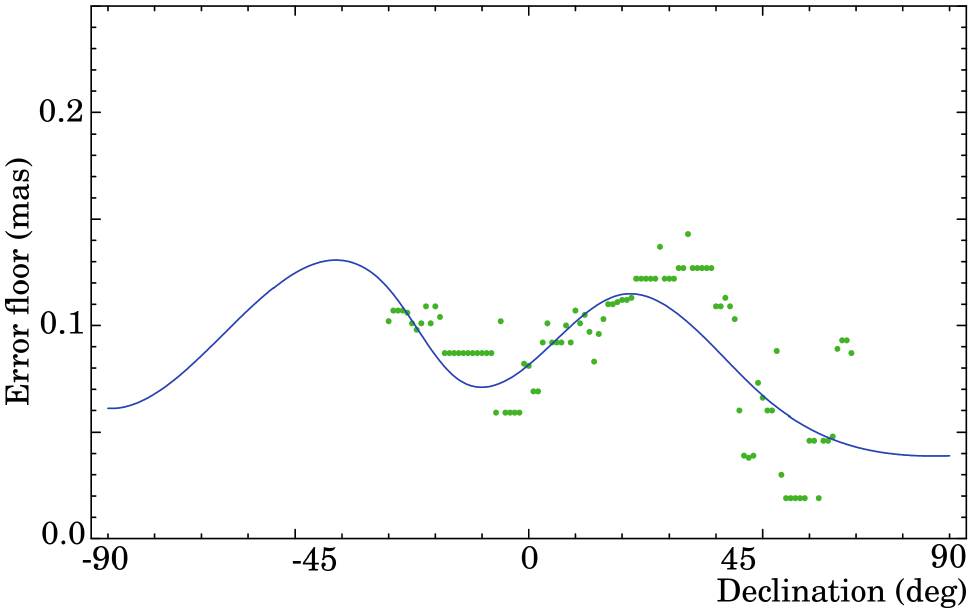}
   \hspace{0.039\textwidth}
   \includegraphics[width=0.48\textwidth]{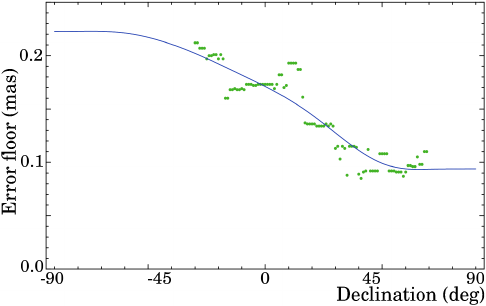}
   \caption{Estimates of the error floor from the global decimation 
            solutions as a function of declination for right
            ascensions (left) and declinations (right). The thin 
            blue line approximates them with a smoothing spline.
           }
   \label{f:error_floor}
\end{figure}

  To evaluate the error floor, we explored its declination dependence. We 
discarded observations with errors exceeding 0.4~mas, ordered positional errors 
over declination, applied scaling factors, computed the histograms for 
the range of declinations $[\delta_l, \delta_u]$, fitted to them the error floor, 
shifted the declination window by $\Delta\delta$, and repeated the process. 
We ran this process from declinations $-40^\circ$ with a window $20^\circ$ and 
a step of $1^\circ$. We do not have enough information to derive
the error floor for the sources with declinations $<-45^\circ$. These sources
were observed with the arrays at the Southern Hemisphere, and therefore,
they were observed in a more favorable conditions than low declinations
sources observed with the northern arrays. Therefore, the error floor should 
be less than the maximum. In the absence of information about the error floor
we elected to use the upper limit arguing that the overestimation of the 
error floor inflicts less harm than underestimation. The results of fitting 
the error floor with a fixed scaling factor are shown in 
Figure~\ref{f:error_floor}.

   Not surprisingly, the error floor for declination errors at low 
declinations is greater. The sources in the declination range 
$[-40^\circ, 0^\circ]$ were mainly observed with VLBA in 
a disadvantageous configuration at systematically low elevations
and are supposed to be affected by mismodeling the atmospheric path 
delay to a greater extent than observations of high declination 
sources. The origin of the wiggling pattern of the error floor 
estimates for right ascension errors is not clear. 

\begin{figure}[h!]
   \includegraphics[width=0.48\textwidth]{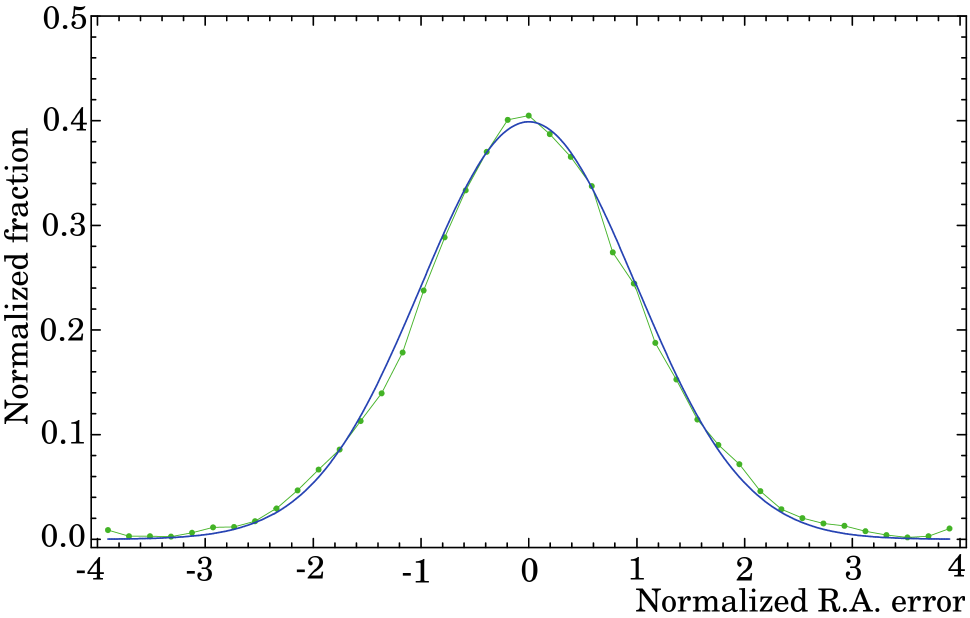}
   \hspace{0.039\textwidth}
   \includegraphics[width=0.48\textwidth]{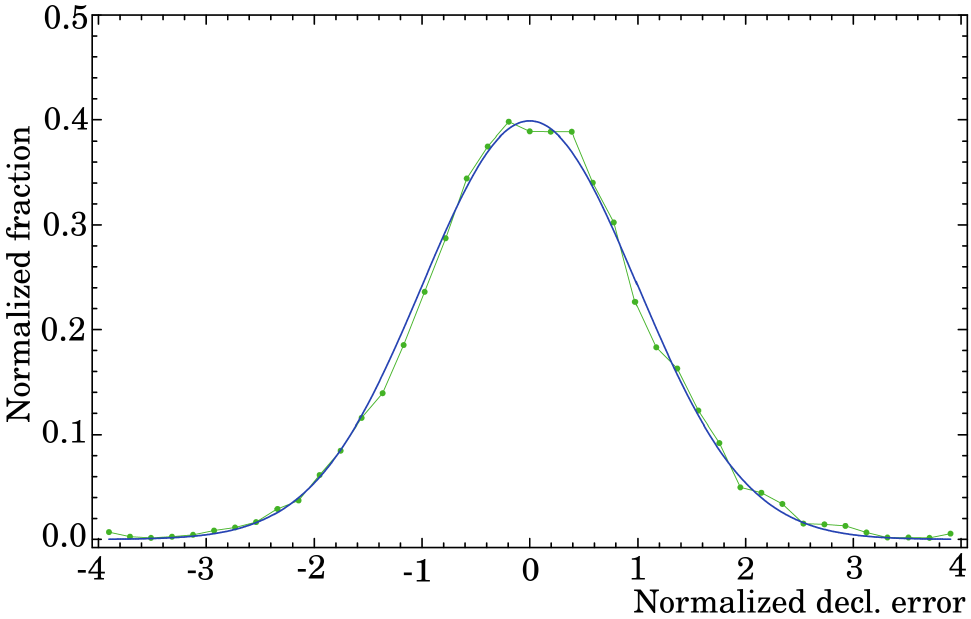}
   \caption{The histogram of the final normalized source positional errors
            for right ascensions scaled by the $\delta\cos$ factor 
            (left) and declination (right) after applying error 
            re-scaling (green points). The smooth blue curve is a
            Gaussian function shown as a reference.
           }
   \label{f:final_err_hist}
\end{figure}

   We smoothed the declination dependence of the error floor estimates 
with a spline by applying constraints on first and second derivatives. 
The error floor parameters tabulated with a step of $1^\circ$ are shown
in Table~\ref{t:err_floor_mod} in the Appendix. We used this 
scale-floor model for error re-scaling in our further 
analysis. Figure~\ref{f:final_err_hist} shows the histogram of the 
normalized source positional errors after applying the re-scaling model. 
Compare them with the distribution of original errors shown in 
Figure~\ref{f:raw_err_hist}.

\subsection{Positional errors from single-band experiments}

  The analysis above considered the dual-band solutions. Single-band 
solutions suffer from mismodeling path delays in the ionosphere. 
In fused solutions single-band observables are properly downweighted,
their share usually in the range of 10--20\%, and the ionospheric
biases of global TEC model were adjusted. However, position estimates 
derived from processing single-band experiments are affected by mismodeling
ionospheric contribution to a greater extent because no adjustments of
ionospheric bias is possible. Since the ionosphere has a bulge in low 
latitudes, ionosphere errors are declination-dependent. Observations of 
southern sources with the Northern Hemisphere arrays, such as VLBA, are 
systematically made at low elevations. Therefore, declination-dependent 
errors in modeling path delay in the ionosphere will be correlated with 
elevation-dependent errors of modeling path delay in the neutral atmosphere 
when observing sources at declinations in the range of 
$[-40^\circ, 0^\circ]$. This cross-talk is expected to lead to 
declination-dependent systematic errors.

  Such errors were investigated in full detail in \citet{r:radf}. In 
particular, it was revealed that the differences between dual-band and 
K-band source position estimates can be characterized by three constituents: 
1)~the common intrinsic noise with a second moment of $\sim\! 0.050$~mas per
component, 2)~the Gaussian noise along jet directions with the second 
moment 0.09--0.12~mas; and 3)~the noise in declination that causes a bias
of 0.050~mas and the rms at a level of 0.07~mas at declinations $>0^\circ$
that monotonically grows with a decrease in declination and reaches 0.3~mas
at declination $-45^\circ$. These three constituents close the error
budget. 

\begin{figure}[h!]
   \centerline{\includegraphics[width=0.61\textwidth]{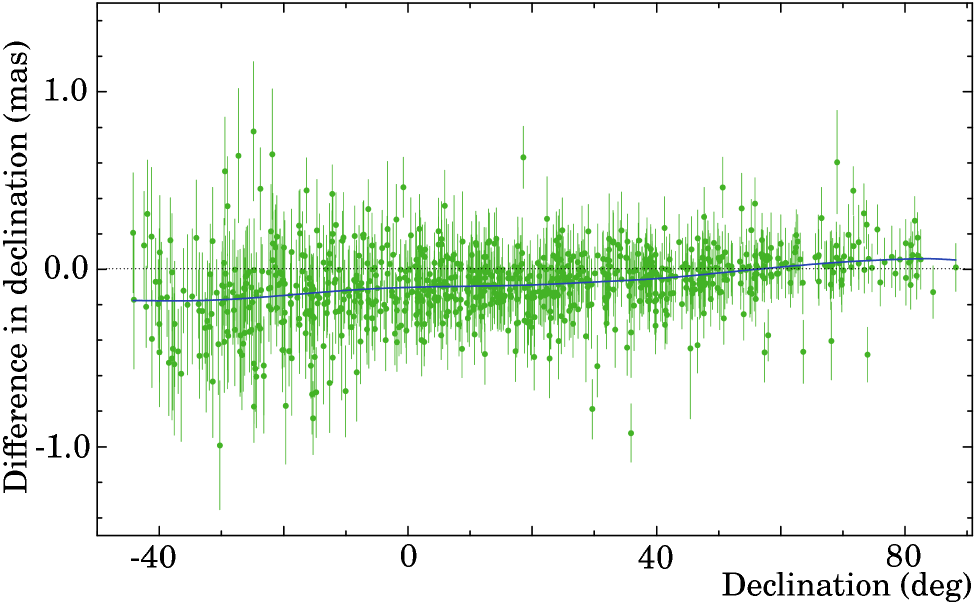}}
   \caption{The differences in declinations from the K-band solution
            with respect to the dual-band solution. The thick blue line
            shows smoother differences. The dash black line shows 0.0.
           }
   \label{f:decl_diff_rfc_k_xs}
\end{figure}

  Figure~\ref{f:decl_diff_rfc_k_xs} shows the differences in declinations 
from the K-band solution with respect to the dual-band solution. Only
sources with declination positional errors less than 0.3~mas are shown.
The thick line shows the result of smoothing using B-splines. We can see 
a negative bias that is growing with a decrease in declinations. 
The overall bias averaged over 833 sources is $-0.074$~mas. A similar 
comparison, but using a different dataset and a different set of
estimated parameters in \citet{r:radf} revealed a bias $-0.042$~mas. 
Is that bias an indication of some deficiency in our solution?

\begin{figure}[h!]
   \includegraphics[width=0.48\textwidth]{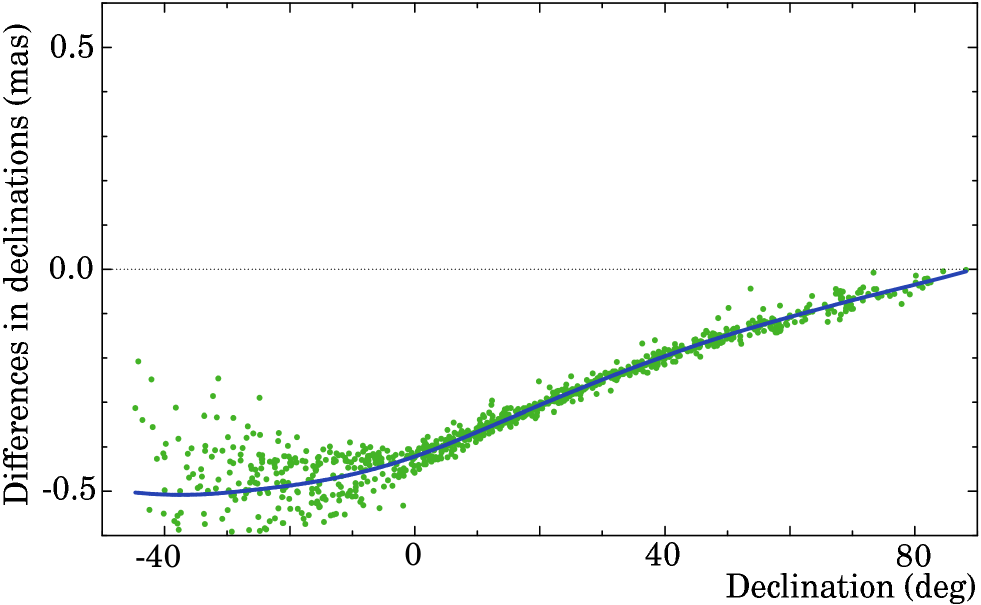}
   \hspace{0.039\textwidth}
   \includegraphics[width=0.48\textwidth]{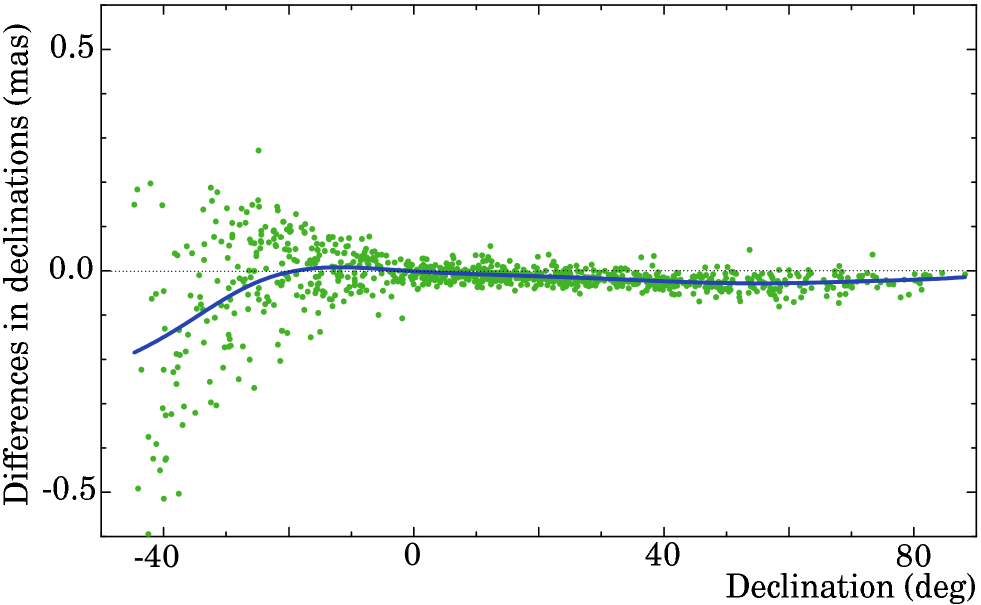}
   \caption{The differences in declinations from K-band solutions
            when the ionospheric contribution is applied with respect
            to solutions when it is not applied. {\it Left: }
            station positions are estimated.
            {\it Right: } station positions are not estimated.
           }
   \label{f:dd_full_iono}
\end{figure}

  It is instructive to visualize the impact of including the ionospheric
contribution in a solution. Surprisingly, the impact noticeably depends 
on whether station positions are estimated or kept fixed.
Figure~\ref{f:dd_full_iono} shows the differences in declinations when
the contribution of the ionosphere to path delay is included in a data
reduction model. The left plot shows the differences when station
positions were estimated, and the right plot shows the differences
when stations are kept fixed. Estimating additional parameters, station 
positions and velocities, makes a solution less robust. Station position
estimates absorb in part the contribution of the ionosphere. A distortion
in station position causes a distortion in declination estimates. That 
is why we have chosen not to estimate station positions in single-band 
solutions.

\begin{figure}[h!]
   \includegraphics[width=0.48\textwidth]{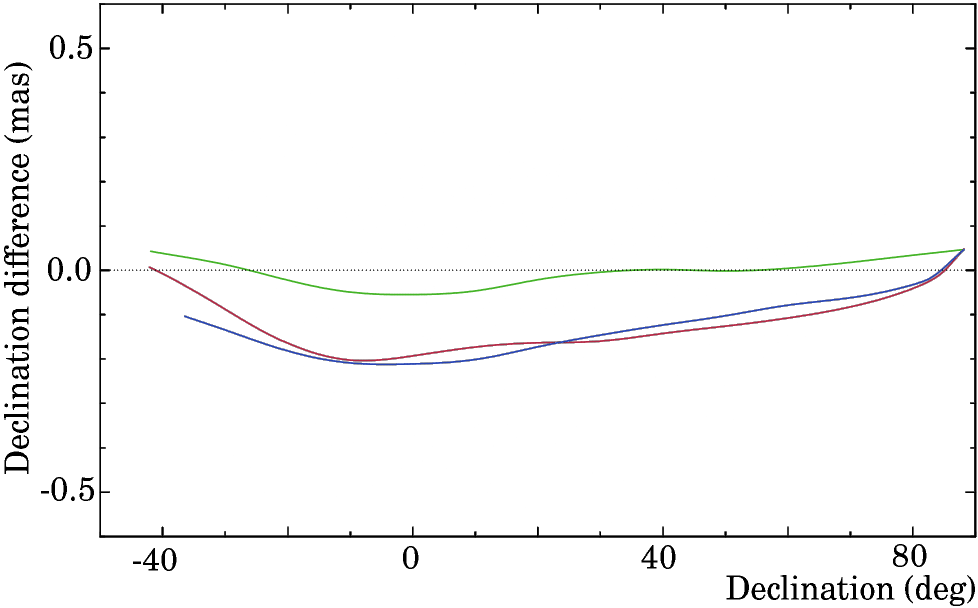}
   \hspace{0.039\textwidth}
   \includegraphics[width=0.48\textwidth]{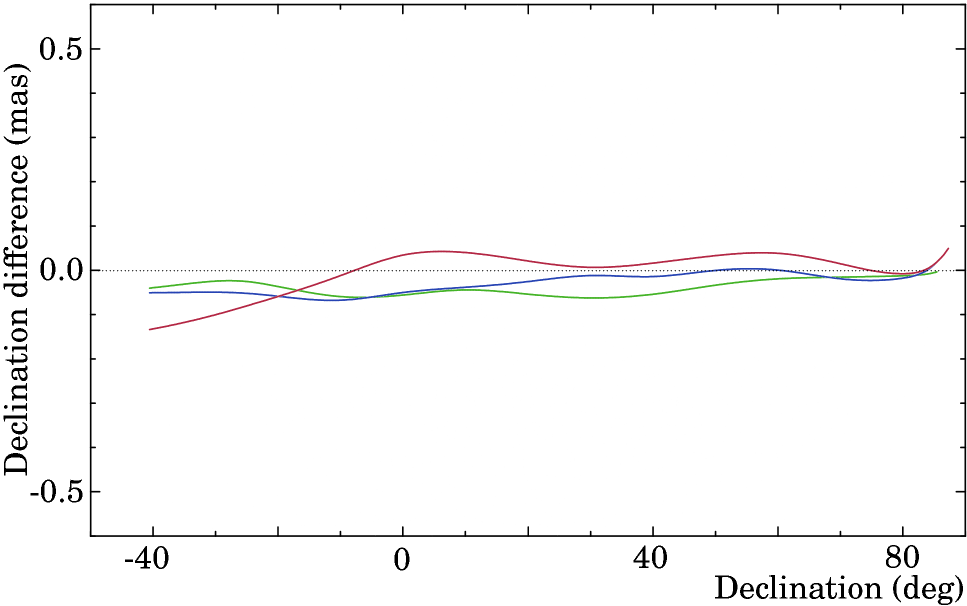}
   \caption{The smoothed differences in declinations from three 
            solutions using subsets of data as a function of 
            declination. {\it Left: } the differences in
            K-band solutions. {\it Right: } the differences in 
            dual-band solutions.
           }
   \label{f:delta_delta_segs}
\end{figure}

  In order to investigate the declination biases further, we ran 
three solutions using three data subsets. These subsets
corresponded to 1)~campaigns bj083 and ud001 in 2015.5--2018.5, 
2)~ud009 in 2018.7--2021.5, and 3)~ud015, bp251, s20tj, a20, a21, a22, 
and a23; in 2021.5--2023.5. Station positions were not estimated. 
The smoothed differences in declinations are shown in the left plot 
of Figure~\ref{f:delta_delta_segs}. The upper green line in these
plots corresponds to differences in solutions 1 and 2, blue and red 
lines correspond to differences in solutions 1 and 3 and in solutions 
2 and 3, respectively. The average biases are $-0.011, -0.146, 
-0.151$~mas. It is worth mentioning that subsets 1 and 2 were observed 
during the minimum of the solar activity and the subset 3 was observed 
during the maximum. As expected, the differences in solutions 1 and 2 
are the smallest. It is instructive to note that the declination bias 
is at a level of 0.2~mas. Is the declination biases a feature that is 
specific for K-band solutions?

   We performed a similar test for three subsets of dual-band solutions:
uf001, ug002, and ug003. Station positions were kept fixed. The differences 
in declinations between these three solutions are shown in the right plot of 
Figure~\ref{f:delta_delta_segs}. The average biases are 0.022, $-0.021$,
and $-0.040$~mas, respectively. We see that declination biases at 8~GHz are 
0.030--0.100~mas, which is significantly lower than at 23~GHz. We conclude 
that the declination biases caused by mismodeling path delay in the neutral 
atmosphere are enhanced by the unaccounted ionospheric contribution. We 
should also note that no biases in right ascension have been found. 
Decimation tests of the K-band solution show the presence of the extra 
variance on par with that found the dual-band solutions. Compare 
Figure~\ref{f:error_floor} with Figure~16 in \citet{r:radf}.

\begin{figure}[h!]
   \includegraphics[width=0.48\textwidth]{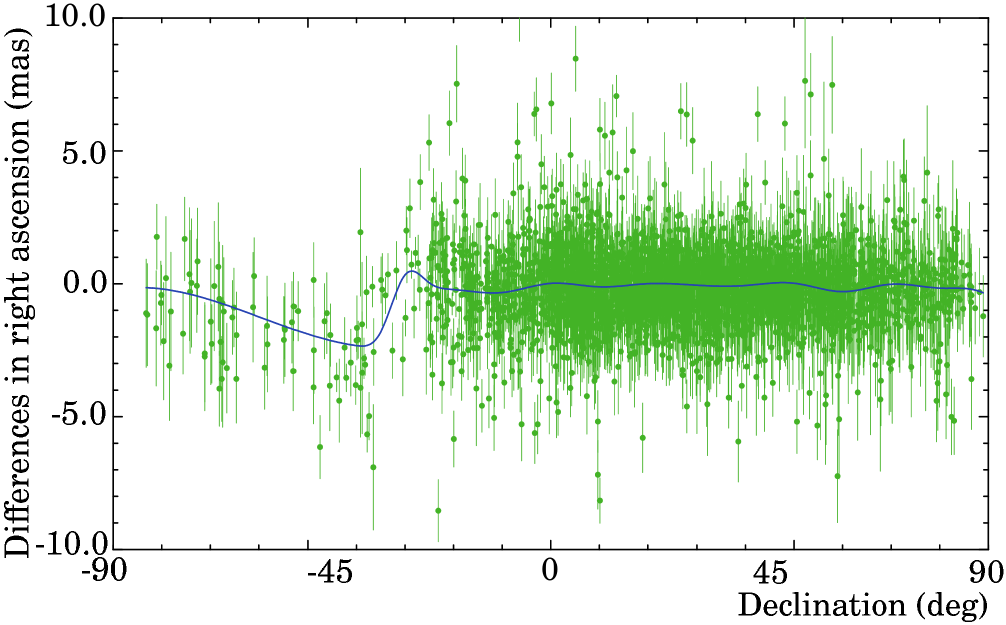}
   \hspace{0.039\textwidth}
   \includegraphics[width=0.48\textwidth]{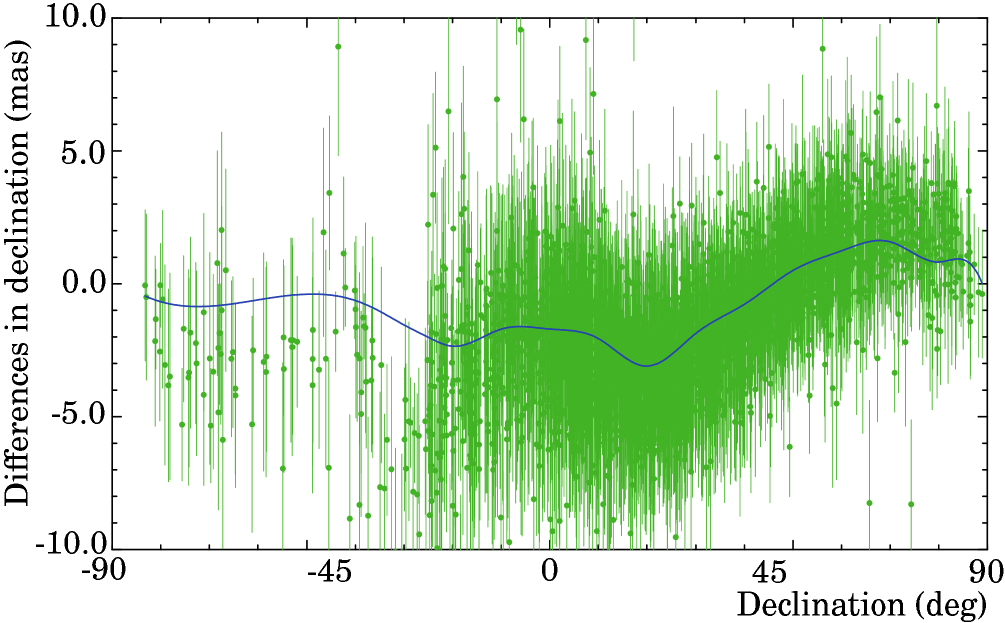}
   \caption{The differences in source position estimates from the S-band
            solution with respect to the dual band solution.
            Only sources with positional errors $< 3$~mas are shown.
            {\it Left: } the differences in right ascension scaled
            by $\cos\delta$. {\it Right: } the differences in 
            declination.
           }
   \label{f:diff_s}
\end{figure}

\begin{figure}[h!]
   \includegraphics[width=0.48\textwidth]{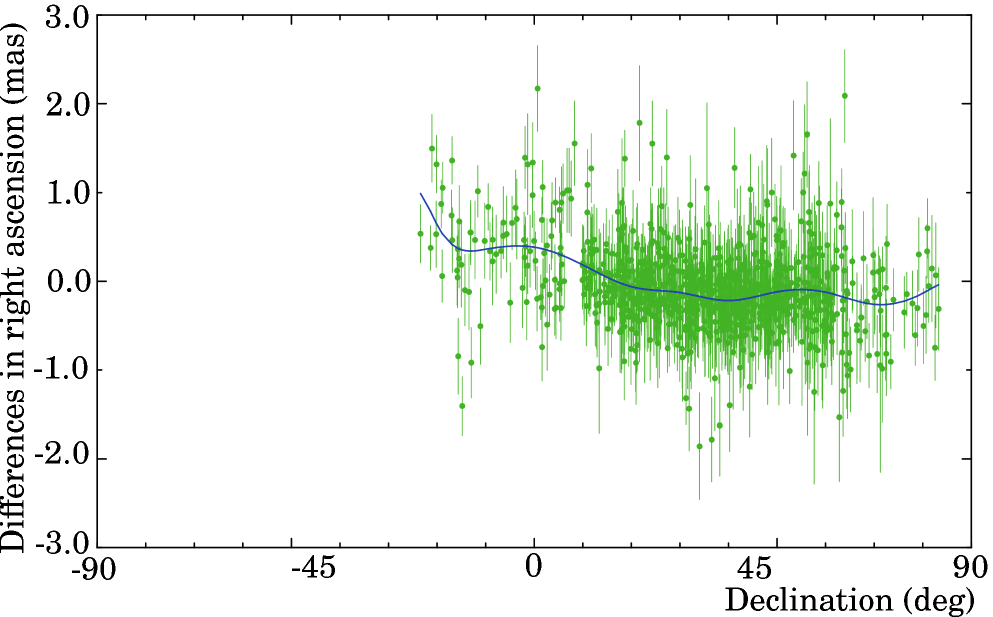}
   \hspace{0.039\textwidth}
   \includegraphics[width=0.48\textwidth]{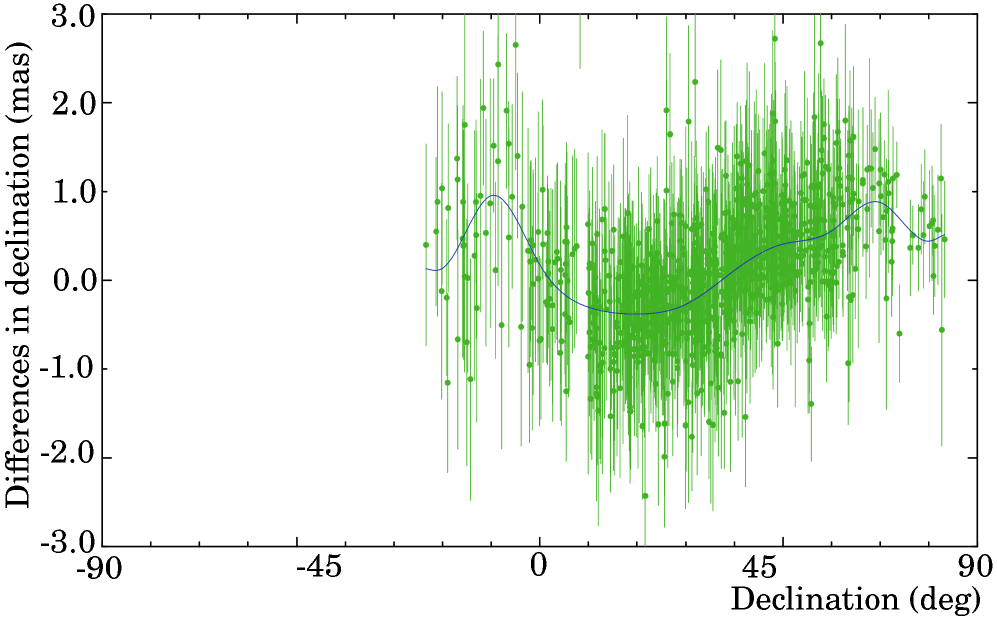}
   \caption{The differences in source position estimates from the C-band
            solution with respect to the dual band solution.
            Only sources with positional errors $< 1$~mas are shown.
            {\it Left: } the differences in right ascension scaled
            by $\cos\delta$. {\it Right: } the differences in 
            declination.
           }
   \label{f:diff_c}
\end{figure}

\begin{figure}[h!]
   \includegraphics[width=0.48\textwidth]{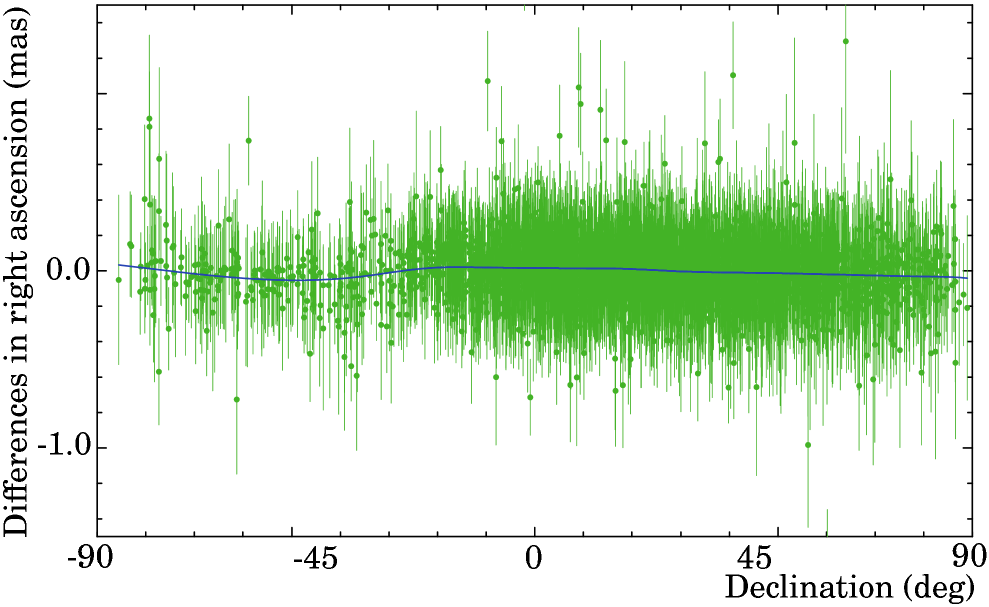}
   \hspace{0.039\textwidth}
   \includegraphics[width=0.48\textwidth]{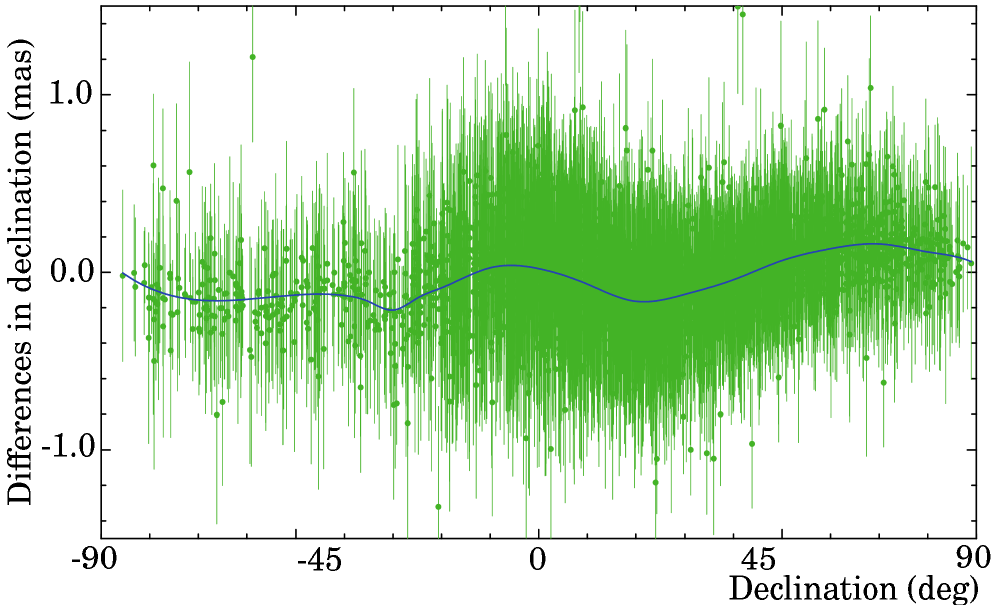}
   \caption{The differences in source position estimates from the X-band
            solution with respect to the dual band solution.
            Only sources with positional errors $< 0.5$~mas are shown,
            solutions using subset of data as a function of 
            {\it Left: } the differences in right ascension scaled
            by $\cos\delta$. {\it Right: } the differences in 
            declination.
           }
   \label{f:diff_x}
\end{figure}

  Figures~\ref{f:diff_s},~\ref{f:diff_c},~\ref{f:diff_x} show the 
source position differences derived from single-band data with respect
to the positions from dual-band data at S, C, and X-band, respectively.
We see the declination biases with the maximum approximately 3~mas 
at S-band, 0.7~mas at C-band, and 0.2~mas at X-band. The bias scales
approximately as reciprocal to a square of the observing frequency.
Differences in right ascension do not exhibit biases. We should stress
that the maxima in biases of single band solutions are comparable with
reported position uncertainties.

\subsection{Comparison with \Gaia}
\label{s:gaia}

  The only absolute astrometry catalogue with an accuracy comparable 
with VLBI is that produced by \Gaia mission \citep{r:gaia_edr3}. Most of 
the targets of \Gaia mission are stars, but a number of AGNs were detected
as well. We perform a procedure of matching common objects between
two catalogues using the approach described in \citet{r:gaia1}. 
We determined the list of 12,864 common sources in both catalogues 
with the probability of false associations not exceeding 0.001.
The procedure of cross association is described in section 
\ref{s:cross}.

\begin{figure}[h!]
   \centerline{\includegraphics[width=0.616\textwidth]{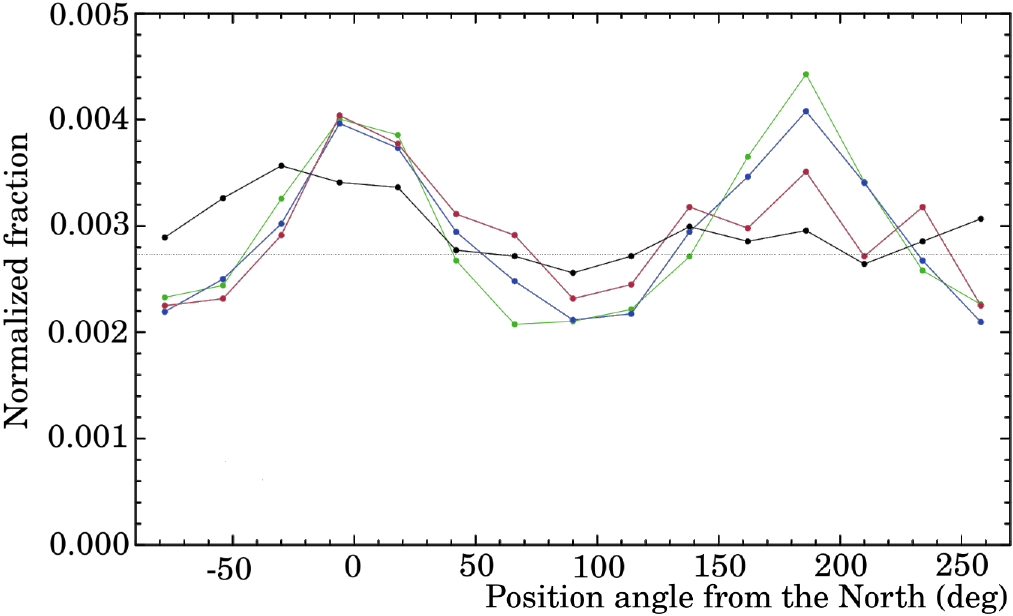}}
   \caption{The normalized distributions of position angles of the 
            differences in source coordinates from \Gaia with respect
            to RFC counted from the north celestial pole (green circles)
            for four subsets. Each subset corresponded to sources that 
            had a certain range of the number of observations used 
            in a solution. The green line: 3 to 40; the blue line: 
            40 to 80; the red line: 81 to 120; and the black line $>120$
            The thin dashed red line shows a uniform distribution.
           }
   \label{f:rfc_dual_gaia_hist}
\end{figure}

  A cursory examination of plots of position differences of the common 
sources did not reveal significant biases in source positions between 
these two catalogue along right ascension and declinations. The declination 
bias is $-0.025$~mas. However, a close examination revealed that the 
differences from \Gaia positions with respect to RFC positions favor the 
declination direction. Figure~\ref{f:rfc_dual_gaia_hist} shows the histogram 
of the position angles of the \Gaia positions with respect to the RFC 
positions. It turned out the less observations a given source had, the 
more noticeable peaks around $0^\circ$ and $180^\circ$ are in the histogram. 
The sources that had fewer than 30--60 observations were mostly observed 
in 1 to 2~scans with the VLBA. positional errors of these sources along 
declination are greater because the extension of the VLBA along longitude. 
The geometry of the VLBA has less impact on positions of those sources 
that were observed in many scans. We interpret it as observing 
in more than 1--2 scans provides a better sampling of different projections
of baseline vectors on the source direction, which makes estimate of source 
position more stable, in a similar way as more scans makes a better 
$uv$-coverage, which improves image fidelity.

\begin{figure}[h!]
   \includegraphics[width=0.48\textwidth]{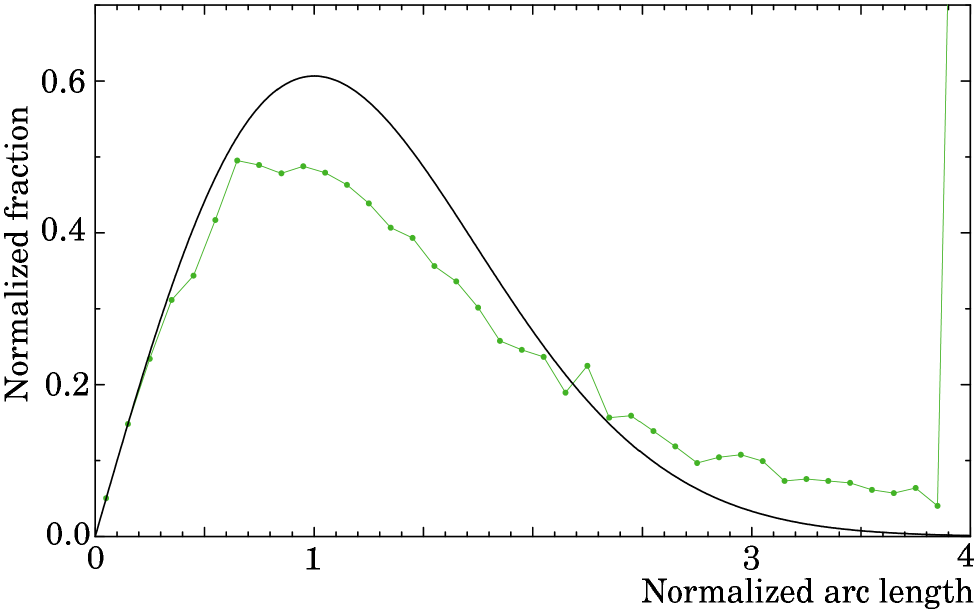}
   \hspace{0.039\textwidth}
   \includegraphics[width=0.48\textwidth]{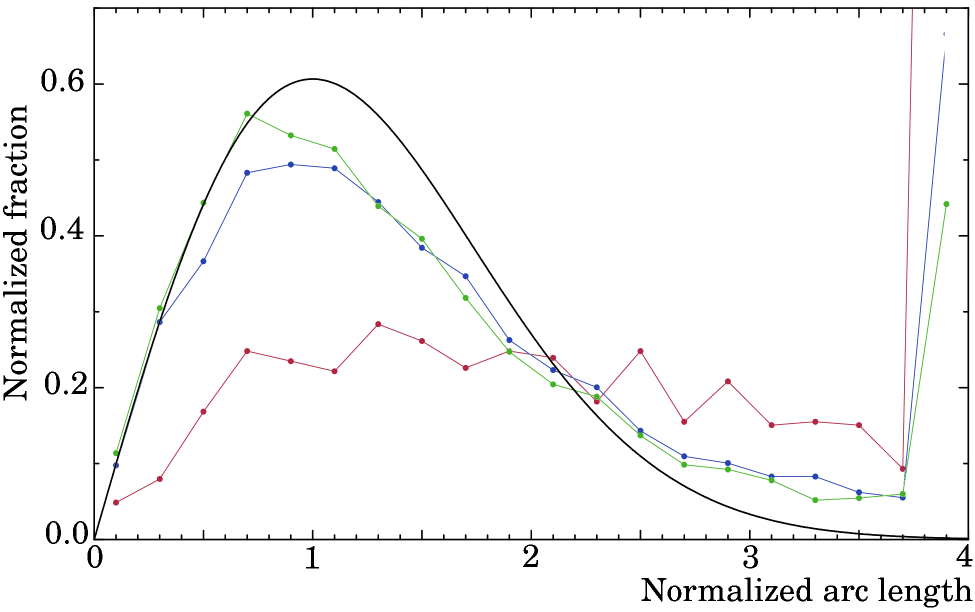}
   \caption{The distributions of the normalized arc lengths between RFC
            and \Gaia position estimates. A Rayleighian distribution with 
            $\sigma=1.0$ is shown with a black thick line for reference. 
            {Left: } all the matching sources. {Right: } the dataset was 
            split into three subsets according to $\chi^2/{\rm ndf}$ \Gaia\
            variable: the green line: $\chi^2/{\rm ndf} < 1.1$, the blue 
            line: $1.1 < \chi^2/{\rm ndf} < 2.0$, and the red line: 
            $\chi^2/{\rm ndf} > 2.0$. The last bin collects all the 
            normalized arc lengths $\geq 4$.
           }
   \label{f:rfc_gaia_distr_all}
\end{figure}

   In order to assess the level of the agreement between the RFC and 
the \Gaia Data release~3 \citep{r:gaia_edr3}, we computed arc lengths 
between VLBI and \Gaia positions along those 12,864 matches. We computed 
the uncertainties of these arc lengths based on reported uncertainties in 
right ascension and declination from both VLBI and \Gaia solutions and 
correlations between them  \citep[see][for details of the 
computation]{r:gaia4}. The left panel of Figure~\ref{f:rfc_gaia_distr_all} 
shows the distribution of arc lengths among all matching sources. We see 
the distribution has a certain deviation from the Rayleigh distribution. 
What is the origin of this deviation?

   It was noted in \citet{r:gaia4} that the statistics of \Gaia source 
positions depend on parameter {\sf astrometric\_chi2\_al} from the \Gaia 
catalogue, which is the ratio of $\chi^2$ per degree of freedom. We split 
the dataset into three brackets of 
$\chi^2/{\rm ndf}$: $< 1.1$ (46\% matches); [1.1, 2.0] 
(43\% matches); and $> 2.0$ (11\% matches) and computed the arc length 
distribution for each dataset. The results are shown in the right 
panel of Figure~\ref{f:rfc_gaia_distr_all}. The distribution from 
the subset with $\chi^2/{\rm ndf} < 1.1$ is the closest to the 
Rayleighian distribution, while the distribution from the subset with 
$\chi^2/{\rm ndf} > 2.0$ is strongly non-Rayleighian. Can the remaining 
difference be explained by omission of some error scaling factor in the 
VLBI solution?

\begin{figure}[h!]
   \centerline{\includegraphics[width=0.616\textwidth]{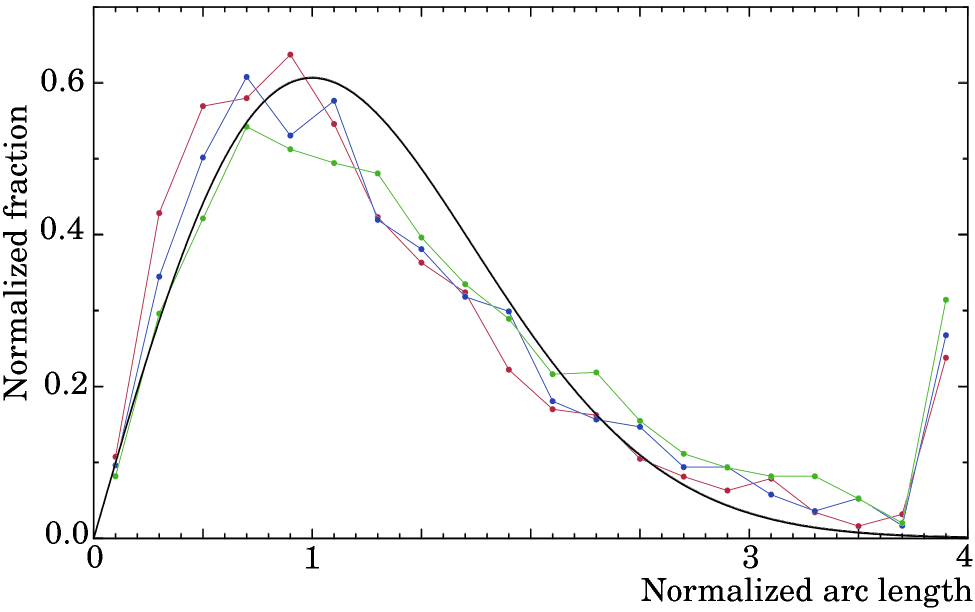}}
   \caption{The distributions of the normalized arc lengths between \Gaia 
            and RFC position estimates from a subset of matching sources 
            with differently re-scaled errors. A Rayleighian 
            distribution with $\sigma=1.0$ is shown with a black thick 
            line for reference. 
            The green line shows the distribution without re-scaling.
            The blue line shows the distribution with a re-scaling factor of 1.2.
            The red line shows the distribution with a re-scaling factor of 1.4.
           }
   \label{f:rfc_gaia_distr_limited}
\end{figure}

   To investigate the validity of the RFC source positional errors further,
we formed a subset of data with a)~$0.9 < \chi^2/{\rm ndf} < 1.2$; 
b)~excluding those sources that have the position angle of differences
VLBI minus \Gaia with respect to the jet direction less than $30^\circ$;
c)~the semi-major axis of the positional error ellipse $<0.2$~mas 
either VLBI or \textit{Gaia}. It was found in \citet{r:gaia1,r:gaia2}
that VLBI/\Gaia positions differences favor the jet direction 
\citep{r:jet_dirs}. Strong pieces of evidence were presented in 
\citet{r:gaia3,r:gaia4,r:gaia5,r:lambert24} in favor of the explaining
this phenomena with the presence of bright optical jets that are shorter 
than the \Gaia point spread function, but are still long enough to 
affect positions of source centroids in the optical range. Exclusion of 
these objects eliminates the origin of the discrepancies that is not related 
to astrometry errors. We excluded sources with small position 
errors to avoid a dichotomy whether the error re-scaling models should
be additive or multiplicative since we cannot discriminate them. We divided 
\Gaia sources uncertainties by $\chi^2/{\rm ndf}$  following the line of
evidence presented in \citet{r:gaia4}. Then we computed three distributions 
by scaling VLBI errors by a factor 1.0, 1.2, and 1.4. The distributions 
are shown in Figure~\ref{f:rfc_gaia_distr_limited}.

  We see that re-scaling VLBI errors even by a factor of 1.2 increases the 
disagreement with the \Gaia catalogue. When we increase the scaling factor,
the distribution of normalized arc lengths is shifted to the left. Although
we excluded a number of sources in these tests, it did not cause a selection
bias for the validation of VLBI error model, because the exclusion criteria
were based entirely on properties of the \Gaia sample.

  Presented results allow us to conclude that a)~the deviation of the 
normalized arc length distribution of RFC/\Gaia EDR3 position differences
is explained in part by unaccounted \Gaia positional errors and by the presence
of milliarcsecond scale source structure; b) the upper limit of biases
in the RFC error model is 20\%, which is in line with \citet{r:gaia_edr3}.

\subsection{Impact of source structure}

  Equation~\ref{e:e27} relates the position of a point with coordinates 
$\vec{s}$. In general, the complex coherence function $\Gamma_{12}$ 
according to the Van~Zitter--Zernike theorem is:
\beq
   \begin{array}{lcl}
      V_{12}(b_x,b_y,f) &=& e^{\tss 2\pi i \, f \tau_0} \, V_s(b_x,b_y,f), \\
      V_s(b_x,b_y,f)    &=& \dss \dintinf B(x-x_0,y-y_0,f) \, e^{\tss -2\pi i (x u + y v )} dx\,dy,
   \end{array}
\eeq{e:e27}
  where $f$ is the circular reference frequency of received signal,
$\tau_0$ is the geometric delay to the nominal reference point on 
the source $(x_0,y_0)$, $B$ is a two-dimensional function of the 
brightness distribution, which depends on angular coordinates $x$, $y$
along right ascension and declination and on frequency,
and $u=b_x \, f/c$, $v = b_y\, f/c$ are scaled projections of the 
baseline vector $\vc{b} = \vc{r}_1 - \vc{r}_2$ to the plane that is 
perpendicular to the image plane at the image nominal reference point. 

  When $B(x,y) = \delta(x,y)$, the integral in the lower formula in
equation~\ref{e:e27} becomes 1 and the source structure contribution 
$\tau_{\rm gr,str} \Frac{1}{2\pi} \, \Frac{\partial{}}{\partial{f}} \, \arg(V_s)$ 
becomes zero. The integral will be constant also if a source has a circular 
symmetry. In all other cases the source structure  contribution will not be 
zero. \citet{r:tho80} and later \citet{r:cha90} considered a simple
case of a two=component model. Although the contribution of group delay 
for that simple case can be written in a close form analytically, 
the contribution of source structure to source position
does not have a simple analytical form. It depends not only on the source 
brightness distribution, but also on the geometry of the network and on 
the observing schedule. 

  When the source brightness distribution is well known, source 
contribution to group delay can be computed from the integral in
expression~\ref{e:e27}. Although the feasibility of this approach 
was established in \citet{r:zep88}, source structure contribution
is not yet applied on a routine basis because of logistical 
difficulties in developing of the infrastructure that would support 
the synthesis of reliable images and identifying the reference point
on the image in an automatic fashion using all the data. The 
RFC catalogue, like all other prior VLBI astrometry catalogues, does 
not apply source structure contribution, and the omission of 
such a correction propagates as a systematic source-specific error 
to reported positions. 

  We can evaluate the magnitude of these errors via simulation, 
comparison of source positions with and without applied source structure
contribution, and via comparison of source positions derived without
applying the source structure contribution at difference frequencies. 
The simulation study of \citet{r:pla16} demonstrated that the 
contribution of source structure to delay at 8~GHz affects source positions
in the range of 10--80~$\mu$as for most of the sources. In \citet{r:gaia3} 
we processed a dataset of 29~active galactic nuclei observed under 
MOJAVE program \citep{r:mojave_apjs} at 15~GHz, computed source structure 
contribution to group delay from images, applied it to data analysis, 
derived source positions, and compared source position estimates with 
and without accounting to source structure. Position differences were 
in the range from 0.01 to 2.40~mas with the median of 0.06~mas, which is 
consistent with simulation. 

  In \citet{r:radf} we computed the position angles of the differences in 
source coordinates derived from X/S data with respect to source coordinates 
derived from K-band data, subtracted the position angle of jet direction,
and built a histogram of the resulting position angles counted from jet 
directions. That histogram showed  two peaks along and opposite to the 
jet direction. Detailed modeling of the histogram allowed us to estimate 
the magnitude of the systematic position differences along the jet direction 
in a form of a Gaussian distribution. The second moment of the Gaussian 
distribution was in the range of 90~$\mu$as (low limit) to 120~$\mu$as 
(upper limit). Since the jet axis is the intrinsic property of the source 
that  describes the asymmetry of the brightness distribution, these 
systematic errors are undoubtedly associated with the presence of source 
structure and/or core shift. All these estimates are consistent with each 
other and indicate that {\it on average}, the unaccounted source structure 
contribution is close to 70~$\mu$as per source position component,
right ascension and declination.

  Words ``on average'' are essential in the above mentioned estimate.
Comparison of X/S and K-band catalogues in \citet{r:radf} showed that
the position differences derived form X/S data versus derived K-band data
exceeded $3\sigma$ for 6\% sources and $5\sigma$ for 2\% sources.
There are sources with differences in position estimates derived from 
observations at different frequencies that exceed the average difference
by orders of magnitude. Examples of such sources were reported for the first 
time in \citet{r:vgaps}. Later, more sources like those have been found 
\citep[for instance,][]{r:obrs2,r:tit22,r:min22}. All of this sources have 
a striking feature: they have two or more compact components with 
a comparable brightness. The most prevailing class of such sources show 
core-jet morphologies with a compact bright hot spot in the jet. 
The second, more rare class is gravitational lenses \citep{r:grav_lenses}. 
And finally, there are several sources that are genuine binary systems,
such as J0405$+$3803 \citep{r:bin_agn}, J0749+225A/J0749+225B 
\citep{r:shen21}, and J2157+662A/J2157+662B (I. Duev et al., in preparation) 
of gravitationally bounded AGNs. Establishing the nature of these sources 
is in general not an easy task, and it requires supporting argumentation. 
This task goes beyond the scope of the current study and we reserve it 
for future publications. Here we collectively call these sources visually 
binary.

  The fringe fitting process, as implemented, implicitly assumes that
the source brightness distribution is a $\delta$ function. Morphologies 
in the form of a compact core and an extended asymmetric feature do not 
affect the efficiency of the fringe fitting process in a measurable way. 
However, morphologies in the form of more than one compact components do 
affect the fitting. It is advantageous to apply an a~priori source brightness 
distribution model during fringe fitting in processing VLBI observations
of gravitational lenses \citep{r:porcas04}. We did not do that step to 
keep data analysis uniform. Sources with more than one compact 
component will be reprocessed in the future.

  Since with rare exceptions the spectral index of a core is flat, 
but the spectral index of a hot spot in a jet is steep 
\citep[e.g.,][]{r:mojave_spectra,r:plavin_cs}, situations when the hot 
spot is brighter at a low frequency, but the core is brighter at a higher 
frequency, are not uncommon. There are instances that the position estimate 
at a low frequency corresponds to one component (usually a hot spot in 
a jet), but at higher frequencies it corresponds to another component 
(usually a core). Such sources are called colloquially ``flip-floppers.''
Such a feature affects source position estimates from single-band data: the 
reported position is close to the specific source component. Such 
a feature affects the source position estimate from dual-band data as 
well, but in a way that is a little bit less obvious: the dual-band 
position estimate is shifted by a factor of $a=f^2_l/f^2_u$ along 
the line connecting two components in the direction opposite to the 
second component. 

  In \citet{r:radf} we found that the total number of outliers between 
dual-band, quad-band, and single-band observations at 23~GHz exceeding 
$3\sigma$ was at a level of 6\% and exceeding $5\sigma$ was at a level 
of 2\%. For most of these cases a close examination of images easily 
revealed peculiar source structure, f.e., the presence of a second 
component. It follows from this comparison that for 94\% of the sources
the contribution of source structure on source position is not 
detected, and for 2\% of the sources it is definitely dominates 
the error budget.

  We note that even if the source structure effect can be compensated for,
this does not mitigate the problem of the absence of a stable reference 
point in extragalactic radio sources. Physically, we expect the nuclei 
of active galaxies, i.e.\ supermassive black holes and accretion disks 
surrounding them, as well as the true base of AGN jets, to have stable 
positions \citep{r:blandford19}. However, they are not observable with 
ground-based VLBI arrays at centimeter wavelengths. The next subsection 
discusses in detail properties of the bright apparent jet base that
is also called ``the core'' --- the feature that is actually observed.

\subsection{Impact of the core-shift on reported source positions}
\label{s:cs}
  
  AGN morphology at milliarcsecond resolutions is typically characterized 
by an opaque core and an optically thin jet that may or may not show bright 
features on VLBI images \citep{r:mojave_spectra,r:mojave_apjs,r:blandford19}. 
Due to the limited dynamic range, the jet may not be detected, but the core 
as a partially resolved feature is almost always present. The size of the 
visible core is determined by the area for which the optical depth due to 
synchrotron self-absorption is close to unity \citep{r:BK79}. The center of 
the core is shifted along the jet with respect to the central engine due to 
the synchrotron opacity. For simplicity, we assume here that the distance 
between the AGN central engine and the true physical jet base is equal 
to zero. 

Since the synchrotron opacity depends on frequency $f$, the apparent core-shift 
is frequency dependent \citep{r:lob98}. Considering the core-shift frequency 
dependence is described as a power law, the source position 
$\vec{S}(f) = \vec{S}_0 + \vec{h} \, f^{a}$, where $h$ is usually
aligned with the jet direction. In the first approximation, the fringe phase 
is the product of the travel distance difference and frequency divided by the 
speed of light. Phase delay is the ratio of the fringe phase to the reference 
frequency and therefore, is equal to the travel distance difference divided 
by $c$, while group delay is a partial derivative of fringe phase over frequency:
\beq
   \tau_{\rm gr} = \Frac{1}{c} \, 
                   \der{\bigl( f \, (\vec{S}_0 + \vec{h} \, f^{a}) \cdot \vec{b}\bigr)}{f}
                   \enskip + \enskip O(c^2)
                   \enskip = \enskip
                   \Frac{1}{c} \, \vec{S}_0 \cdot \vec{b} \enskip + \enskip
                   \Frac{1}{c} \, (1+a)(\vec{h} \cdot \vec{b}) \, f^a
                   \enskip + \enskip O(c^2),
\eeq{e:e28}
  where $\vec{b}$ is the baseline vector in the inertial coordinate system.

  When the energy density of the relativistic particles and magnetic field 
is approximately equal, the so-called equipartition condition, and jet geometry 
is conical, this dependence is predicted to be $f^{-1}$ \citep{r:K81,r:lob98}.
Observations, in generally, confirm that 
\citep[e.g.,][]{r:kov08,r:sok11,r:abe18}. However examples of deviations of 
the power law from $-1$ are also known \citep[e.g.,][]{r:kutkin14,r:chamani23}. 

  To make the situation even more complicated, according to 
\citet{r:plavin_cs,r:chamani23}, core-shift varies with time on a scale from 
months to years, and these variations are related to a flaring activity.
During flares the density of charged relativistic 
particles moving through the jet changes.
These disturbances cause a violation of the equipartition condition, and 
as a result, break the $-1$ power~law.
Moreover, due to the basic causality arguments and 
the finite speed of plasma propagation along the jet, any plasma 
disturbance moving along the jet will break the $-1$ power~law \citep{r:plavin_cs}.

\newcommand{\NNote}[1]{\Rdb{#1}}
  When the power~law index in the core-shift  versus frequency dependence 
is $-1$, as we see from equation~\ref{e:e28}, core-shift has no impact 
on astrometric position derived from group delays which in this case 
pinpoints the true base of AGN jet as it was pointed out by \citet{r:por09}. 
We underline that this is the case for all presented solutions: fused, 
dual-band, and single-band. The core-shift impacts source positions derived 
from group delays only when the power~law index deviates from $-1$, 
the closer the power~law index to $-1$, the less is its impact.

  We do not apply correction of the core-shift to data reduction. 
Therefore, its contribution manifests as a noise which is challenging 
to evaluate due to the complex effects discussed above.
To date, the core-shift was measured for more than a hundred AGNs 
by \citet{r:kov08,r:sok11,r:MOJAVE_IX} and its variability was systematically 
studied in 40 AGNs \citep{r:plavin_cs}. Namely, it was found that the typical
differential core-shift between 2 and 8~GHz was on a level of 0.4--0.5~mas, 
but it may reach 1.5~mas \citep{r:kov08,r:plavin_cs}. According to 
\citet{r:plavin_cs}, the typical differential core-shift rms between 2 and 
8~GHz is 0.18~mas. \citet{r:kov08} provided a theoretical estimate of the 
typical core-shift between 8~GHz and the true jet base: 0.1~mas. However,
when source positions or source position differences are derived from 
{\it phase delays}, the core-shift will affect them at that level.

In order to estimate the magnitude of the core-shift impact on RFC 
source positions, we need to utilize epoch-specific variability information 
on both as well as the core-shift value and its power~law, which are 
currently poorly known. The scarcity of such measurements prevents us from 
making quantitative estimates. In \citet{r:radf} we have established that 
the extra noise along the jet between 8 and 24~GHz positions has a second 
moment of 0.09~mas. We attribute this to both the core-shift and source 
structure contribution. This estimate can be considered as a conservative 
upper bound of the differential core-shift impact between 8  and 24~GHz.

\subsection{Impact of scattering in the interstellar medium}

\begin{figure}[h!]
   \includegraphics[width=0.21\textwidth]{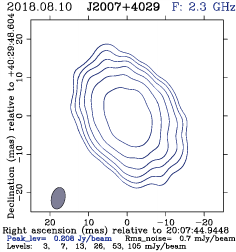}
   \includegraphics[width=0.23\textwidth]{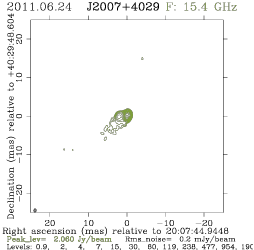}
   \includegraphics[width=0.269\textwidth]{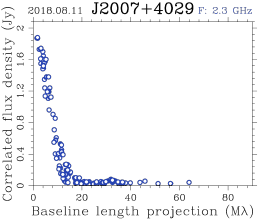}
   \includegraphics[width=0.271\textwidth]{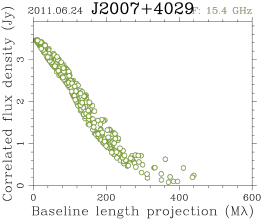}
   \caption{Example of a source that is affected by scattering
            in the interstellar medium. The first plot from the left:
            scattered image at 2.3~GHz. The second plot: image of the 
            same source at 15.4~GHz where scattering is negligible.
            The third and the fourth plots show the calibrated visibility
            amplitude as a function of the projected baseline length.
            See details in \citet{r:korukova23}.
           }
   \label{f:rfc_scattered}
\end{figure}

  A source positional error from a given observation is reciprocal to the 
projected baseline length. Due to refractive scattering in the interstellar 
medium \citep{r:puskov15,r:kor22}, the correlated flux density at long 
baselines at low frequencies is substantially reduced, sometimes by 
one order of magnitude, and a corresponding image reveals a smooth 
elliptical Gaussian shape --- see Figure~\ref{f:rfc_scattered} as an 
example.

  Scattering in the interstellar medium affects mainly compact AGNs 
within $\pm10^\circ$ of the Galactic plane. Group delay uncertainties are 
increased by the same factor, which increases the positional error. Correlated 
signal at long baselines may fall even below the detection limit, and 
therefore, such observations will not be used in data analysis, which 
increases the source positional error even further. The scattered source 
size it typically depends on frequency as $f^{-2}$, while the 
intrinsic opaque core size depends on frequency as $f^{-1}$ 
\citep[see discussion in][]{r:lob98,r:puskov15,r:kor22}. 
As a result, the higher the frequency, the lower the influence of 
scattering is expected on astrometric accuracy. 

  We should notice that correlated amplitudes in Figure~\ref{f:rfc_scattered} 
have a steep drop in the range form 0 to 15~$M\lambda$, but then do not vanish, 
and stay of 1--2\% of the peak level in the range of 20 to 64~M$\lambda$. 
This pattern is due to the presence of refractive scattering substructures 
in an AGN, originally discovered for Sgr\,A${}^\star$ with the 
VLBA+GBT \citep{r:gwinn14} and found also in quasars \citep{r:johnson16} 
using Space VLBI observations with \textit{RadioAstron}\citep{r:radik}. 
Although in these cases the scattering is broad enough to be completely 
resolved out, substructures with smaller size are compact enough to 
provide a measurable flux density at long baselines with a random, 
noise-like character. According to \citet{r:gwinn14}, the substructure 
is expected to be fixed for a time of an observing session, 4 to 24~hour 
and average out over times longer than that for Galactic rotation to 
carry the line of sight across the scattered image, i.e.\ a few weeks. 
Although these observations at long baseline will be used in astrometric 
solutions, the position estimates will be related either to the 
scattering substructure of an AGN if data from only one epoch are used, 
or will be affected by a jitter due to changing substructures when data 
from many epochs are used in data analysis.

  Additionally, a lens-like structure in the ionized component of the 
Galactic interstellar medium may rarely produce multiple images of a compact
AGN due to refractive plasma effects. This can be observed at parsec scales. 
Such a phenomenon can affect even high radio frequencies and, consequently, 
degrade astrometric solutions, see an example for the quasar 2023+335 
at 15~GHz in \citet{r:pushkarev13}. 

\subsection{Comparison with ICRF3}

  Since most VLBI data are publicly available, there are other groups that 
use the data and derive source positions. It is instructive to compare 
our catalogue with other solutions based on a subset of the full 
collection of VLBI data. We selected for comparison the ICRF3 catalogue 
\citep{r:icrf3}. The ICRF3 catalogue lists positions of 4536 sources 
derived from the dual-band solution and 824 sources derived from K-band. 
It also contains positions derived from 32~GHz VLBI data, but since that 
catalogue is based entirely on proprietary data, we exclude it from 
consideration. All the sources from the ICRF3 catalogue are found in the 
RFC, and therefore, we can consider it as a subset of the RFC, 
approximately 1/5 of the total RFC source count. The ICRF3 includes 
{\it some} astrometric programs \citep[see for details][]{r:icrf3}, 
while we aimed to use {\it all} suitable observations in our work.

  The two catalogues, RFC and ICRF3 (dual-band version), are rotated 
against each other at angles of  0.024,  0.048, and $-0.042$~mas 
along x-, y-, and z-axes. The declination bias defined as the weighted
mean declination difference, which is $-0.020$~mas. The catalogues have 
a different error model. The ICRF3 catalogue was derived from group 
delays computed by the AIPS and Fourfit software 
\citep{r:aips,r:fourfit_ivs}, which did not account for the phase 
noise and therefore, their estimates of group delay uncertainties were 
smaller and less realistic than those derived by \PIMA \citep{r:vgaps}. 
The ICRF3 catalogue adopted a multiplicative scaling factor of 1.5 and 
an error floor of 0.030~mas based on the local decimation, while RFC used 
scaling factors 1.08 and 1.16 for right ascensions and declination and 
the declination-dependent error floor in a range from 0.06 to 0.2~mas
based on results of global decimation.

\begin{figure}[h!]
   \centerline{\includegraphics[width=0.616\textwidth]{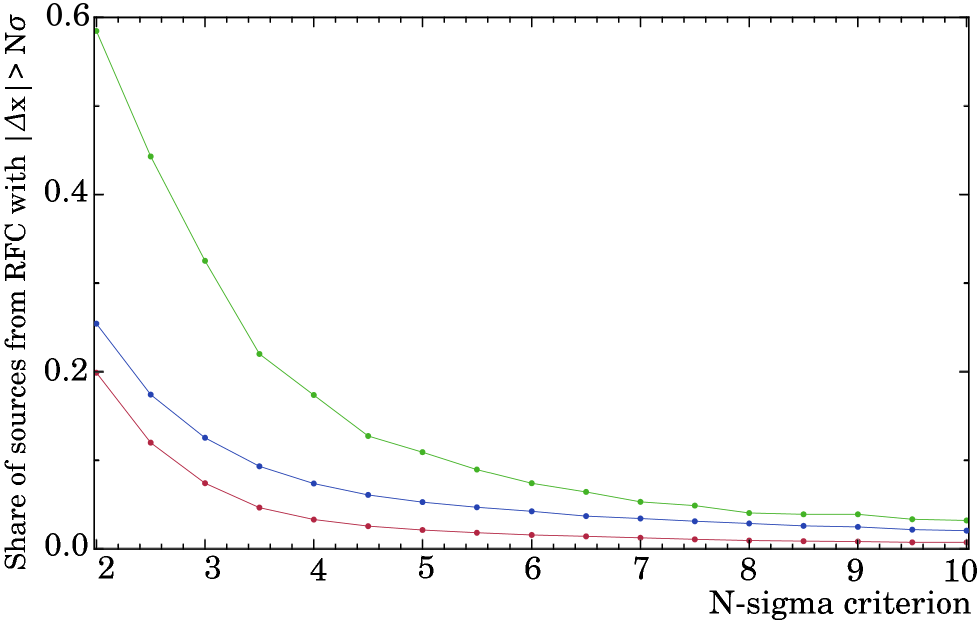}}
   \caption{The distribution of the share of sources with
            position differences between the RFC and ICRF3 catalogues
            significant at a given $N\sigma$ level. 
            The upper green line
            corresponds to $\sigma = \min(\sigma_{\rm rfc},\sigma_{\rm icrf3})$.
            The middle blue line corresponds to $\sigma = \sigma_{\rm rfc}$.
            The low red line corresponds to 
            $\sigma = \max(\sigma_{\rm rfc},\sigma_{\rm icrf3})$.
           }
   \label{f:rfc_icrf3_nsig}
\end{figure}

  Figure~\ref{f:rfc_icrf3_nsig} shows the share of sources with the 
differences $|\Delta x$ in the positions between two catalogues as a function
of the significance level defined as $|\Delta x| < N \sigma$.
In order to account for the disparity of the error models, three 
distributions were plotted: 
$|\Delta x| < N \min(\sigma_{\rm rfc},\sigma_{\rm icrf3})$ (upper plot), 
$|\Delta x| < N \sigma_{\rm rfc}$ (middle plot), and 
$|\Delta x| < N \max(\sigma_{\rm rfc},\sigma_{\rm icrf3})$ (low plot). 
It is remarkable that one third of the sources have differences at the 
$3\sigma$ level when the minimum between the uncertainties of the two 
catalogues is considered. What is the origin of these differences?

\begin{figure}[h!]
   \includegraphics[width=0.384\textwidth]{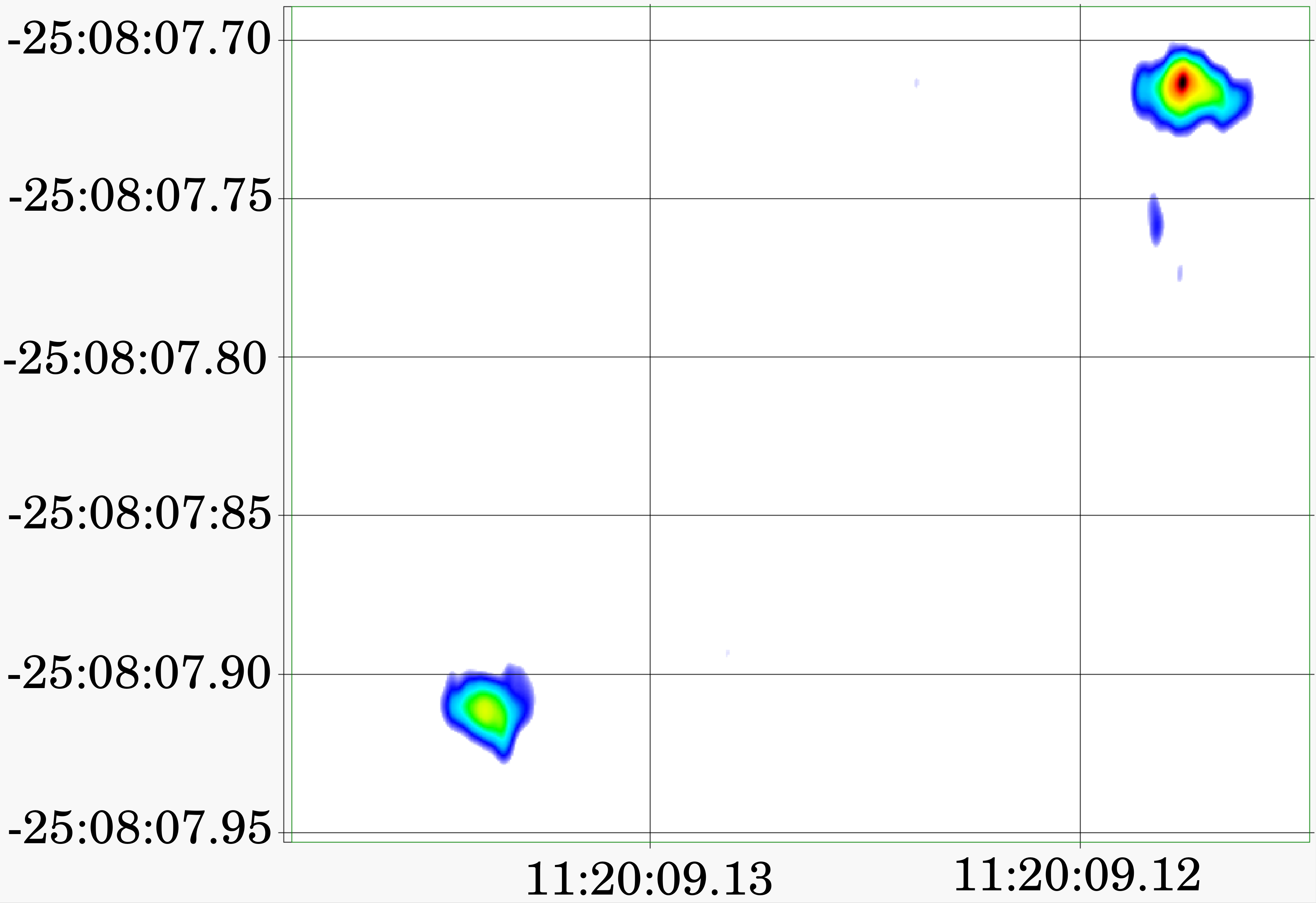}
   \hspace{0.039\textwidth}
   \includegraphics[width=0.46\textwidth]{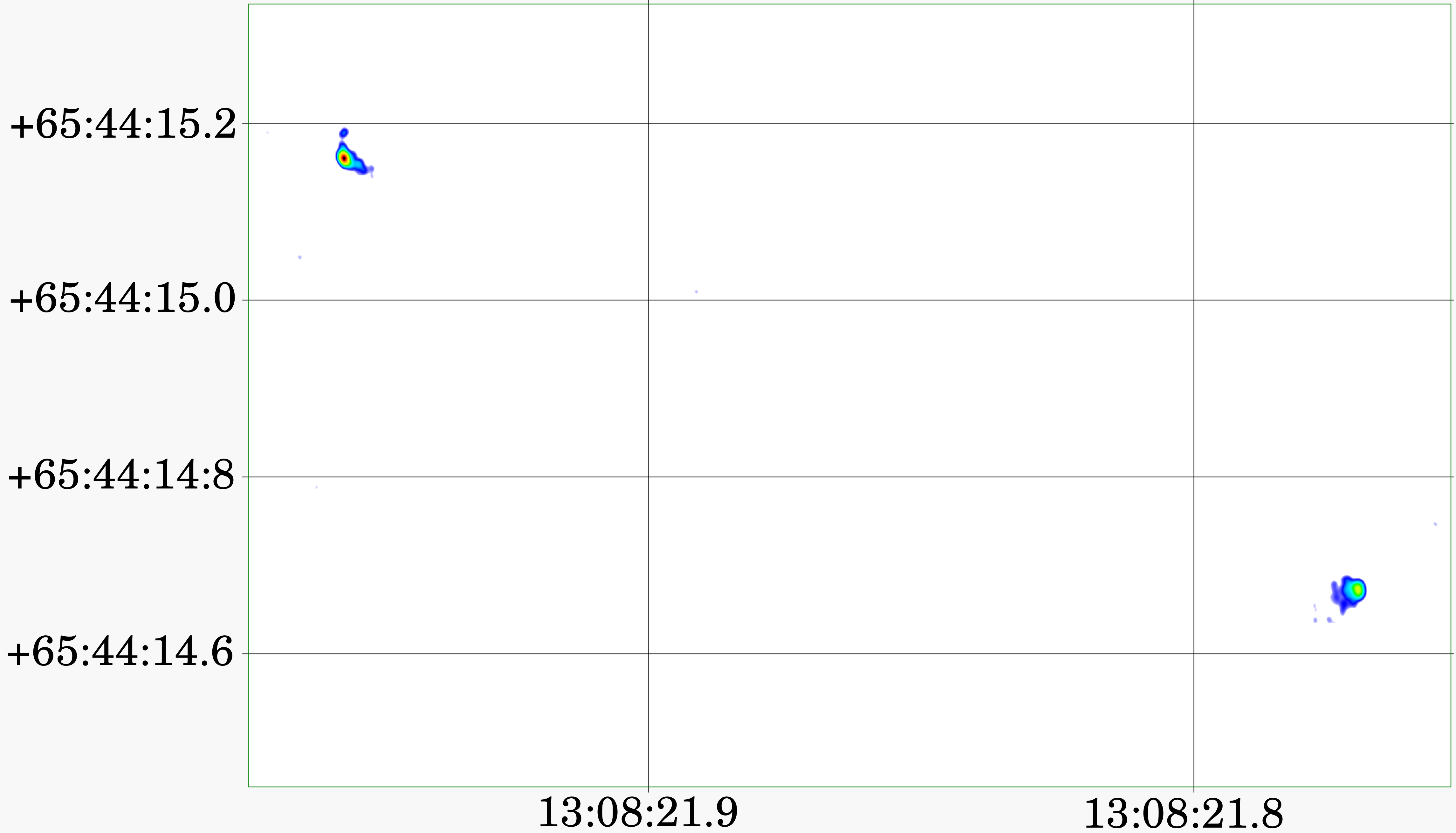}
   \caption{Example of S-band images of sources with large differences 
            between RFC and ICRF3 catalogues. \\
           {\it Left: } RFC~J1120$-$2508.
           {\it Right: } RFC~J1308$+$6544.
           }
   \label{f:rfc_icrf3_maps}
\end{figure}

  The position uncertainties in the RFC and ICRF3 catalogues heavily depend
on the number of observations. For all the sources, the RFC uses all the 
observations the X/S ICRF3 catalogue used and many other observations the 
ICRF3 did not use. For frequently observed sources for which ICRF3 positional 
errors $< 0.1$~mas, the RFC uncertainties are greater by the factor of 3.54 
for right ascension and 2.12 for declinations. For the sources with ICRF3 
positional errors $< 0.2$~mas, the RFC uncertainties are greater by factors 
of 2.90 and 1.76 because of differences in the error scaling model: the RFC 
uncertainties account 
for the red noise while the ICRF3 uncertainties do not. From the other hand, 
for those sources for which the ICRF3 errors exceed 0.5~mas, the RFC errors 
are 0.35 and 0.57 of the ICRF3 errors for right ascensions and declinations. 
Among the sources with ICRF3 errors greater than 1.0~mas, their RFC errors 
are 0.28 and 0.38 of ICRF3 errors, i.e.\ the positional accuracy is better
by a factor of 3--4 due to including more observations in analysis. Within
each bin in Figure~\ref{f:rfc_icrf3_nsig}, about 60\% of the sources have 
significant differences because their ICRF3 position uncertainties are too 
small, and remaining 30\% have significant differences because the ICRF3 
solution used only a small share of the observation that the RFC solution 
used. Sources with differences more than 
$3 \times \max(\sigma_{\rm rfc},\sigma_{\rm icrf3})$ fall into two 
categories: the sources with many more used observations in the RFC than 
in the ICRF3 and peculiar sources, such as visual doubles. As an example, 
Figure~\ref{f:rfc_icrf3_maps} shows images of two sources that have 
position differences more than $25\sigma$.

  The uncertainties of K-band positions are about a factor of 2 smaller
in the RFC catalogue than in the ICRF3 catalogue owing to much more 
observations used. The declination bias of our K-band source positions
with respect to ICRF3 is $-0.075$~mas. There is no declination bias of 
the ICRF3 K-band catalogue with respect to the ICRF3 dual-band catalogue. 
Since the ICRF3 catalogue used sixteen K-band experiments that are not 
publicly available, we cannot reproduce the K-band ICRF3 solution exactly. 
Therefore, the origin of the bias is not firmly established. We should 
note that in general, the declination bias in source positions derived 
from K-band is not stable since it depends on the total electron content
in the ionosphere. We found comparable biases when we processed subsets 
of K-band data (see left plot in Figure~\ref{f:delta_delta_segs}). This 
helps to explain the discrepancies of the ICRF3 K-band catalogue against 
the RFC catalogue.

  The 25\%, 50\%, and 75\% quartiles of the semi-major error ellipse axes
of the dual-band version of the ICRF3 are 0.14, 0.22 and 0.41 mas 
respectively. The same statistics of the RFC catalogue {\it among the list
of sources present in the ICRF3} are 0.16, 0.22, and 0.29, respectively.
 We see that the first quartile of the positional error of the ICRF3 
catalogue is 15\% smaller because it does not account for the red noise, 
the second quartile, i.e.\ the median are the same, and the positional 
accuracy of the RFC is 38\% greater in the third quartile because the RFC is
based on more observations than the ICRF3. Table~\ref{t:rfc_icrf3_comp}
shows the number of sources with positional errors less or greater than
some limit. The number of sources with positional errors less than 
0.21~mas is the same in both catalogues, but then this number from the RFC
grows much faster.

\begin{table}[h!]
   \caption{The number of sources with the semi-major error ellipse 
            axis within certain error ranges in the ICRF3 and RFC catalogues.
           }
   \begin{center}
      \begin{tabular}{rrr}
          \hline\hline
           Error  & ICRF3 & RFC    \\
           \ntab{c}{(mas)}  & \ntab{c}{\# src} & \ntab{c}{\# src} \\
          \hline
          $< 0.2$ & 2022  &  1887  \\
          $< 0.3$ & 2899  &  3896  \\
          $< 0.5$ & 3653  &  5658  \\
          $< 0.7$ & 3961  &  6797  \\
          $< 1.0$ & 4177  &  8547  \\
          $< 1.5$ & 4319  & 11690  \\
          $< 2.0$ & 4376  & 14111  \\
          $> 2.0$ &  160  &  7797  \\
          \hline
      \end{tabular}
   \end{center}
   \label{t:rfc_icrf3_comp}
\end{table}

\section{A brief overview of the image results}
\label{s:image_res}

  A catalogue that has only positions and does not provide information
about source flux density has a limited use. We have imaged most of the 
sources. The current data release has information about historic median 
correlated flux densities for 21,730 sources, or 99\%. For most 
of the VLBA experiments we processed the records of system temperature 
measurements, flagged out outliers, interpolated them for missing values, 
computed the a~priori 
system equivalent flux density (SEFD), and calibrated fringe amplitudes. 
Then we applied group delay and phase delay rate determined by the fringe 
fitting procedure, calibrated the complex bandpass, estimated and applied 
amplitude renormalization, determined time intervals when antennas
did not point on sources and flagged them out, flagged the observations 
that have been deselected during the astrometric analysis, averaged fringe 
visibilities over time within a scan and over frequency within each IF for 
each source and each band, and combined averaged visibilities for a given 
source over a certain period time in the range from one experiment 
(4--24 hours) to 4 months. Since most of the sources were observed 
in one or several scans, combining data densifies the $uv$-coverage, and 
therefore, improves image fidelity. From the other hand, combining data 
over a longer period of time may distort an image because of source 
variability. The process of generation of a~priori calibrated time- and 
frequency-averaged visibilities from survey data was performed 
with \PIMA for most of the campaigns, except VCS1--6, NPCS, 
and some other experiments that were calibrated with AIPS and 
imaged with Difmap \citep{r:difmap}. We produced images performing the 
hybrid synthesis technique in a batch mode. This follows the approach 
suggested by \citet{r:difmap_autoimaging} utilizing the software package 
Difmap. Finally, we have screened all the images and reprocessed manually 
those images that showed anomalies, usually related to flagging of data 
with poor amplitude calibration. This procedure is discussed in more 
details in \citet{r:wfcs}.

  We put all the images, a~priori calibrated visibility data, also known as 
uva-data, self-calibrated visibility data after the hybrid image reconstruction
algorithm, also known as uvs-data, in a publicly accessible 
Astrogeo VLBI FITS image database 
\protect\dataset[\doi{10.25966/kyy8-yp57}]{\doi{10.25966/kyy8-yp57}}.
In addition to our own work, the database has the contributions 
from many scientists who decided to make images that they have synthesized
publicly available. Here is the list of contributors in the alphabetic
order:
Alessandra Bertarini,
Nicholas Corey,
Yuzhu Cui,
Xuan He,
Dan Homan,
Laura Vega Garcia,
Jose-Luis Gomez,
Leonid Gurvits,
Svetlana Jorstad,
Tatiana Koryukova,
Sang-Sung Lee,
Rocco Lico,
Elisabetta Liuzzo,
Matt Lister,
Alan Marsher,
Christopher Marvin,
Alexandr Popkov,
Alexandr Pushkarev,
Eduardo Ros,
Tuomas Savolainen,
Kirill Sokolovsky,
An Tao,
Greg Taylor,
Alet de Witt,
Minghui Xu, and
Bo Zhang.

\begin{figure}[h!]
   \includegraphics[width=0.40\textwidth]{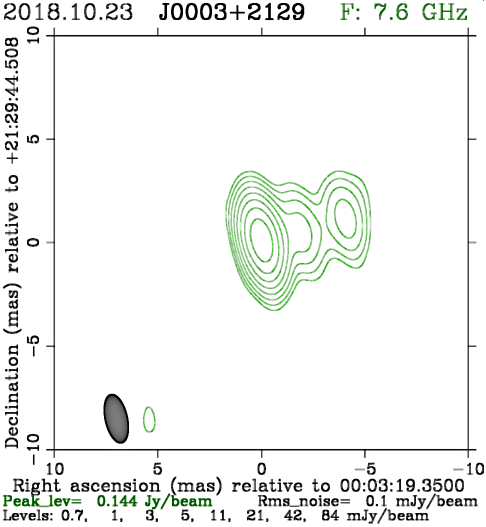}
   \hspace{0.039\textwidth}
   \includegraphics[width=0.525\textwidth]{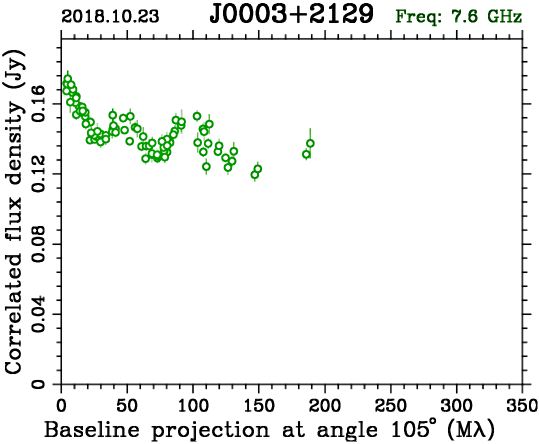}
   \caption{Example of an image and a plot of the correlated flux
            density with respect to the baseline projection length.
            The image presented in the left plot is a result of 
            a convolution of the CLEAN component model and the clean 
            beam that is represented by an elliptical Gaussian with the
            full width half maximum size shown in the left bottom 
            corner. The first contour in the image plot is shown at the 
            5~times the image rms level. The correlated flux density data 
            points presented in the right plot are calculated as an 
            average per baseline, frequency, and scan, and the error 
            bars accounts for the thermal noise only.
           }
   \label{f:rfc_image_example}
\end{figure}

  Most of the contributors used AIPS \citep{r:aips} for a~priori calibration. 
By September 24, 2024, the database had 125,623 images of 20,472 sources and 
this number is growing. For many sources images at different bands and at 
different epochs are available. An example of an image and a plot of the 
correlated flux density is shown in Figure~\ref{f:rfc_image_example}.

  The quality and fidelity of source images varies strongly. For sources 
with 4--6 detections an image has very limited fidelity, often follows
the shape of a clean beam, and does not reveal much structure. Estimates 
of a typical source size and compactness are still possible from calibrated 
visibility data. On the other hand, images from hundreds of observations
have a dynamic range over 1:1000 and provide fine details. 
The typical image noise rms was 0.3--0.5~mJy for sources observed in 
1995--2010. Due to the use of wider bandwidths, the image noise rms 
reduced and 0.1--0.2~mJy for sources observed in 2015--2022.

\subsection{Correlated flux density at short medium, and long baselines}

  The full information about source brightness distributions is contained
in images. In order to characterize the source strength in a concise form,
we computed the median correlated flux density in three baseline projection
length ranges: short ($< 1000$~km), medium (1000 to 5000~km), and long
($> 5000$~km). We used calibrated visibility amplitudes after self-calibration 
and imaging while taking into account data flagging and station-based 
amplitude correction. The visibility data were averaged in frequency over 
all IFs and in time over all time epochs of a given image. 
The correlated flux densities were computed for each band separately.
Used band names are presented in Table~\ref{t:bandnames}. When images 
of a given source and a given band are available for more than one epoch, 
we computed the median flux density values over all visibility datasets 
related to reconstructed images.

A radio interferometer does not provide the total flux density.
The correlated flux density corresponds to a fraction of the total
flux density of components with a typical sizes less than 1--5~mas
at long baselines, 5--20~mas at medium baselines, and 20--100~mas at 
short baselines, with an extended emission from components larger than
100--200~mas being lost.

\begin{table}
   \caption{Bands names used in this work.}
   \Note{
   \begin{center}
       \begin{tabular}{ll}
           \hline\hline
            Band & Freq. range           \\
                 & \ntab{c}{GHz}         \\
           \hline
            S    & \phantom{0}2.2--2.4   \\
            C    & \phantom{0}4.1--5.0   \\
            X    & \phantom{0}7.3--8.8   \\
            U    & 15.2--15.5 \\
            K    & 22.0--24.2 \\
           \hline
       \end{tabular}
   \end{center}
   }
   \label{t:bandnames}
\end{table}

  We did not image sources from certain campaigns, for instance many segments 
of LCS1 and LCS2 campaigns because of scarcity of calibration information. 
Instead, in most of these cases we performed the non-imaging analysis described 
in \citet{r:lcs2}. That procedure included three steps: computation of the 
a~priori SEFD using recorded system temperature and prior antenna gain 
measurements, computation of multiplicative station-dependent gain corrections
using publicly available images of calibrator sources, correcting the a~priori
SEFD, and then computation of the median correlated flux densities.
  
  These correlated flux density estimates are helpful for a coarse 
assessment of the SNR in future observations. A caution should be exercised
in using these estimates. First, for sources with a strong jet or an 
asymmetric core the correlated flux density strongly depends on the 
orientation of the baseline projection, whether along the jet or across the 
jet. In extreme cases the difference can reach a factor of ten. Second, source 
variability, typical for VLBI-selected AGNs, is expected to change the flux 
density.

\section{The catalogue}
\label{s:the_cat}

\begin{figure}[h!]
   \includegraphics[width=1.01\textwidth]{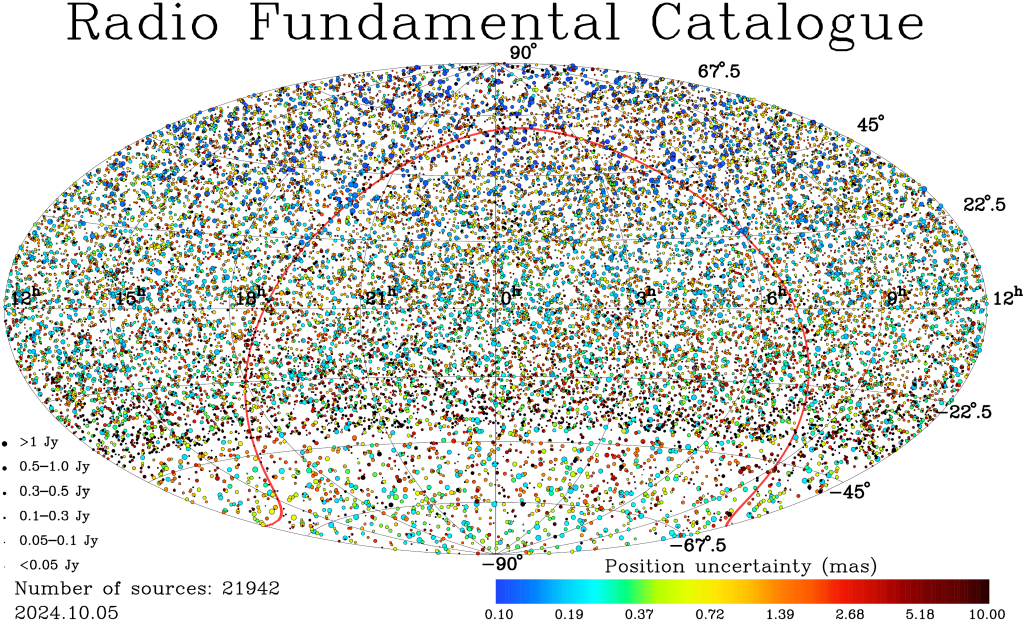}
   \caption{The distribution of RFC sources over the celestial sphere
            in the equatorial coordinate system. The circle size 
            corresponds to flux density at 8~GHz (or at lower
            frequencies if a source was not detected at 8~GHz). The 
            circle color corresponds to the semi-major axis of the 
            positional error ellipse. The red line shows
            the Galactic plane.
           }
   \label{f:rfc_map}
\end{figure}

  The catalogue consists of three main ASCII tables: the master table, 
a table with the multi-band positional offsets, and a table with source 
associations, as well as eight auxiliary tables. Since the 
number of columns in the three main tables is too large to fit 
a page width, we do not show them and instead present a description 
of the variables in the Appendix. The distribution of sources in the 
catalogue over the celestial sphere in the equatorial coordinate 
system using the equi-area Hammer projection is shown in 
Figure~\ref{f:rfc_map}. 

  The main master Table~\ref{t:main} presents the source positions from the 
fused dataset, re-scaled position uncertainties, correlations between 
right ascension and declination estimates, the number of observations, 
scans, and sessions used for data analysis, the weighted mean epoch of 
observations, and three estimates of correlated flux density as a median 
in three ranges of baseline length projections at S, C, X, U, and 
K-bands. See Table~\ref{t:bandnames} for the corresponding 
band names and frequencies. The epoch of the catalogue to account for 
precession-nutation: J2000.0, i.e.\ 2000.01.01, 12 UTC. The epoch of the 
catalogue to account for galactic aberration, proper motion, and 
parallax is 2016.0. 

  Positions derived from the dual-band datasets that include both X/C 
and X/S data, as well as single band positions at S, C, X, and K-band 
are presented in the multi-band Table~\ref{t:multi} 
in a form of displacements along right ascension and declination with 
respect to the positions derived from the fused dataset. Their 
displacements, position uncertainties, correlation between right 
ascension and declination estimates, number of sessions, scans, 
and observations are presented in the multi-band table as well. 

  The association Table~\ref{t:cross-match} provides the results of 
cross-matching of the RFC sources with other catalogues, as well as
redshifts, source types, and jet directions from literature and 
NASA/IPAC Extragalactic Database 
(NED)\footnote{\url{https://ned.ipac.caltech.edu/classic}}\citet{r:ned} 
when available. For completeness, we present an 
auxiliary Table~\ref{t:nondet} of 20,000 sources that were observed 
in wide-band high sensitivity VLBI programs
at 4--8~GHz, but have not been detected at baselines 
with projection lengths 100--5000 km. The table 
shows the upper limit of the correlated flux density of observed 
sources that are {\it expected} to be detected within one 
arcminute of the pointing direction. That limit is set to 20\% 
above the flux density of the faintest sources detected in an
experiment that observed a given source. Although a lack of detection 
should not be construed as an absence of a VLBI source above the 
specified flux density, this table provides a hint that it 
is unlikely that strong emission from a mas-scale structure is present 
in the field.

\begin{figure}[h!]
   \includegraphics[width=0.48\textwidth]{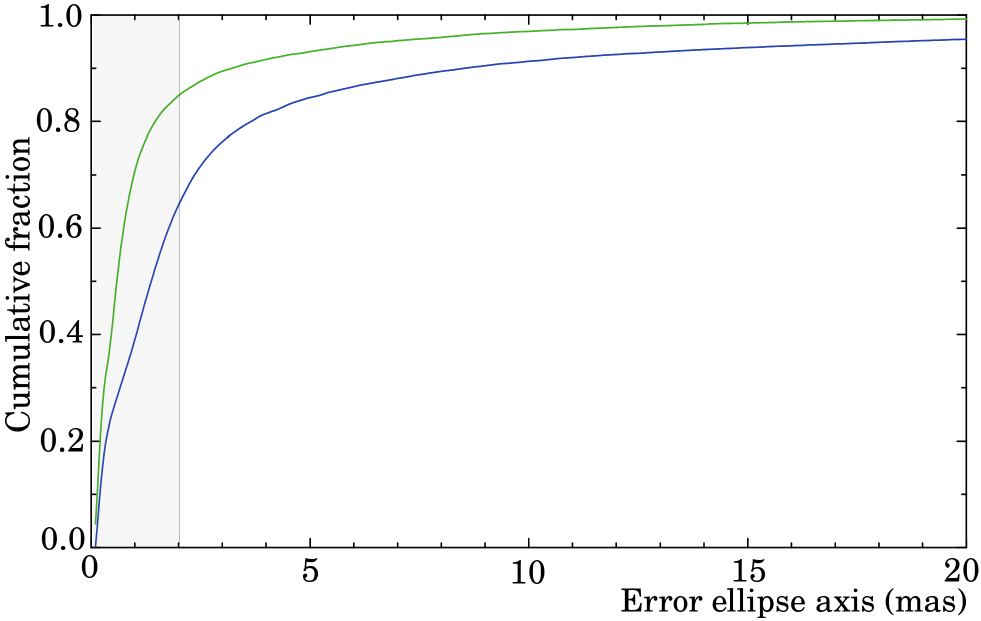}
   \hspace{0.019\textwidth}
   \includegraphics[width=0.48\textwidth]{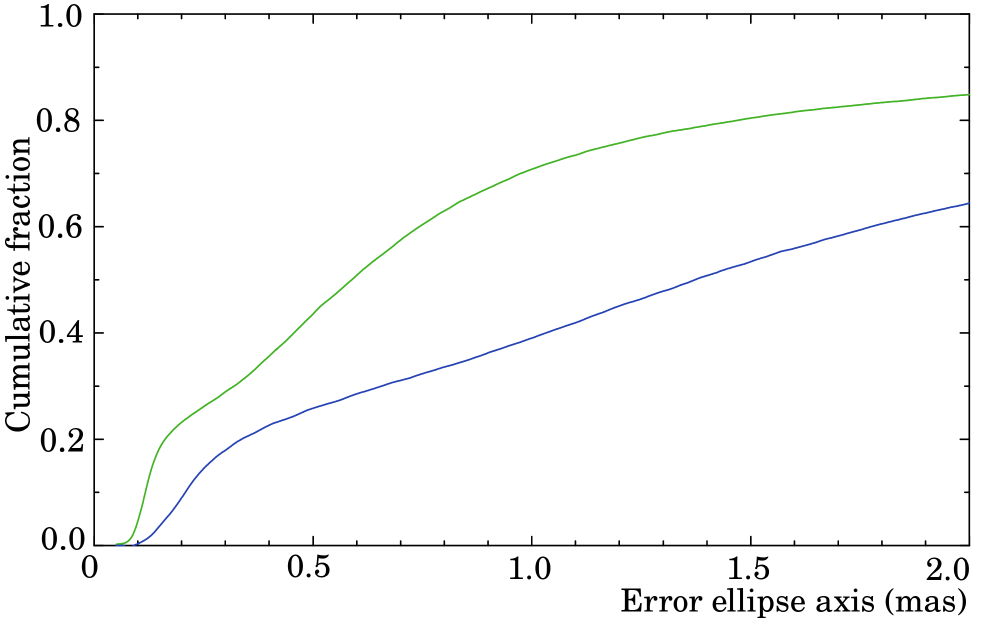}
   \caption{The cumulative error distributions. The upper green line:
            semi-minor positional error ellipse axes.  The lower blue line:
            semi-major positional error ellipse axes. Right: zoom of the area 
            that is marked with a gray color on the left panel.
           }
   \label{f:rfc_error_distr}
\end{figure}

  Since source positional errors range from 0.090 mas to 1480~mas --- four 
order of magnitude, RFC astrometry errors are characterized by the 
cumulative distributions of semi-minor error ellipse axes and semi-major 
error ellipse axes shown in Figure~\ref{f:rfc_error_distr} with green and 
blue colors, respectively. Table~\ref{t:quart} shows three quartiles
of the RFC semi-minor and semi-major error ellipse axes.

\begin{table}[h!]
   \caption{Three quartiles of RFC semi-minor and semi-major 
            positional error ellipse axes in mas.
           }
   \begin{center}
      \begin{tabular}{rrr}
          \hline\hline
               & $\sigma_{\rm min}$ & $\sigma_{\rm maj}$ \\
          \hline
          25\% & 0.23 & 0.47 \\
          50\% & 0.59 & 1.37 \\
          75\% & 1.16 & 2.83 \\
          \hline
      \end{tabular}
   \end{center}
   \label{t:quart}
\end{table}

\begin{figure}[h!]
   \includegraphics[width=0.48\textwidth]{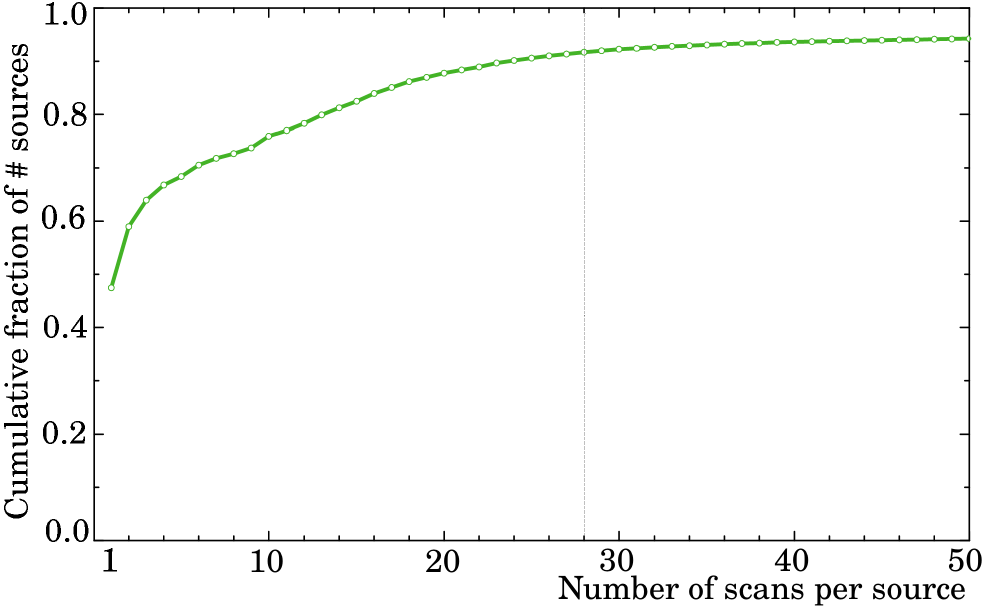}
   \hspace{0.039\textwidth}
   \includegraphics[width=0.48\textwidth]{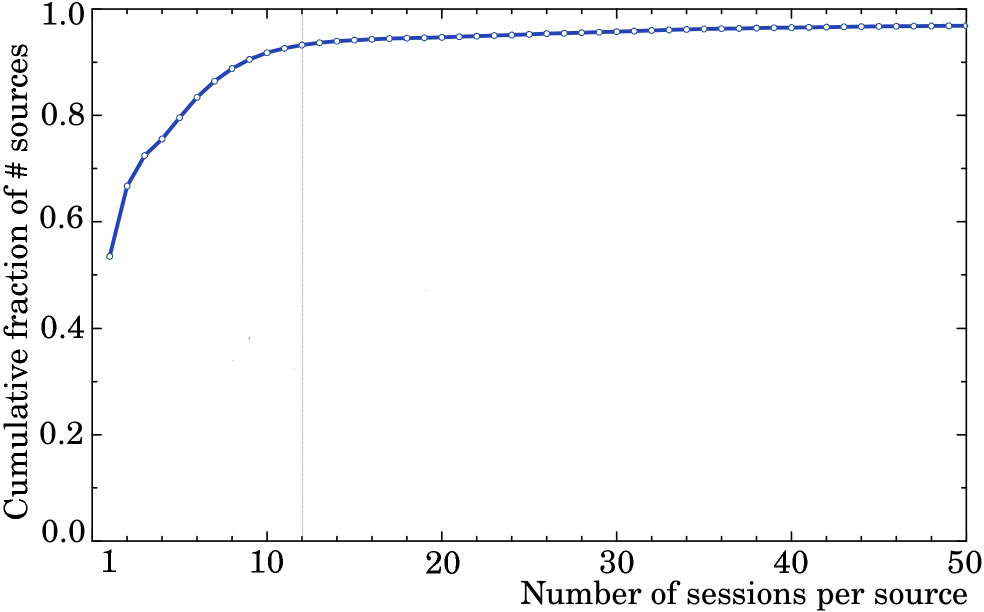}
   \caption{{\it Left:}  the cumulative distribution of the 
                         number of scans per source.
            {\it Right:} the cumulative distribution of the 
                         number of observations per source.
           }
   \label{f:rfc_num_distr}
\end{figure}

  Positional accuracy depends on source correlated flux density, array 
sensitivity, integration time, array geometry, and the number of 
observations. Left plot in Figure~\ref{f:rfc_num_distr} shows the statistics 
of the number of scans in which individual sources have been observed. 
We see that roughly 50\% of the sources have been observed in 
one scan. The cumulative distribution of the number of scans flattens beyond 
28 scans. There are 1981 sources or 8\% that have been observed and detected 
in 28 or more scans. There are 1483 sources, or 7\%, that  have been observed 
in 12 or more sessions. Right plot in Figure~\ref{f:rfc_num_distr} 
show the cumulative distribution of the number of sessions per source. The 
thin vertical lines show the cutoff --- 28 scans and 12 sessions beyond which 
the cumulative fraction of the number of scans and the number of sessions grow 
very slowly. These cutoffs roughly correspond to the contribution of geodetic 
sessions that observe a small subset of sources very often, up to 100,665 scans 
of RFC~J0555+3948 and 12,299 sessions of J1800+7828. From the other hand, 
approximately 1/2 of sources were observed only in one scan and one 
session; 2/3 sources were observed in 4 scans or 2 sessions.

\begin{figure}[h!]
   \includegraphics[width=0.53\textwidth]{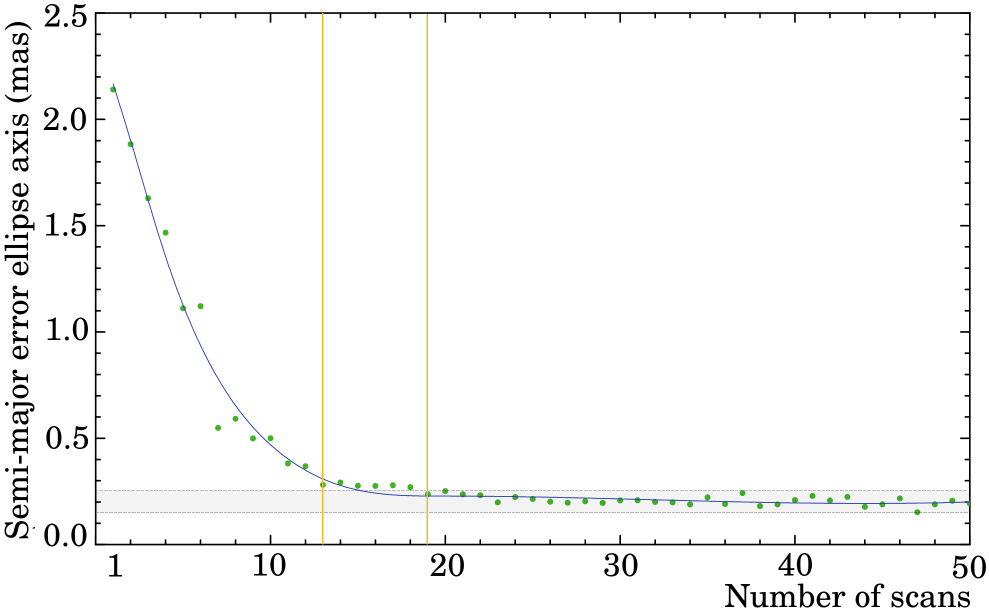}
   \hspace{0.015\textwidth}
   \includegraphics[width=0.444\textwidth]{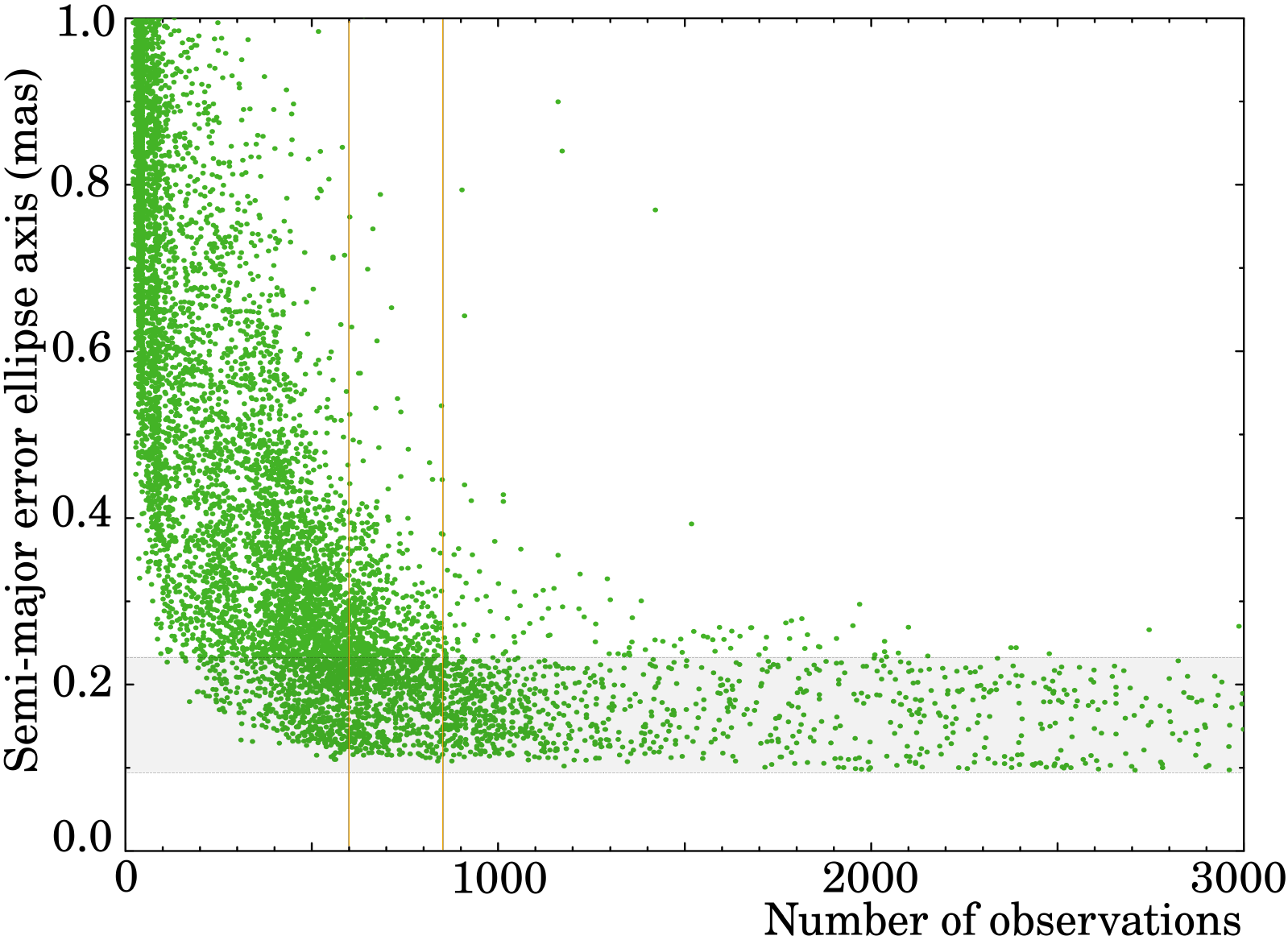}
   \caption{{\it Left:}  the dependence of the median source 
            positional error on the number of scans.
            {\it Right:}  the dependence of the median source 
            positional error on the number of observations.
            The gray area denotes the range of position 
            errors dominated by the contribution of systematic 
            errors. Two vertical bars denote the range of the 
            number of scans or the number of observations 
            in a transition zone from random-error dominated
            to the systematic-error dominated.
           }
   \label{f:error_nsca_nobs}
\end{figure}

   An increase of the number of scans and the number of observations
reduces the position uncertainties. Figure~\ref{f:error_nsca_nobs}
demonstrates the dependence of semi-major positional error ellipse axes 
on the number of scans and the number of observations. We see that beyond 
some limit using more observations does not improve accuracy because 
of the presence of red noise. Vertical lines in plots show the 
transition zone from random error dominated to the systematic error 
dominated: 13--19 scans and 600--850 observations. 

  It should be noted that the numbers above characterize the observing 
programs, not the intrinsic property of the VLBI technique. Observations 
at a different network with a different recording rate would have 
ended up with other limits of positional errors and the number of scans 
and observations that are needed to reach these limits.

\begin{figure}[h!]
   \includegraphics[width=0.48\textwidth]{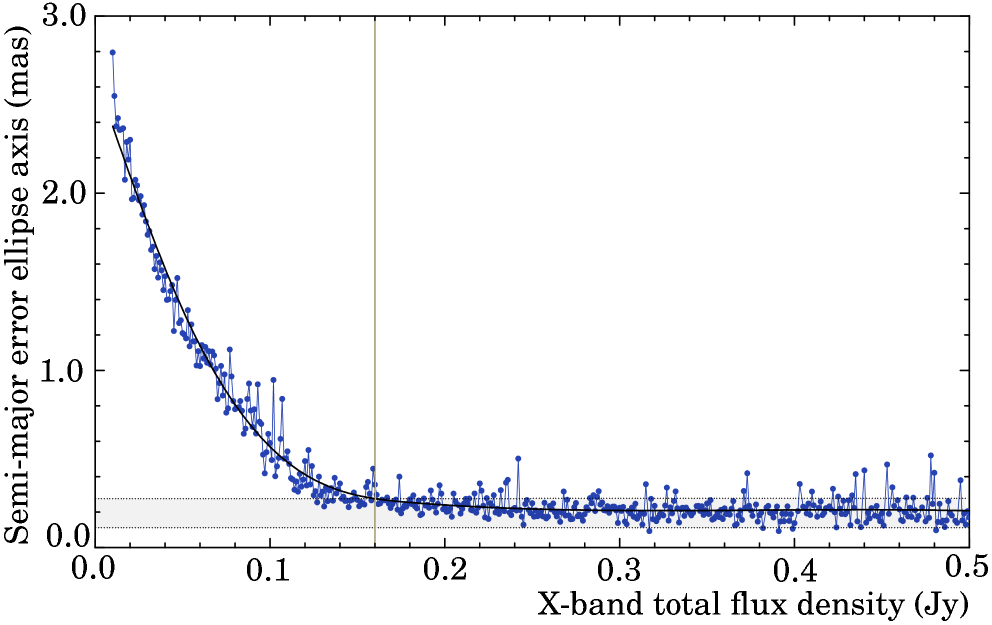}
   \hspace{0.039\textwidth}
   \includegraphics[width=0.48\textwidth]{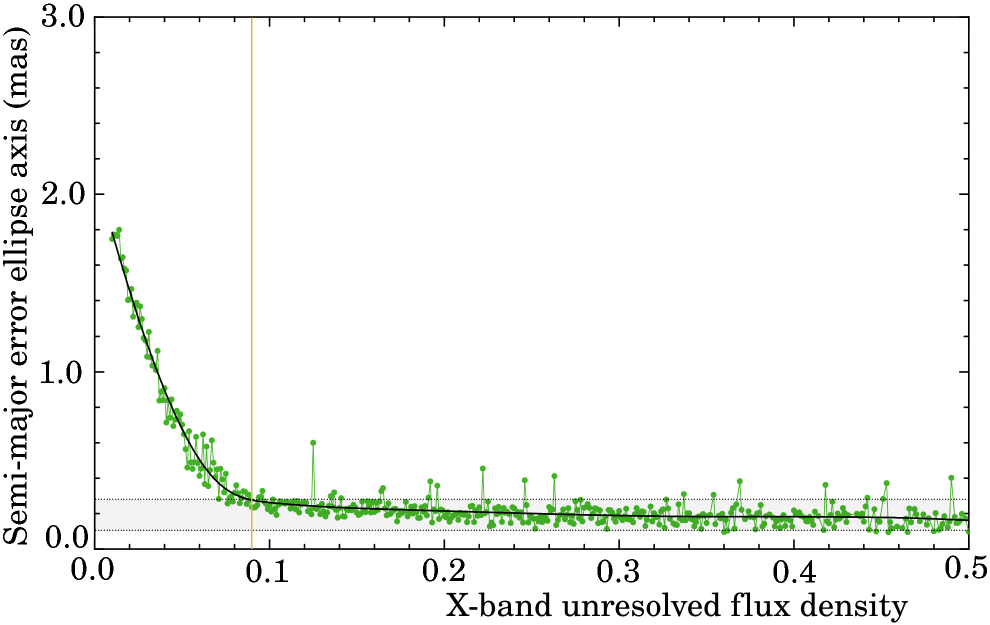}
   \caption{{\it Left:}  the dependence of the median semi-major 
                         positional error ellipse on the median flux 
                         density at baselines shorter than 1000 km.
            {\it Right:} the dependence of the median semi-major 
                         positional error ellipse on the median flux 
                         density at baselines longer than 1000 km.
           }
   \label{f:rfc_errors_flux}
\end{figure}

  Figure~\ref{f:rfc_errors_flux} shows the dependence of the position
uncertainties in a form of the semi-major error ellipse axes on the 
correlated flux density at baselines shorter than 1000~km and at
baselines longer than 5000~km. The gray area corresponds to the 
range of position uncertainties where systematic errors dominates
the error budget. The vertical lines denote the flux density
that corresponds to a transition from the regime when the 
contribution of systematic errors dominates to the regime when they 
do not impact source positions. The blue and green points in 
Figure~\ref{f:rfc_errors_flux} correspond to the median correlated flux 
density at 8~GHz in 10~mJy wide bins. The black line shows the
result of smoothing. It follows from these plots that all the sources 
with the correlated flux density at short baselines greater than 0.16~Jy 
and at long baselines greater than 0.09~Jy at 8~GHz have position 
uncertainties dominated by systematic errors. We note that the RFC catalog 
contains the majority of extragalactic radio sources on the sky with total 
parsec-scale flux density at 8~GHz above 0.1~Jy because all the 
sources at declinations $>-40^\circ$ from VLASS catalogue brighter than 
0.1~Jy have been observed. Detailed characteristics of the catalog 
completeness will be analyzed and presented in Paper~II.

\begin{figure}[h!]
   \centerline{\includegraphics[width=0.61\textwidth]{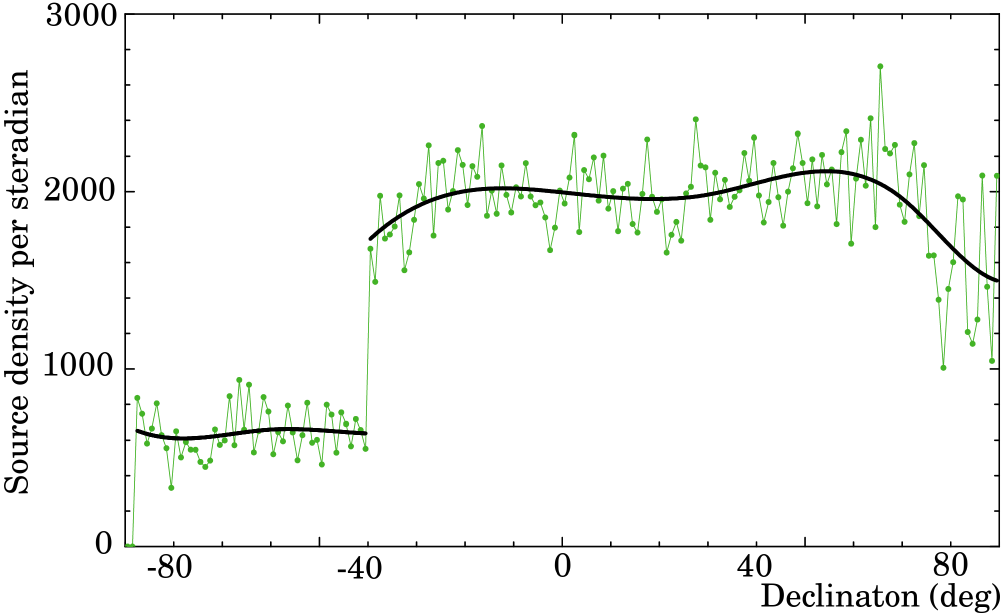}}
   \caption{\Note{The RFC source density as a function of $1^\circ$ wide declination 
            stripe zones. The black line shows the source density smoothed 
            using B-spline. Note that the zone area in the stripes depends 
            on $\cos\delta$ which affects statistics close to the poles.}
           }
   \label{f:rfc_decl_dens}
\end{figure}

  We can see in Figure~\ref{f:rfc_map} that the number of sources at 
declinations $<-40^\circ$ is low. Figure~\ref{f:rfc_decl_dens} shows 
the source density as a function of declination. The density has a jump 
at $<-40^\circ$. The average source density is 1950 objects per steradian 
at $\delta > -40^\circ$ and it drops by a factor of 3.0 to 640 objects 
per steradian at $\delta < -40^\circ$. This jump is due to the cutoff 
in declination in VLBA surveys, because below that declinations the 
zone mutual visibility of VLBA becomes small. The decrease of sky
density at $\delta > 80^\circ$ is probably due to a statistical
fluctuations since the area in high declination zones shrinks as
$\cos\delta$. We can also notice that the average positional accuracy in 
the zone $[-40^\circ, -30^\circ]$ is worse than in other zones because 
these source were mainly observed with a part of VLBA without northern 
stations, which detrimentally affected the positional accuracy along 
the declination axis.

\subsection{Naming convention}

We followed the IAU convention\footnote{\url{https://cds.unistra.fr/Dic/iau-spec.html}}
for naming the sources. The RFC designators refer to the J2000.0
epoch, while the common names refer to the B1950.0 epoch following the 
tradition established in the 1980s. We assigned new designators and common 
names for those sources whose positions determined from VLBI 
observations were not reported in the literature. We did not change names 
for the sources which positions derived from VLBI observations have already 
been published. We used the same approach for 10-character long RFC 
designators (14 character long counting RFC prefix with a following
blank) and 8-character long common names. We call those RFC designators 
that use only the coordinate system prefix J, digits, sign $-$ or $+$, 
and common names that use digits and sign $-$ or $+$ canonical. When two 
or more sources occupy the area in the sky that makes names ambiguous, 
we alter the last character of the name and assign it letter 
A, B, C, etc. We assigned a letter to a weaker source or a weaker source 
component. For those multi-component sources that have a sub-component 
designator known in literature we followed the published notation for 
sub-components. We call these names as well as names that correspond to 
a wrong position non-canonical. The Appendix contains 
Tables~\ref{t:ncj}--\ref{t:ncb} with 137 non-canonical RFC designators 
and 427 non-canonical common names.

\subsection{Multiple sources}

  More than one source can be detected in an antenna beam. The 
component separation procedure that was briefly described in 
section~\ref{s:filtering} allows us  to detect reliably additional
strong components at separations of 300~mas and greater. It works 
less efficiently at separations of 100--300~mas and usually fails at 
separations shorter than 100~mas. When a source has the second 
component at an angular distance less than $1''$, we always 
check the image. In rare cases when we cannot identify 
the second component in the images, we dropped unconfirmed components 
from the catalogue.  We also dropped the second components with 
a separation less than 100~mas in rare cases when the 
component separation procedure ended up with two components. 
Images of many sources show more than one component, and we set the 
borderline at 100~mas to discriminate sources with multiple components 
that are identified on images from objects in which the second or
third components have been determined during the astrometric analysis.
The success of the component separation procedure depends on the 
ratio of peak flux density and the number of observations. If the second 
component is weaker than a factor of 5 and there were fewer than 
30 observations of that source used in the solution, it is usually not 
separated and not reported in the catalogue as a separate source, 
but it still can be identified on images.

  Table~\ref{t:pairs} in the Appendix presents 104 pairs 
of sources listed as separate objects in the RFC that are closer 
than one arcminute. Some of these pairs are known in the literature
as gravitational lenses, some are known as AGNs at different redshifts
that are just projected close to each others are not gravitationally
bound, some sources in gravitationally bound systems, and some of 
these pairs are different component of the same AGN. We do not 
attempt to establish the nature of these sources in this study
deferring it for future publications.

\subsection{Non-AGNs and nearby sources}

  All but 24 sources in the RFC are AGNs. There are 22 stars, including
one pulsar, two supernova remnants in the M82 galaxy, and the weak 
compact object in the center of our Galaxy, Sgr\,A*. We applied data for 
parallaxes of radio stars from the \Gaia EDR3 catalogue when they were 
available and estimated proper motions of Galactic objects. 
The estimates of proper motions and their uncertainties from our 
main solution are presented in Table~\ref{t:prp}. Proper motion
uncertainties were not re-scaled since not enough statistics are collected 
to make a judgment about validity of uncertainties. Since parallaxes from 
\Gaia were used for processing observations of radio stars, the 
estimates of proper motions are not fully independent with respect to 
proper motions from \Gaia. 

\begin{table}[h!]
    \caption{Proper motion of Galactic objects}
    \begin{center}
       \begin{tabular}{llrrrrr}
           \hline\hline
           \ntab{c}{(1)} & \ntab{c}{(2)} & \ntab{c}{(3)} & \ntab{c}{(4)} & \ntab{c}{(5)} & \ntab{c}{(6)} & \ntab{c}{(7)} \\
           \hline
           RFC J0240$+$6113 & star   & $    -1.27 $ & $    0.10 $ &  0.21 & 0.12 & $ -0.223 $ \\
           RFC J0326$+$2842 & star   & $    52.18 $ & $ -101.86 $ &  0.71 & 0.67 & $  0.599 $ \\
           RFC J0336$+$0035 & star   & $   -31.22 $ & $ -157.88 $ &  0.23 & 0.33 & $  0.111 $ \\
           RFC J0535$-$052E & star   & $     0.85 $ & $    2.71 $ &  1.58 & 4.94 & $ -0.569 $ \\
           RFC J0535$-$0523 & star   & $     4.41 $ & $   -3.33 $ &  0.20 & 0.38 & $  0.259 $ \\
           RFC J0835$-$4510 & pulsar & $   -71.45 $ & $   29.69 $ &  0.41 & 0.18 & $ -0.237 $ \\
           RFC J0930$+$4429 & star   & $    -2.23 $ & $   -0.72 $ &  3.20 & 3.07 & $ -0.116 $ \\
           RFC J1055$+$6028 & star   & $   -73.01 $ & $   -6.96 $ &  2.01 & 1.05 & $  0.715 $ \\
           RFC J1331$+$1712 & star   & $     0.26 $ & $    1.25 $ &  3.06 & 4.16 & $  0.047 $ \\
           RFC J1406$+$3539 & star   & $    -0.21 $ & $   -0.11 $ & 13.00 &13.01 & $  0.000 $ \\
           RFC J1500$-$0831 & star   & $   -65.62 $ & $   -5.31 $ &  0.19 & 0.34 & $ -0.235 $ \\
           RFC J1501$+$5619 & star   & $     0.06 $ & $   -2.86 $ & 12.68 &12.16 & $  0.007 $ \\
           RFC J1534$+$2330 & star   & $    11.41 $ & $   -0.67 $ &  5.90 & 1.47 & $ -0.111 $ \\
           RFC J1553$-$2358 & star   & $   -14.78 $ & $  -24.45 $ &  0.12 & 0.13 & $  0.210 $ \\
           RFC J1614$+$3351 & star   & $  -322.20 $ & $  -87.39 $ &  0.11 & 0.13 & $  0.173 $ \\
           RFC J1745$-$2900 & AGN    & $    -3.58 $ & $   -5.63 $ &  0.07 & 0.13 & $  0.177 $ \\
           RFC J1818$-$1214 & star   & $    -2.78 $ & $  -12.33 $ &  5.10 & 6.99 & $ -0.667 $ \\
           RFC J1826$-$1450 & star   & $     8.31 $ & $   -8.39 $ &  0.17 & 0.35 & $ -0.119 $ \\
           RFC J1911$+$0458 & star   & $    -0.09 $ & $    0.10 $ &  0.11 & 0.24 & $ -0.281 $ \\
           RFC J2032$+$4057 & star   & $    -3.14 $ & $   -3.61 $ &  0.57 & 0.47 & $  0.536 $ \\
           RFC J2053$+$4423 & star   & $    35.95 $ & $   -0.04 $ &  1.46 & 0.56 & $ -0.393 $ \\
           RFC J2349$+$3625 & star   & $    -0.60 $ & $  -44.54 $ &  1.59 & 1.33 & $  0.089 $ \\
           RFC J2355$+$2838 & star   & $   646.03 $ & $   34.56 $ &  0.54 & 0.63 & $ -0.025 $ \\
           RFC J0955$+$6940 & SN?    &     n/a      &    n/a    &  n/a    & n/a  & n/a    \\
           RFC J0955$+$6901 & SN     &     n/a      &    n/a    &  n/a    & n/a  & n/a    \\
           \hline
       \end{tabular}
    \end{center}
    \tablecomments{Column descriptions:
             (1) the RFC source name;
             (2) the object type: star, pulsar, or supernova;
             (3) the proper motion in right ascension in mas/year;
             (4) the proper motion in declination in mas/year;
             (5) the uncertainty of the proper motion in right ascension 
                 in mas/year without the $\cos\delta$ factor;
             (6) the uncertainty of the proper motion in declination 
                 in mas/year;
             (7) the correlation between proper motions in right 
                 ascension and declination.
    }
    \label{t:prp}
\end{table}

\subsection{Cross-matching}
\label{s:cross}

  We performed cross-matching the RFC list against 14 large surveys
listed in Table~\ref{t:matches}. We evaluated the sky density of a given
survey as a number of sources per steradian. For some catalogues the density 
was considered constant. For other catalogues that have a large number of 
objects, such as \textit{Gaia}, PanSTARRS (PS1) and ALLWISE, we computed the 
density on a 3D grid. The first two dimensions are along right ascension 
and declination, and the third dimension is over magnitude. Knowing the 
sky density of surveys, we computed the a~priori probability of finding 
a source at a random direction assuming the distribution of sources over
the sky or the grid element is uniform. We report all associations with 
the probability of false positive below 0.01.

  This statistical association process is based on the following assumptions:
a)~an RFC source and its counterpart from a survey have the same position
affected only by random positional errors with the second moments reported
as uncertainties and b)~the sources are distributed uniformly with a known
density. Violation of these assumptions cause either missing a counterpart
or reporting spurious counterparts. 

\begin{figure}[h!]
   \includegraphics[width=0.47\textwidth]{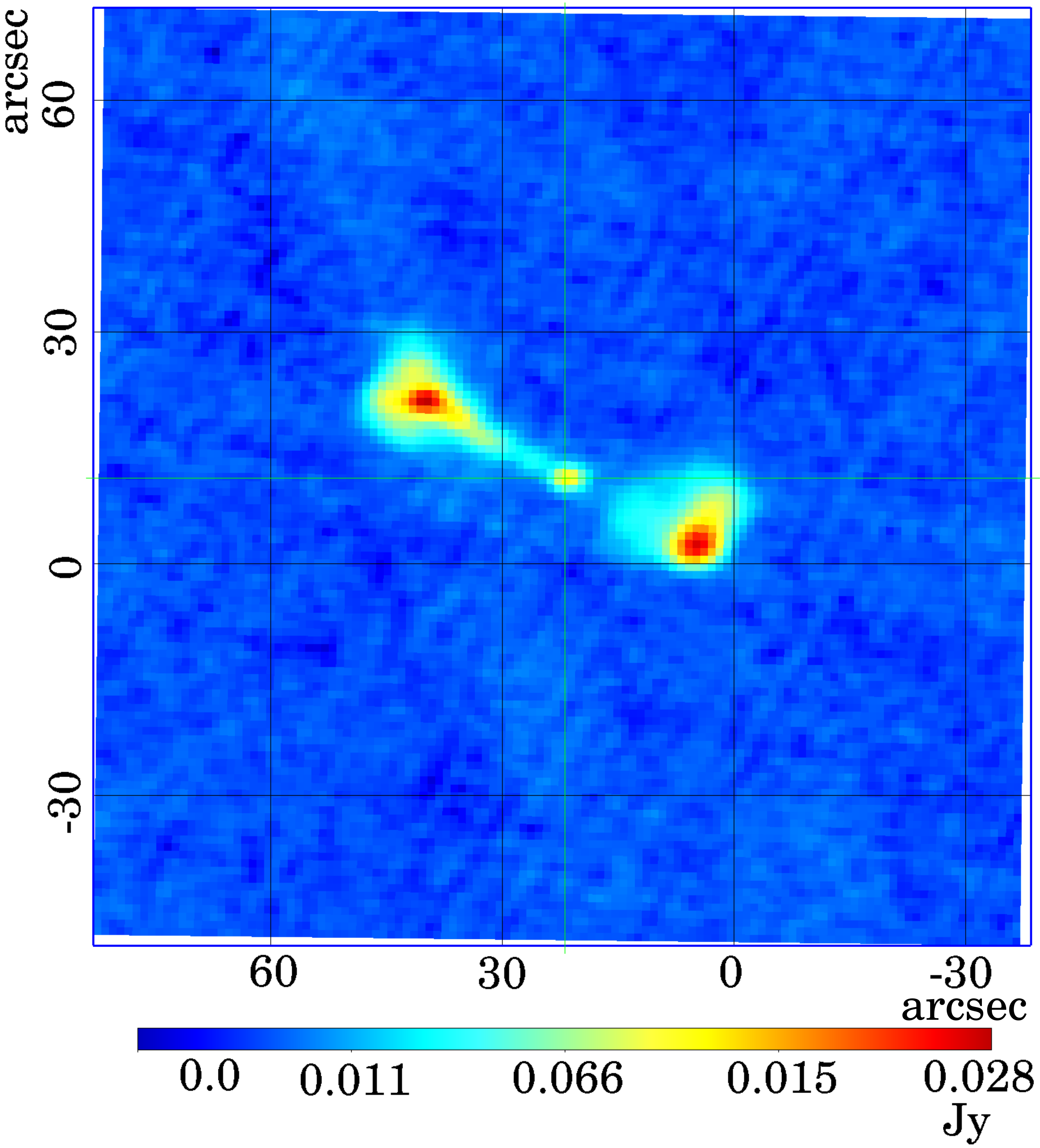}
   \hspace{0.039\textwidth}
   \includegraphics[width=0.48\textwidth]{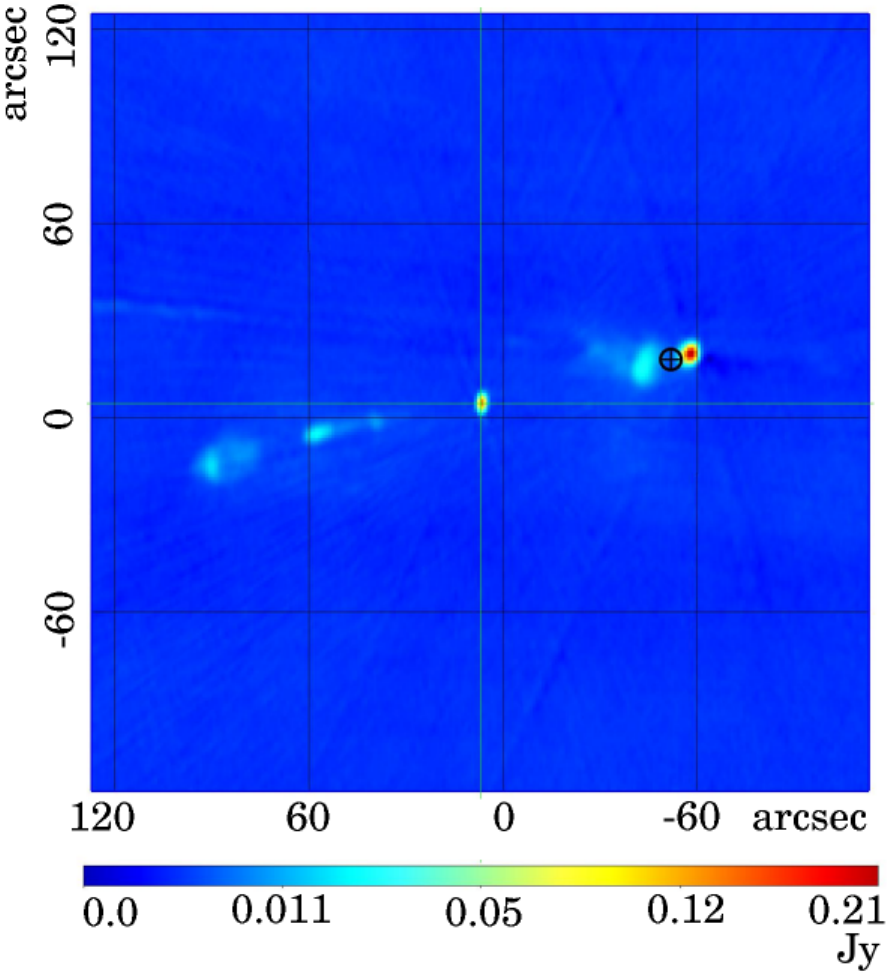}
   \caption{An example of VLASS images of areas in the vicinity of RFC sources.
            The thin green lines in the image centers show the RFC position.
            {\it Left: } RFC~J0328$+$5509.
            {\it Right:} RFC~J2137$-$1432. $\oplus$ character denotes the 
                         AT20G position.
           }
   \label{f:rfc_matches}
\end{figure}

  Figure~\ref{f:rfc_matches} demonstrates the difficulty of cross-matching.
Source J0328$+$5509 is extended at its VLASS image. The compact component
is between two bright extended radio lobes, and the algorithm that identifies
point sources may miss it by considering it as a part of a source with 
a complex structure. Source J2137$-$1432 has emission at 22~GHz associated 
with the extended feature of a complex multi-component radiogalaxy.

\begin{figure}[h!]
   \centerline{\includegraphics[width=0.616\textwidth]{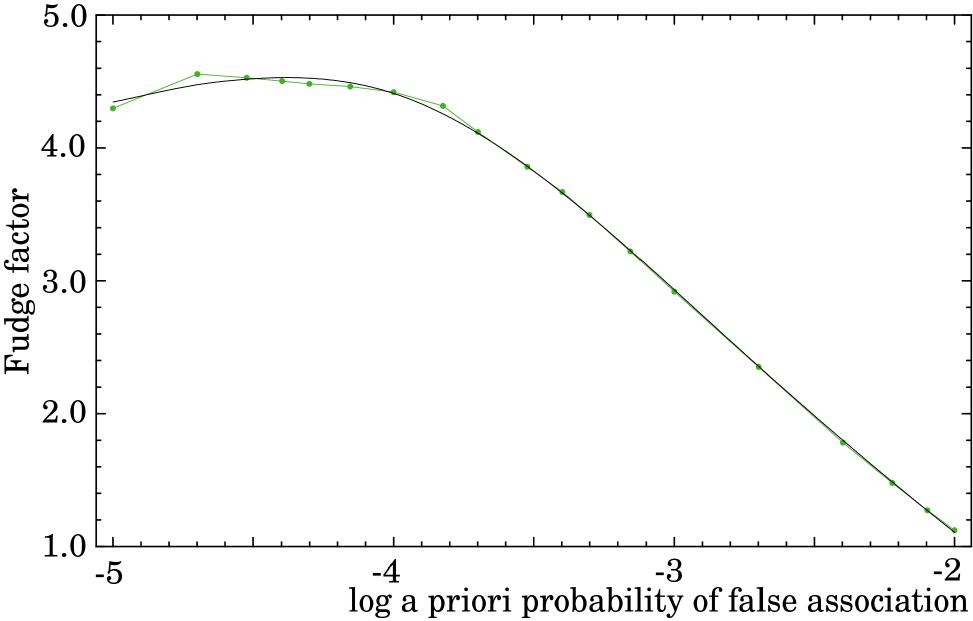}}
   \caption{The empirical fudge factors to the a~priori estimate of the 
            probability of false association between \Gaia EDR3 and 
            the RFC catalogue.
           }
   \label{f:rfc_gaia_prob_fudge}
\end{figure}

   To assess the impact of clustering, i.e.\ a deviation of the source 
distribution from uniform, and determine the a~posteriori probability of 
false association, we rotated the RFC catalogue at random angles within 
20--$40'$ and computed the number of associations with the rotated catalogue.
Comparing the numbers of these spurious associations with their 
mathematical expectations, we determined the fudge factor as the ratio of 
associations to their mathematical expectations based on the a~priori 
probability. We repeated this procedure 1024 times on a grid of the 
a~priori probability of false association. Interpolating this fudge factor 
as a function of the a~priori probability of false association, we are able 
to generate a list of matches with rather accurate probability of false 
association. Figure~\ref{f:rfc_gaia_prob_fudge} shows these fudge factors 
for the \Gaia EDR3 catalogue.

   Because of poor localization of \Fermi LAT sources, we used a more 
sophisticated approach for establishing their association 
\citep[see for full details][]{r:aofus1}. We determined the likelihood 
ratios defined as the ratio of the probability that a radio counterpart 
will be found inside a circle of a given radius to the probability 
of finding a background radio  source with a given flux density or 
greater outside the same circle.

  We report the probabilities of false associations for all the surveys,
and set a lower limit of $1\cdot 10^{-5}$ when our computations show
lower probability, arguing that the assumptions that we used for 
cross-matching are broken for assessing very low probabilities of 
false association. The summary of cross-matching statistics of the RFC 
sources are given in Table~\ref{t:matches}. 
For \textit{Gaia}, PanSTARRS, and ALLWISE the reported probabilities 
are a~posteriori corrected for clustering. In addition, we added the 
jet direction angles determined by \citet{r:jet_dirs}, as well as 
redshift and source type as collected by the NED. We did not
validate redshifts and source types, but provided them as is. We should 
note that a caution should be exercised in using association for 
the population analysis considering all mentioned factors that can skew 
the statistics.

\begin{table}
   \caption{Summary of the statistics of association of the RFC sources 
            with 14~large surveys, as well as with the catalogue of jet 
            angles, redshifts, and source types from NED.
           } 
   \begin{center}
      \begin{tabular}{ll@{\:}rrrrll}
          \hline\hline
          Catalogue    & \nntab{c}{Decl. range}       & \ntab{l}{Numb}  & Area & \ntab{l}{Assoc.} &
          Range & Reference \\
                       &                            & & \ntab{l}{assoc} &      & \ntab{l}{fraction} & & \\
          \hline
          NVSS         & $[-40\degr, $ & $  90\degr]$ & 20186 & 0.82 & 98 \% & 1.4~GHz  & \citet{r:nvss}                    \\
          VLASS        & $[-40\degr, $ & $  90\degr]$ & 20298 & 0.82 & 99 \% & 2--4~GHz & \citet{r:vlass}                   \\
          SUMSS        & $[-90\degr, $ & $ -30\degr]$ &  2918 & 0.25 & 96 \% & 843~MHz  & \citet{r:sumss1,r:sumss2}         \\
          TGSS         & $[-53\degr, $ & $  90\degr]$ & 17249 & 0.90 & 81 \% & 150 MHz          & \citet{r:tgss}                    \\
          AT20G        & $[-90\degr, $ & $   0\degr]$ &  4359 & 0.50 & 46 \% & 5, 8, and 22~GHz & \citet{r:at20g,r:atpmn}   \\
          \Gaia EDR3   & $[-90\degr, $ & $  90\degr]$ & 13320 & 1.00 & 60 \% & 400--1000~nm     & \citet{r:gaia_edr3}       \\
          PanSTARRS    & $[-30\degr, $ & $  90\degr]$ & 14532 & 1.00 & 76 \% & 445--1020~nm     & \citet{r:ps_dr1,r:ps1_db} \\
          ALLWISE      & $[-90\degr, $ & $  90\degr]$ & 15116 & 1.00 & 69 \% & 3.6--22~$\mu$m   & \citet{r:allwise}         \\
          2MASS        & $[-90\degr, $ & $  90\degr]$ &  5877 & 1.00 & 26 \% & 1.4--2.2~$\mu$m  & \citet{r:2mass}           \\
          GALEX        & $[-90\degr, $ & $  90\degr]$ &  6606 & 1.00 & 30 \% & 134--283~nm      & \citet{r:galex_guvcat}    \\
          2RXS         & $[-90\degr, $ & $  90\degr]$ &  2640 & 1.00 & 12 \% & 0.1 to 2.4 kev   & \citet{r:2rxs}            \\
          XMMSL        & $[-90\degr, $ & $  90\degr]$ &  1657 & 1.00 &  7 \% & 0.1 to 12 kev    & \citet{r:xmmsl2}          \\
          1eRASS       & $[-90\degr, $ & $  42\degr]$ &  5649 & 0.50 & 56 \% & 0.2--5.0~kev     & \citet{r:1erass}          \\
          FERMI LAT    & $[-90\degr, $ & $  90\degr]$ &  3272 & 1.00 & 14 \% & 50 MeV to 1 TeV  & \citet{r:fermi-4fgl}      \\
          jet angle    & $[-90\degr, $ & $  90\degr]$ &  9207 & 1.00 & 42 \% & n/a              & \citet{r:jet_dirs}        \\
          redshift     & $[-90\degr, $ & $  90\degr]$ &  8885 & 1.00 & 40 \% & n/a              & NED \citep{r:ned}         \\
          source type  & $[-90\degr, $ & $  90\degr]$ & 10067 & 1.00 & 45 \% & n/a              & NED \citep{r:ned}         \\
          \hline
      \end{tabular}
   \end{center}
   \tablecomments{The associations are considered established if the 
                  probability of false association is below 0.01. The column
                  Area shows the fraction of the celestial sphere a given 
                  catalogue covers. The column Assoc. fraction shows the
                  fraction of the RFC sources that are associated within 
                  the area covered by a given catalogue.
   }
   \tablenotetext{(a)}{ \hspace{0.5em}
                        1eRASS covers the area with galactic latitudes $b<0^\circ$
                      }
   \label{t:matches}
\end{table}

\section{Discussion}
\label{s:disc}

  An absolute astrometry catalogue does not depend on the a~priori positions
of observed sources and the a~priori positions of any other sources. 
A change in the a~priori source positions, station positions, and the Earth 
orientation parameters does not affect the results. To check our software, 
we performed a test and added the zero mean random noise with the second 
moments of 100~mas in source positions, 10~cm in station positions and 
50~mm/yr to station velocities, and then we ran the solution. The results
did not change within rounding errors. The ability to make a totally 
reference frame free solution is a substantial advance in 
the field.

The term ``fundamental catalogue'' coined by \citet{r:fk1} was
originally applied for a catalogue of observations of celestial objects
made by absolute methods in right ascensions and declinations. In the
past, fundamental catalogues were constructed by weighted combinations 
of {\it results} presented in individual absolute astrometry 
catalogues that were considered the most precise \citep{r:fricke85}. 
Combining catalogues poses a significant challenge to account for 
differences in data reduction, in systematic errors of individual
catalogue, and in assigning correct weights of individual catalogues.
This procedure is not transparent, brings an element of subjectivity,
and is not equivalent to the best fit of data in the least squares
sense. We have overcome these difficulties by combining
{\it all observations} in one least squares solution, leaving 
no data point behind. We argue that this approach provides the result 
that is closer to the ideal of the fundamental astronomy than what 
can be achieved by combining individual catalogues.

  An absolute position catalogue has three free rotations, and these 
rotations cannot be determined from observations in principle. The 
observations determine a family of solutions that are transformed 
to each other via a 3D rotation. For convenience, we selected the 
orientation of the published RFC release to preserve the continuity 
to the previous versions of the RFC and to the ICRF families of 
catalogues and its precursors that can be traced back to the 1980s.
In particular, the rotation of the RFC with respect to the ICRF3 
catalogue is within 0.05~mas. This is a factor of 2--4 
less than the contribution of systematic errors in position 
uncertainties, which are negligible for most of the applications. 
We should note that the rotation angles are not uniquely defined: 
they depend on the subset of common sources used for computation 
of the rotation and on assigned weights. We should point out that 
the relative rotation of two absolute catalogues is a quantity that 
does not have a physical meaning. Since the absolute orientation 
of a catalogue cannot be measured, any inference based on 
measurable quantities cannot depend on a specific choice of 
the catalogue orientation. 

  Absolute astrometry observing programs are rather demanding. The 
observing programs should be designed in such a way that source
positions, station positions, and the Earth orientation be 
estimated with the highest accuracy. This can be achieved 
because a)~source positions, station positions, and the Earth
rotation evolve as slow continuous functions; and 
b)~observing sessions have a significant fraction of overlaps.
An overlap in observed sources means that a fraction of sources 
are observed in different experiments at different epochs, and 
these sources tie the system of equations together, 
provided source position evolution is negligible. An overlap in 
observed stations means arrays of stations co-observed with
other stations at different epochs. Since the model of station 
position evolution is adjusted, these observations tie all arrays 
together. An overlap in the Earth orientation means the common 
stations from all sub-arrays participate in observations dedicated 
to determination of the Earth orientation within several days
of astrometric observations, and the parameters of the Earth
orientation tie together geodetic and astrometric observing
sessions since the Earth rotation is sufficiently smooth.
We exploited the overlaps in the RFC solution explicitly by
estimating source positions, parameters of the model of 
station evolution, and parameters of the mathematical model of
the Earth orientation in a single global solution.

  An absolute position catalogue defines the reference. That 
enables differential observations that are much less demanding
for their design. Differential observations allow us to determine 
a positional offset of a target with respect to the reference 
sources, also known as calibrators. Knowing position of a reference 
source, we get the position of the target. As we mentioned
in the Introduction, the positional error of the target is the sum 
in quadrature of the uncertainty of the target-calibrator position 
offset and the reference source position uncertainty.
Therefore, characterizing positional errors 
of absolute astrometry catalogue is very important. We should emphasize
that the RFC as well as all other VLBI absolute catalogues 
is derived from analysis of group delays. The impact of the core-shift 
on group delays and phase delays is different, and at frequencies below 
10~GHz these differences are greater than other systematic errors. 
When a goal of differential astrometry observations is to analyze 
just target-calibrator positional offsets, 
for instance for determination of proper motions and parallaxes, 
this distinction can be omitted, provided the core-shift remains
stable. However, for differential astrometry programs at frequencies 
below 10~GHz that require high positional accuracy, the core-shift should 
be determined. The technique of core-shift determination for the 
purpose of improving differential astrometry results is described 
in \citet{r:hao24}.

\begin{figure}[h!]
   \includegraphics[width=0.48\textwidth]{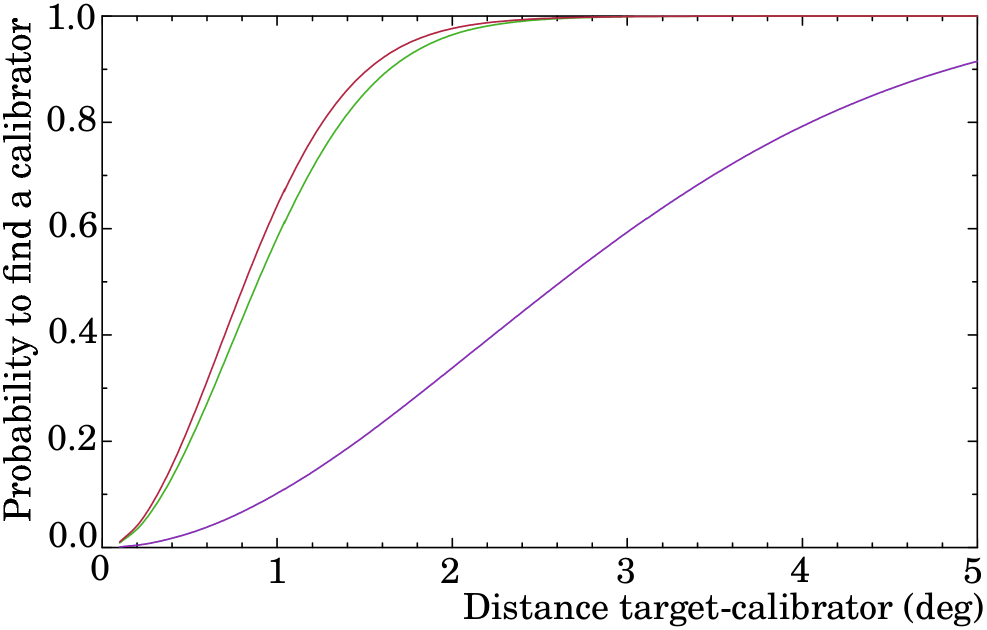}
   \hspace{0.039\textwidth}
   \includegraphics[width=0.48\textwidth]{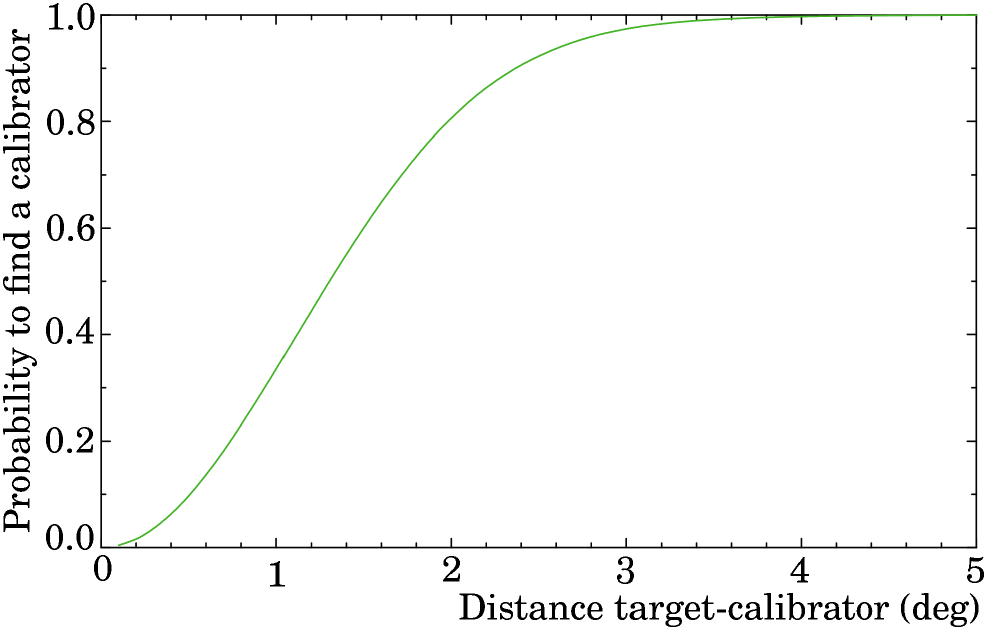}
   \caption{The probability to find a phase calibrator as a function 
            of the angular distance target-calibrator for three
            bands: C, X, and K. 
            {\it Left: } zone $[-40^\circ, +90^\circ]$. The upper
            red curve is for C-band, the middle green curve is for
            X-band, and the low purple curve is for K-band
            {\it Right: } zone $[-90^\circ, -40^\circ]$ for
            X-band.
           }
   \label{f:prob_cal}
\end{figure}

The Radio Fundamental Catalogue provides a rich list of
objects for phase referencing. The specific criteria for 
which a given source should satisfy to be considered as a phase 
calibrator depend on the application. As an example, we considered 
the following criteria: 1)~flux density $>30~$mJy at medium or long 
baselines and 2)~semi-major error ellipse $<3$~mas. We computed the 
probability to find a phase calibrator at a grid of 
$0.25^\circ \times 0.25^\circ$ as a function of the angular distance 
between target and calibrator on the entire celestial sphere for 
bands C, X, and K. Due to scarcity of 22~GHz flux density information, 
we dropped the first criterion for that band. We show the dependencies 
of these probabilities versus angular distance at zones 
$[-40^\circ, +90^\circ]$ and $[-90^\circ, -40^\circ]$ in 
Figure~\ref{f:prob_cal}. The probability to find a phase calibrator 
within $2^\circ$ at C or X band in the first zone is 96--98\% 
and in the second zone is 81\%.

  Since there is no position catalogue with an accuracy significantly
better than the RFC, we have to evaluate its positional accuracy indirectly.
We showed that the decimation test provides rather different results 
depending on the order of splitting the observations into subsets. 
We interpret this as a consequence of the presence of correlations in 
the noise. White noise is an idealization that has a limited range 
of applicability. Figures~\ref{f:floor_scale_ra}--\ref{f:floor_scale_dec}  
represent the manifestation of the red noise in the data. We used
a sophisticated scheme for ad~hoc weights update at four levels:
1)~the additive weight correction during fringe fitting; 
2)~the additive elevation-dependent noise in group delays;
3)~the additive elevation-independent noise in group delays;
4)~rescaling source positions uncertainties. All these empirical 
weight updates are made to compensate the impact of the deficiency
of our error model. As we showed, one cannot reliably discriminate
the error floor from the error scaling factor. We opted to using a 
hybrid scheme that adjusts error floor using the sources with
small positional errors and the scaling factor for the rest of
the sources. Here we argue that the scaling factor should not 
depend on position statistics. Although we can easily interpret
the additive noise floor as a manifestation of errors that 
are independent of the observations, e.g.\ atmospheric path delay, or
the source structure contribution, we cannot offer a simple
interpretation for the scaling factor. We consider that relatively
small scaling factors og 1.08 and 1.16 for right ascension and 
declination is a sign of an improvement of the error model.
 
  The presence of a relatively large declination-dependent error 
floor in the range of 0.04 to 0.22~mas is somewhat disappointing. It is 
greater than the 0.03 and 0.05~mas floor reported in literature in the past
\citep{r:icrf3}. This does not reflect a deficiency in our data analysis 
strategy that would have added an extra noise, but reflects our diligence 
in accounting for the red noise that was not considered in prior 
publications. The presence of the red noise has a critically important 
implication. When only the white noise affects observations, position 
accuracy improves indefinitely with an increase of the number of 
observations. In the presence of correlation, or using an equivalent 
formulation, the red noise, positional accuracy is approaching a limit 
with an increase of the number of observations, and upon reaching that 
limit does not grow any more.

  A detailed investigation of the origin of the red noise goes
well beyond the scope of this article. Three factors certainly
play a role: 1)~deficiency of the modeling path delay in the 
neutral atmosphere; 2)~lack of modeling source structure 
contribution; and 3)~mismodeling path delay in the ionosphere when 
processing single-band delay data. In \citet{r:radf} we provided 
a detailed argumentation in favor of these factors.

It was known for decades that positions of radio sources
exhibit changes at a level of tenths of a milliarcsecond 
\protect\citep[e.g.][]{r:ma91}. These changes are not entirely random, 
but have a systematic component that in some cases is related to
a flaring activity. This systematic component makes position changes 
correlated. The disparity of the global versus local floor-scale diagrams
in Figures~\ref{f:floor_scale_ra}--\ref{f:floor_scale_dec} is explained 
by these correlations. Although coordinate estimates determined over 
a short period time may characterize the source position over that 
period more precisely when very bright sources are observed, determination 
of epoch-based positions and their covariances poses a number of challenges. 
An epoch-based positional offset depends not only the contribution of random 
errors affecting a given source, but also on positional offsets of other 
sources observed in that experiment, which cause an additional network
jitter.  Position estimates are connected either explicitly through 
epoch-based net-rotation constraints, or implicitly by fixing positions of 
some sources. A typical number of observations in a given astrometry session
is several hundreds, compare with a total number of 22 thousands sources. 
The observed jitter will be a superposition of the jitter caused by source 
structure, a network position jitter, and the jitter caused by propagation 
delay. The problem of separation of these contributions is not yet solved, 
as we can see from the recent work of \citet{r:cig24}: the excessive 
noise in source position time series over declination is roughly twice 
larger than over right ascension, which implies that {\it on average}, 
the observed jitter is dominated by mismodeling atmospheric path 
delay. The Radio Fundamental Catalogue provides the time averaged positions 
for the reported weighted mean epochs and serves as a reference for 
characterizing deviations with respect to the mean.

  Comparison of the RFC with \Gaia shows a significant improvement
in the agreement of the normalized position differences with respect 
to early publications. We attribute this improvement to enhancements
in the VLBI source positional error model{, which made it more realistic}. 
Analysis of the position difference leads us to the conclusion that both 
RFC and \Gaia EDR3 error models are accurate to at least the 20\% level. 
We cannot estimate the accuracy of the error model with a greater 
confidence because the \Gaia positional errors are affected by the presence 
of optical jets that systematically shift the position of a \Gaia 
centroid along the jet directions with respect to the center engine. 
Our comparison revealed the presence of the extra noise in \Gaia EDR3 
reflected as an increase of the $\chi^2$/ndf quantity, and the presence of 
the elevation-dependent noise in the RFC that is more noticeable
for sources that had fewer than 120 observations.

\subsection{The historical context of VLBI surveys}

  Since the first pioneering work of \citet{r:first-vlbi-cat} that has 
demonstrated the use of VLBI for astrometry, further progress in VLBI 
astronomy evolved following four routes.

  The first route is running pathfinder surveys. The goal of these 
surveys was to find compact sources and measure their correlated flux 
densities. Using the Deep Space Network,
\citet{r:morabito82,r:morabito83,r:wehrle84,r:morabito85,r:preston85,r:morabito86}
observed over 1500 targets at 2.3 and 8.6~GHz and have detected over 1000 
objects. Using these surveys, positions of 323 sources were determined with 
an accuracy of 300--1000~mas. These surveys were not designed for astrometry 
and used a narrow bandwidth. This explains their low positional accuracy. 
Visibility data from these campaigns were not made publicly available and are 
not used in the RFC. The development of the VLBA in the 1990s made it possible
to run large VLBI surveys. It was realized that when observations are made using 
a spanned bandwidth over 500~MHz, positional accuracy at the milliarcsecond
level can be achieved. 

  The development of the Mark~III recording system \citep{r:mark3} allowed
astronomers to record a spanned bandwidth of 360~MHz and later 720~MHz. 
That made it possible to re-observe the sources detected in pathfinder surveys 
and determine their positions with a sub-milliarcsecond level of accuracy. 
\citet{r:cma86} published the first absolute astrometry catalogue based 
on a wide-band system using observations collected under geodetic VLBI 
programs. Following that route, a number of observing sessions with 
expanded source lists were conducted. The outcome of these experiments 
were catalogues of sources with a milliarcsecond level precision that were 
disseminated in the IERS annual reports \citep{r:arias95}. Based on these 
observing programs, the ICRF1 absolute astrometry catalogue was published 
\citep{r:icrf1}. Following that route, a large number of observing programs 
was organized. Several astrometric multi-program solutions were published, 
among them, ICRF2 \citep{r:icrf2}, ICRF3 \citep{r:icrf3}, 
and WFCS \citep{r:wfcs}.

  With the development of the hybrid imaging algorithms
\citep{r:rw78,r:cotton79}, it became possible to use VLBI for making images 
of observed radio sources. VLBI image surveys starting with the work of 
\citet{r:pearson88}, which marked the third route of  evolution of VLBI 
astronomy. A number of imaging survey followed. We mentioned here the two 
largest surveys: 
VIPS \citep{r:vips} and 2\,cm~VLBA / MOJAVE \citep{r:2cmVLBA,r:mojave_apjs}, 
with 1119 and 623 sources, respectively. Although astrometric VLBI 
observations can be used for imaging \citep[e.g.,][]{r:piner12,r:pushkarev12},
and imaging experiments could be used for astrometry and geodesy 
\citep[see, for example,][]{r:kra21}, these programs ran rather independently. 
In imaging experiments IFs are usually allocated contiguously, and that 
significantly reduces the astrometric accuracy. Astrometric programs often 
did not have a good amplitude calibration, what made imaging challenging and 
resulting flux density scale inaccurate. We should note that since 2020s all 
these three routes have a tendency to converge since the total bandwidth 
substantially increased, and even contiguous IF allocation provides wide 
enough spanned bandwidth for the precise group delay determination.

\begin{figure}[h!]
   \centerline{\includegraphics[width=0.616\textwidth]{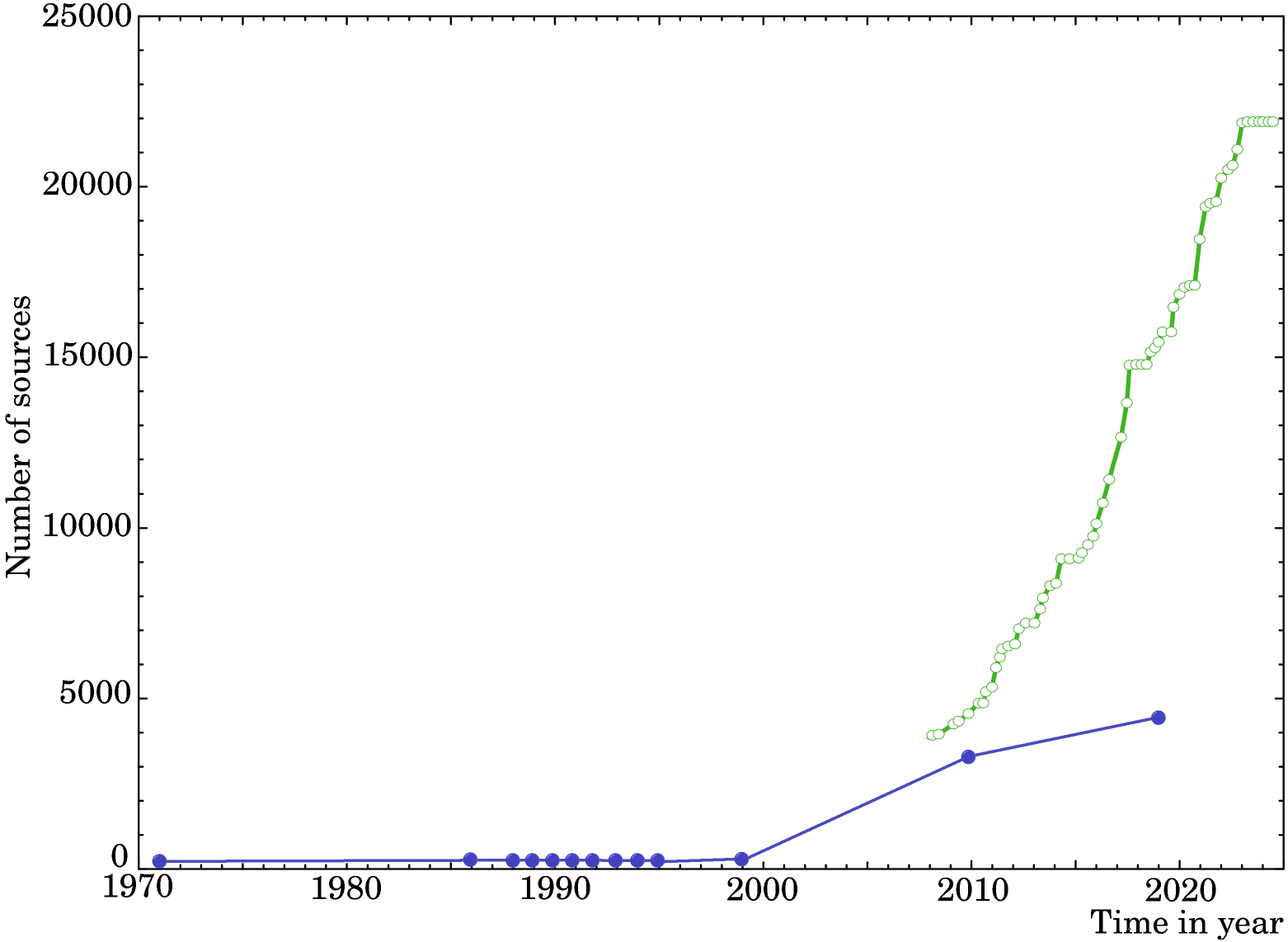}}
   \caption{The evolution of the VLBI absolute astrometry catalogues. 
            The green circles show releases of the RFC. The blue circles
            show the historical VLBI position catalogues: 
            \citep{r:first-vlbi-cat} followed by \citep{r:cma86}, the IERS 
            series of source catalogues, then ICRF1, ICRF2, and ICRF3.
           }
   \label{f:rfc_history}
\end{figure}

In Figure~\ref{f:rfc_history} we showed the growth of the number of sources 
in the RFC online releases (green line) and in other multi-program 
catalogues (blue line) shown as a reference.

  For completeness, we should mention the fourth route: differential
astrometry. Observing a pair of a calibrator and a target, atmospheric 
path delay contribution can be reduced roughly as a target-calibrator 
separation expressed in radians. This allows one to achieve the accuracy 
of the displacement of a target with respect to a calibrator at the level 
of 0.03~mas \citep{r:honma_redi14}. A target can be very weak, since 
it still can be detected using a long integration time. \citet{r:mjive20} 
published a large mJIVE-20 survey of 4336 sources detected using 
differential astrometry at 1.4~GHz. Only 73 sources from  mJIVE-20 are 
common with the RFC. The median reported positional accuracy of mJIVE-20 
is 0.7~mas. Systematic errors caused by the core-shift at L-band are 
expected to contribute at the level of 2~mas \citep{r:sok11,r:hao24}. 
We do not include this program in the RFC because mJIVE-20 does not 
fall into a category of absolute astrometry.

\subsection{Future development}

  Undoubtedly, applying the source structure contribution in data 
reduction and refining models of atmospheric path delay will
improve accuracy. Although as \citet{r:tho80} showed, computation of 
source structure contribution from source images is relatively 
straight-forward, logistically, this problem is very time consuming. 
That is why little progress in this area has been achieved for over 
four decades. Nevertheless, the availability of over $10^5$ VLBI 
images brings certain optimism that some day in the future applying 
source structure contribution will become routine. In \citet{r:radf} 
we showed that {\it on average}, the source structure contributes to 
source positional errors at a level of 0.05--0.07~mas per component,
right ascension and declination.

  We should acknowledge that our use of a sophisticated error 
re-scaling scheme is a reflection of the basic flaw in the current 
paradigm of space geodesy that uses the a~priori weight matrices 
with zero off-diagonal terms, which implies the noise is 
uncorrelated, i.e.\ white. We envisage that the full covariance 
matrices of the noise in group delay observables will be used in the 
future in line with ideas of \citet{r:halsig19} or similar. That will 
allow us to develop a robust error model without resorting to the use 
of the empirical floor and scaling factors. This model, in turn, 
will allow us to predict better positional errors from planned 
observations and optimize the observing strategy accordingly.

  We do not envisage that the number of sources with positions determined
with VLBI in the absolute astrometry mode will be improved significantly, 
i.e.\ by a factor of 1.5 or greater in the next 10--15~years because there 
is no planet-wise resource to perform such a program. Therefore, we 
anticipate that future observing programs will focus primarily on sources 
whose positions we want to improve. Beyond the 10--15~year horizon, new
high sensitivity instruments, such as SKA and ngVLA will become 
operational and likely new observational programs for a support of these
facilities will be initiated, either in the absolute or differential
astrometry mode. 

  Re-observations of all the sources which were observed within one scan 
only in pathfinder VLBI surveys in four to six scans would improve their 
position uncertainties one order of magnitude down to the 0.2--0.5~mas level
and significantly improve fidelity of reconstructed images needed for the 
source structure modeling. Because of a lack of 
resources to re-observe all the RFC sources with a positional accuracy worse 
than 0.2~mas before the SKA and ngVLA era, the priorities should be 
established. As we have shown, additional observations improve position 
accuracy only up to a certain limit. Unfortunately, this consideration was 
not always taken into account in the past, and we hope the situation will 
be improved in the future. We see several areas where new observations 
will have a big impact.

  First, there are certain areas on the sky that are more frequently 
observed. These are the ecliptic band and the area close to the Galactic 
plane. The first area is of a great importance for space navigation and for
planned observations of radio beacons on the Moon and other Solar system 
bodies. Many perspective calibrator sources have the positional accuracy 
of a factor of 3 to 10 above the limit set by the presence of the red noise. 
The areas in the Galactic plane are important for measurement of parallaxes 
and proper motions. Observations of targets in the vicinity of the Galactic 
plane at 23~GHz are advantageous because at this frequency the impact of 
scattering in the interstellar medium is significantly reduced.

  Second, the positional accuracy of objects in the Southern Hemisphere not
reachable by the VLBA can be improved by dual-band or quad-band observations
with the LBA, and planned IVS radio telescopes in Thailand, Malaysia,
and Indonesia, as well as the African VLBI network. With the on-going
development of the Thai National VLBI Array (TVA) in Thailand \citep{r:tva} 
and the South-East Asian VLBI Network (SEAVN) new opportunities for 
dedicated astrometric programs targeting the Southern Hemisphere 
are expected to emerge. Currently, most of RFC sources at declinations 
below $-40^\circ$ have been observed at X-band only, which limits 
the positional accuracy.

  Third, a number of sources with a peculiar structure, such as visual 
doubles, or those showing an indication of a complex structure in existing 
images can be re-observed using more scans per source in order to provide 
high fidelity images, to establish their nature, and to improve their 
positions by applying source structure. 
  
\section{Summary and Outcomes}
\label{s:sum}

   We present a catalogue of absolute positions and correlated flux
densities of 21,942 sources detected with VLBI. All of them, but 24,
are AGNs. This is a result of analysis of 17 thousand hours of VLBI 
observations dedicated for astrometry and 194 thousand hours 
dedicated for geodesy since 1980 through 2024. We used virtually all 
suitable publicly available VLBI observations to date in our analysis. 
Source position uncertainties range from 0.09 to 1480~mas with the 
median semi-minor and semi-major error ellipse axes 0.6 and 1.4~mas, 
respectively.

  The Radio Fundamental Catalogue increases the number of sources
reported in prior VLBI catalogues that accumulated data from 
historical observing programs by a factor of 5. It provides positions 
of all the sources reported in prior absolute astrometry VLBI 
catalogues with significantly improved positional accuracy and more 
realistic reported uncertainties. The Radio Fundamental Catalogue is 
accompanied with a collection of over 125,000 images of observed radio 
sources, cross-matches against 14 large surveys, sources properties 
collected by the NED, including redshifts for 1/3 of the objects, and 
jet directions reported in literature.

  An advanced error model that accounts for the contribution of 
red noise is developed for the derivation of the Radio Fundamental 
Catalogue. Comparison of the Radio Fundamental Catalogue with the \Gaia 
EDR3 catalogue in the optical range demonstrates a significant improvement 
in the agreement of the normalized arc length position difference with 
respect to prior publications. This comparison sets the upper limit of 
inaccuracy of the Radio Fundamental Catalogue error model: 20\%. 

  Considering all these factors, we position the Radio Fundamental 
Catalogue as a new standard in VLBI radio astrometry that overrides 
previous catalogues used for realization of the celestial reference 
frame. The Radio Fundamental Catalogue is the most complete catalogue 
that provides milliarcsecond level accurate positions and correlated flux 
densities based on analysis of VLBI observations. The catalogue 
establishes the foundation for space geodesy, space navigation, 
differential astrometry, AGN jet and scattering studies, as well as 
population analysis in radio astronomy.

  The first version of the catalogue became available online on
February 12, 2008. Since then the catalogue is updated on 
a quarterly basis, and it will be updated with a three months cadence
in the future. Each the catalogue release has a notation {\tt rfc\_yyyys}, 
where {\tt yyyy} is the year and {\tt s} is a suffix: {\tt a}, {\tt b}, 
{\tt c}, or {\tt d}. In each update cycle the astrometric solutions
that use all the data since 1980 through present are updated. Each 
update includes fixes in prior experiments and incorporates new 
experiments. The new versions of the catalogue will be available at 
\dataset[\doi{10.25966/dhrk-zh08}]{\doi{10.25966/dhrk-zh08}}.

\begin{acknowledgments}
   This work was done using only the following publicly available 
datasets 
\par\vspace{-2ex}\par
\begin{enumerate}\setlength{\itemsep}{-0.9ex}
  \item collected with the VLBA network of the NRAO and available 
        \url{https://data.nrao.edu/portal/};
  \item collected with the IVS network and available at the NASA Crustal 
        Dynamics Data Informational System (CDDIS) 
        \url{https://cddis.nasa.gov/archive/vlbi/}; 
  \item collected with the LBA network  and available at 
        \url{https://atoa.atnf.csiro.au}
  \item collected with the EVN network and available at 
        \url{http://archive.jive.nl/scripts/portal.php};
  \item collected at the EAVN network and available at 
        \url{https://radio.kasi.re.kr/arch/search.php}
  \item collected at the KVN network and available at 
        \url{https://radio.kasi.re.kr/arch/search.php}
  \item collected at the KaVa network and available at 
        \url{https://radio.kasi.re.kr/arch/search.php}
  \item collected at the VERA network and available at 
        upon request to vera-contact@ml.nao.ac.jp
  \item collected at the CVN network and available at 
        upon request to sfc@shao.ac.cn
\end{enumerate}

  The NRAO is a facility of the National Science Foundation operated under 
cooperative agreement by Associated Universities, Inc. The author 
acknowledges use of the VLBA under the USNO's time allocation for some 
datasets. The Long Baseline Array is part of the Australia Telescope 
National Facility (https://ror.org/05qajvd42) which is funded by the 
Australian Government for operation as a National Facility managed by CSIRO. 
The European VLBI Network is a joint facility of independent European, 
African, Asian, and North American radio astronomy institutes. This work 
is made use of the East Asian VLBI Network (EAVN), which is operated under 
cooperative agreement by National Astronomical Observatory of Japan (NAOJ), 
Korea Astronomy and Space Science Institute (KASI), Shanghai Astronomical 
Observatory (SHAO), Xinjiang Astronomical Observatory (XAO), Yunnan 
Astronomical Observatory (YNAO), National Astronomical Research Institute 
of Thailand (Public Organization: NARIT), and National Geographic Information 
Institute (NGII), with the operational support by Ibaraki University (for the 
operation of Hitachi 32-m and Takahagi 32-m telescopes), Yamaguchi University
(for the operation of Yamaguchi 32 m telescope), and Kagoshima University 
(for the operation of VERA Iriki antenna). We are grateful to the staff of the 
KVN who helped to operate the array and to correlate the data. The KVN and 
a high-performance computing cluster are facilities operated by the Korea 
Astronomy and Space Science Institute (KASI). The KVN observations and 
correlations are supported through the high-speed network connections among 
the KVN sites provided by the Korea Research Environment Open NETwork 
(KREONET), which is managed and operated by the Korea Institute of Science 
and Technology Information (KISTI). We would like to thank all staff members 
of the VERA stations and the Mitaka correlation center for their assistance 
in the observations. This research has made use of data from the MOJAVE 
database that is maintained by the MOJAVE team \citep{r:mojave_apjs}. This 
research has made use of the NASA/IPAC Extragalactic Database (NED), which 
is funded by the National Aeronautics and Space Administration and operated 
by the California Institute of Technology.

L.Y.P was supported in part by the NASA Space Geodesy Project.
Y.Y.K was supported by the MuSES project which has received funding 
from the European Research Council (ERC) under the European Union's 
Horizon 2020 Research and Innovation Programme (grant agreement No 101142396).
We certify that no machine learning or artificial intelligence
techniques were used neither during data analysis, nor in manuscript
preparation. 

  The work on the Radio Fundamental Catalogue commenced in 2000. For the
course of twenty four years we got help and good advice from many colleagues. 
It is our pleasure to acknowledge
Walter Alef,
Karen Baver,
Simone Bernhart,
Alessandra Bertarini,
Chirsian Bizouard,
Johannes B\"{o}hm,
Sergei Bolotin,
Geraldine Bourda,
Walter Brisken,
Yoon Kyung Choi,
Brian Corey,
Nicholas Corey,
Bill Cotton,
Yuzhu Cui,
Adam Deller,
Philip Edwards,
Martine Feissel,
Alan Fey,
Ed Fomalont,
Laura Vega Garcia,
John Gipson,
Jose-Luis Gomez,
Anne-Marie Gontier (deceased),
David Gordon,
Jakob Gruber,
Sergei Gulyaev,
Vadim Gubanov (deceased),
Leonid Gurvits,
Rudiger Haas,
Xuan He,
Dan Homan,
Mareki Honma,
Christopher Jacobs,
Frederic Jaron,
Megan Johnson,
Michael Johnson,
Svetlana Jorstad,
Taehyun Jung,
Nikolai Kardashev (deceased),
Ken Kellermann,
Segei Klioner,
Leonid Kogan (deceased),
Tetsuro Kondo,
Tatiana Koryukova,
Yuri A. Kovalev,
Georgiy Krasinski (deceased), 
Hana Krasna, 
Thomas Krichbaum,
Sergei Kurdubov,
Sang-Sung Lee,
Jeong Ae Lee, 
Rocco Lico,
Elisabetta Liuzzo,
Matt Lister,
Andrei Lobanov,
Chopo Ma,
Lucia Mccallum,
Seiji Manabe,
Dan MacMillan,
Alan Marscher,
Iv\'an Mart\'i-Vidal,
Christopher Marvin,
Leonid Matveenko (deceased),
Alexey Melnkiov,
Cristina Garcia Miro,
Arthur Niell,
Jim Moran,
Ramesh Narayan,
Axel Nothnagel,
Kristina Nyland,
Chris Phillips,
Alexandr Plavin,
Christian Ploetz,
Sergei Pogrebenko,
Alexandr Popkov,
Richard Porcas,
Alexandr Pushkarev,
Cormac Reynolds,
Mar\'ia Rioja,
Eduardo Ros,
Jim Ryan (deceased), 
Tuomas Savolainen,
Fengchun Shu,
Kirill Sokolovsky,
Frank Schinzel,
Harald Schuh,
An Tao,
Greg Taylor,
Paulo Tomassi,
Rene Vermeulen,
Petr Voitsik,
Jan Wagner,
Alet de Witt,
Craig Walker,
Minghui Xu,
Shuangjing Xu,
Elenonora Yagudina, 
J.~Anton Zensus,
and Bo Zhang.

  A significant part of the work was done at nights, weekends, 
and vacation. We thank our families for understanding and ask 
for forgiveness.

\end{acknowledgments}

\appendix

  The Radio Fundamental Catalogue main master Table~\ref{t:main}, the 
Table~\ref{t:multi} with dual-band and single-band source positions, 
and the cross-matching Table~\ref{t:cross-match} are too wide to be 
shown here. The are presented as machine readable tables only.

  The samples of the Table~\ref{t:err_floor_mod} with the parameters of 
the error floor as a function of declination for right ascension, the 
Table~\ref{t:ncj} of non-canonical RFC designator, the Table~\ref{t:ncb} 
of non-canonical common names, the Table~\ref{t:pairs} of close source 
pairs, the Table~\ref{t:nondet} with the list of source that were observed,
but have not detected, the Table~\ref{t:astro} of the names and dates of 
astrometric experiment names, the Table~\ref{t:geod24hr} of geodetic 24~hr 
experiment names, and the Table~\ref{t:geod1hr} of geodetic 1~hr experiment 
names are shown in the appendix. They are published entirely in the 
machine-readable format.

\startlongtable
\begin{deluxetable}{r @{\enskip} l @{\enskip} l p{0.80\textwidth}}
   \tablecaption{Column description of the main master table with 
                 the Radio Fundamental Catalogue\tablenotemark{a}.
                }
      \tablehead{
         \colhead{\#} & 
         \colhead{Unit} & 
         \colhead{Label} &
         \colhead{Description}
      }
      \startdata \setcounter{myitemm}{0}
      \itemm & ---    & Name   & RFC object designator. A 10 character long J2000-name 
                                 with prefix RFC \\
      \itemm & ---    & Comnam & Common name \\
      \itemm & h      & RAh    & Hours of Right Ascension (J2000) \\
      \itemm & min    & RAm    & Minutes of Right Ascension (J2000) \\
      \itemm & s      & RAs    & Seconds of Right Ascension (J2000) \\
      \itemm & ---    & DE-    & Sign of the Declination (J2000) \\
      \itemm & deg    & DEd    & Degrees of Declination (J2000) \\
      \itemm & arcmin & DEm    & Arcminutes of Declination (J2000) \\
      \itemm & arcsec & DEs    & Arcseconds of Declination (J2000) \\
      \itemm & mas    & eRA    & Error in right ascension without $\cos\delta$ factor \\
      \itemm & mas    & eDE    & Error in declination \\
      \itemm & ---    & Corr   & Correlation between right ascension and  declination \\
      \itemm & ---    & Nobs   & Number of observations used in the fused solution \\
      \itemm & ---    & Nsca   & Number of scans used in the fused solution \\
      \itemm & ---    & Nses   & Number of observing sessions used in the fused solution \\
      \itemm & Jy     & FsS    & Median flux density at S-band, [2.2, 2.4]~GHz, at baseline 
                                 projection lengths shorter than 1000 km\tablenotemark{b} \\
      \itemm & Jy     & FmS    & Median flux density at S-band, [2.2, 2.4]~GHz,   
                                 at baseline projection lengths in the range 1000 to 5000 km \\
      \itemm & Jy     & FlS    & Median flux density at S-band, [2.2, 2.4]~GHz,   
                                 at baseline projection lengths longer than 5000 km\tablenotemark{c} \\
      \itemm & Jy     & FsC    & Median flux density at C-band, [4.3, 5.1]~GHz,   
                                 at baseline projection lengths shorter than 1000 km\tablenotemark{b} \\
      \itemm & Jy     & FmC    & Median flux density at C-band, [4.3, 5.1]~GHz,   
                                 at baseline projection lengths in the range 1000 to 5000 km \\
      \itemm & Jy     & FlC    & Median flux density at C-band, [4.3, 5.1]~GHz,   
                                 at baseline projection lengths longer than 5000 km\tablenotemark{c} \\
      \itemm & Jy     & FsX    & Median flux density at X-band, [7.3, 8.6]~GHz,   
                                 at baseline projection lengths shorter than 1000 km\tablenotemark{b} \\
      \itemm & Jy     & FmX    & Median flux density at X-band, [7.3, 8.6]~GHz,   
                                 at baseline projection lengths in the range 1000 to 5000 km \\
      \itemm & Jy     & FlX    & Median flux density at X-band, [7.3, 8.6]~GHz,   
                                 at baseline projection lengths longer than 5000 km\tablenotemark{c} \\
      \itemm & Jy     & FsU    & Median flux density at U-band, [15.2, 15.5]~GHz, 
                                 at baseline projection lengths shorter than 1000 km\tablenotemark{b} \\
      \itemm & Jy     & FmU    & Median flux density at U-band, [15.2, 15.5]~GHz, 
                                 at baseline projection lengths in the range 1000 to 5000 km \\
      \itemm & Jy     & FlU    & Median flux density at U-band, [15.2, 15.5]~GHz, 
                                 at baseline projection lengths longer than 5000 km\tablenotemark{c} \\
      \itemm & Jy     & FsK    & Median flux density at K-band, [23.2, 24.2]~GHz, 
                                 at baseline projection lengths shorter than 1000 km\tablenotemark{b} \\
      \itemm & Jy     & FmK    & Median flux density at K-band, [23.2, 24.2]~GHz, 
                                 at baseline projection lengths in the range 1000 to 5000 km \\
      \itemm & Jy     & FlK    & Median flux density at K-band, [23.2, 24.2]~GHz, 
                                 at baseline projection lengths longer than 5000 km\tablenotemark{c} \\
      \itemm & yr     & MeaEpo & Weighted mean epoch of observations \\
      \enddata
      \par
      \tablenotetext{(a)}{ \hspace{0.5em}
                     Table \ref{t:main} is too wide to be shown here.
                     It is published in the machine-readable format only
                     as file rfc.txt.
                    }
    \tablenotetext{(b)}{ \hspace{0.5em}
       Represent an estimate of the flux density integrated over VLBI image.
    }
    \tablenotetext{(c)}{ \hspace{0.5em}
       Represent an estimate of the unresolved flux density of VLBI image.
    }
    \tablenotetext{(d)}{ \hspace{0.5em}
       $-9.99$ indicates a Null value: no estimate is available.
    }
    \label{t:main}
\end{deluxetable}


\startlongtable
\begin{deluxetable}{r l l p{0.75\textwidth}}
   \tablecaption{Column description of the table with source positions from 
            dual-band, S-band, C-band, X-band, and K-band as differences 
            with respect to the RFC positions derived from analysis 
            of the fused group delays\tablenotemark{a}.
           }
      \tablehead{
         \colhead{\#} & 
         \colhead{Unit} & 
         \colhead{Label} &
         \colhead{Description}
      }
      \startdata \setcounter{myitemm}{0}
      \itemm & --- & Name  & RFC object designator. A 10 character long J2000-name with prefix RFC \\
      \itemm & mas & eRA   & Error in right ascension without $\cos\delta$ factor applied \\
      \itemm & mas & eDE   & Error in declination \\
      \itemm & --- & Nobs  & Number of observations used in the fused solution \\
      \itemm & --- & Nsca  & Number of scans used in the fused solution \\
      \itemm & --- & Nses  & Number of observing sessions used in the fused  \\
      \itemm & mas & dxD   & Offset of the dual-band position along the right ascension 
                             axis without $\cos\delta$ factor applied with respect to the
                             position from the fused solution \\
      \itemm & mas & dyD   & Offset of the dual-band position along the  declination axis 
                             with respect to the position from the fused solution \\
      \itemm & mas & exD   & Uncertainty of the dual-band position along the right ascension 
                             axis without $\cos\delta$ factor applied \\
      \itemm & mas & eyD   & Uncertainty of the dual-band position along the declination axis \\
      \itemm & --- & CorrD & Correlation between right ascension and declination of the 
                             dual-band source position \\
      \itemm & --- & NobsD & Number of observations used in the dual-band  solution \\
      \itemm & --- & NscaD & Number of scans used in the dual-band solution \\
      \itemm & --- & NsesD & Number of observing sessions used in the  dual-band solution \\
      \itemm & mas & dxS   & Offset of the S-band position along the right ascension axis 
                             without $\cos\delta$ factor applied with respect to the
                             position from the fused solution \\
      \itemm & mas & dxS   & Offset of the S-band position along the  declination axis
                             with respect to the position from the fused solution \\
      \itemm & mas & exS   & Uncertainty of the S-band position along the right ascension 
                             axis without $\cos\delta$ factor  applied \\
      \itemm & mas & eyS   & Uncertainty of the S-band position along the declination axis \\
      \itemm & --- & CorrS & Correlation between right ascension and declination of the 
                             S-band source position \\
      \itemm & --- & NobsS & Number of observations used in the S-band  solution \\
      \itemm & --- & NscaS & Number of scans used in the S-band solution \\
      \itemm & --- & NsesS & Number of observing sessions used in the  S-band solution \\
      \itemm & mas & dxC   & Offset of the C-band position along the right ascension axis 
                             without $\cos\delta$ factor applied with respect to the
                             position from the fused solution \\
      \itemm & mas & dyC   & Offset of the C-band position along the  declination axis
                             with respect to the position from the fused solution \\
      \itemm & mas & exC   & Uncertainty of the C-band position along the right ascension 
                             axis without $\cos\delta$ factor applied \\
      \itemm & mas & eyC   & Uncertainty of the C-band position along the declination axis \\
      \itemm & --- & CorrC & Correlation between right ascension and declination of the 
                             C-band source position \\
      \itemm & --- & NobsC & Number of observations used in the C-band  solution \\
      \itemm & --- & NscaC & Number of scans used in the C-band solution \\
      \itemm & --- & NsesC & Number of observing sessions used in the  C-band solution \\
      \itemm & mas & dxX   & Offset of the X-band position along the right ascension axis 
                             without $\cos\delta$ factor applied with respect to the
                             position from the fused solution \\
      \itemm & mas & dyX   & Offset of the X-band position along the  declination axis
                             with respect to the position from the fused solution \\
      \itemm & mas & exX   & Uncertainty of the X-band position along the right ascension 
                             axis without $\cos\delta$ factor  applied \\
      \itemm & mas & eyX   & Uncertainty of the X-band position along the declination axis \\
      \itemm & --- & CorrX & Correlation between right ascension and declination of the 
                             X-band source position \\
      \itemm & --- & NobsX & Number of observations used in the X-band  solution \\
      \itemm & --- & NscaX & Number of scans used in the X-band solution \\
      \itemm & --- & NsesX & Number of observing sessions used in the  X-band solution \\
      \itemm & mas & dxK   & Offset of the K-band position along the right ascension axis 
                             without $\cos\delta$ factor  applied with respect to the
                             position from the fused solution \\
      \itemm & mas & dyK   & Offset of the K-band position along the  declination axis
                             with respect to the position from the fused solution \\
      \itemm & mas & exK   & Uncertainty of the K-band position along the right ascension 
                             axis without $\cos\delta$ factor  applied \\
      \itemm & mas & eyK   & Uncertainty of the K-band position along the declination axis \\
      \itemm & --- & CorrK & Correlation between right ascension and declination of the
                             K-band source position \\
      \itemm & --- & NobsK & Number of observations used in the K-band  solution \\
      \itemm & --- & NscaK & Number of scans used in the K-band solution \\
      \itemm & --- & NsesK & Number of observing sessions used in the  K-band solution \\
      \enddata
    \tablenotetext{(a)}{ \hspace{0.5em}
            Table \ref{t:multi} is too wide to be shown here.
                   It is published in the machine-readable format only
                   as file multi\_band.txt
    }
    \tablenotetext{(b)}{ \hspace{0.5em}
       $-9.99$ indicates a null value: no estimate is available.
    }
    \label{t:multi}
\end{deluxetable}

\par\bigskip\par

\startlongtable
\begin{deluxetable}{r l l p{0.75\textwidth}}
    \tablecaption{Column description of the table with Cross-matching RFC with 
                  14 surveys\tablenotemark{a}.
                 }
    \tablehead{
         \colhead{\#} & 
         \colhead{Unit} & 
         \colhead{Label} &
         \colhead{Description}
    }
      \startdata \setcounter{myitemm}{0}
      \itemm & ---    & Name           & RFC object designator \\
      \itemm & ---    & NVSS\_name     & Name of the association in the NVSS catalogue \\
      \itemm & arcsec & NVSS\_dist     & Angular distance to the association in the NVSS catalogue \\
      \itemm & ---    & NVSS\_pfa      & The probability of falls association with a counterpart from the NVSS catalogue \\
      \itemm & ---    & VLASS\_name    & Name of the association in the  VLASS catalogue \\
      \itemm & arcsec & VLASS\_dist    & Angular distance to the association in the VLASS catalogue \\
      \itemm & ---    & VLASS\_pfa     & The probability of false association with a counterpart from the VLASS catalogue \\
      \itemm & ---    & SUMSS\_name    & Name of the association in the  SUMSS catalogue \\
      \itemm & arcsec & SUMSS\_dist    & Angular distance to the association in the SUMSS catalogue \\
      \itemm & ---    & SUMSS\_pfa     & The probability of false association with a counterpart from the SUMSS catalogue \\
      \itemm & ---    & TGSS\_name     & Name of the association in the  TGSS catalogue \\
      \itemm & arcsec & TGSS\_dist     & Angular distance to the association in the TGSS catalogue \\
      \itemm & ---    & TGSS\_pfa      & The probability of false association with a counterpart from the TGSS catalogue \\
      \itemm & ---    & AT20G\_name    & Name of the association in the  AT20G catalogue \\
      \itemm & arcsec & AT20G\_dist    & Angular distance to the association in the AT20G catalogue \\
      \itemm & ---    & AT20G\_pfa     & The probability of false association with a counterpart from the AT20G catalogue \\
      \itemm & ---    & Gaia\_name     & Name of the association in the  Gaia catalogue \\
      \itemm & arcsec & GAIA\_dist     & Angular distance to the association in the Gaia catalogue \\
      \itemm & ---    & GAIA\_pfa      & The probability of false association with a counterpart from the Gaia catalogue \\
      \itemm & ---    & PS1\_name      & Name of the association in the  PS1 catalogue \\
      \itemm & arcsec & PS1\_dist      & Angular distance to the association in the PS1 catalogue \\
      \itemm & ---    & PS1\_pfa       & The probability of false association with a counterpart from the PS1 catalogue \\
      \itemm & ---    & WISE\_name     & Name of the association in the  WISE catalogue \\
      \itemm & arcsec & WISE\_dist     & Angular distance to the association in the WISE catalogue \\
      \itemm & ---    & WISE\_pfa      & The probability of false association with a counterpart from the WISE catalogue \\
      \itemm & ---    & 2MASS\_name    & Name of the association in the  2MASS catalogue \\
      \itemm & arcsec & 2MASS\_dist    & Angular distance to the association in the 2MASS catalogue \\
      \itemm & ---    & 2MASS\_pfa     & The probability of false association with a counterpart from the 2MASS catalogue \\
      \itemm & ---    & GALEX\_name    & Name of the association in the  GALEX catalogue \\
      \itemm & arcsec & GALEX\_dist    & Angular distance to the association in the GALEX catalogue \\
      \itemm & ---    & GALEX\_pfa     & The probability of false association with a counterpart from the GALEX catalogue \\
      \itemm & ---    & 2RXS\_name     & Name of the association in the  2RXS catalogue \\
      \itemm & arcsec & 2RXS\_dist     & Angular distance to the association in the 2RXS catalogue \\
      \itemm & ---    & 2RXS\_pfa      & The probability of false association with a counterpart from the VLASS catalogue \\
      \itemm & ---    & XMMLS\_name    & Name of the association in the  XMMLS catalogue \\
      \itemm & arcsec & XMMLS\_dist    & Angular distance to the association in the XMMLS catalogue \\
      \itemm & ---    & XMMLS\_pfa     & The probability of false association with a counterpart from the XMMLS catalogue \\
      \itemm & ---    & 1eRASS\_name   & Name of the association in the  1eRASS catalogue \\
      \itemm & arcsec & 1eRASS\_dist   & Angular distance to the association in the 1eRASS catalogue \\
      \itemm & ---    & 1eRASS\_pfa    & The probability of false association with a counterpart from the 1eRASS catalogue \\
      \itemm & ---    & FERMI\_name    & Name of the association in the  FERMI catalogue \\
      \itemm & arcsec & FERMI\_dist    & Angular distance to the association in the FERMI catalogue \\
      \itemm & ---    & FERMI\_pfa     & The probability of false association with a counterpart from the FERMI catalogue \\
      \itemm & deg    & PosAng         & Jet position angle \\
      \itemm & ---    & z              & Redshift \\
      \itemm & ---    & z\_ref         & Astrophysics Data System (ADS) reference to the redshift \\
      \itemm & ---    & type           & Source type according to NED \\
      \enddata
    \tablenotetext{(a)}{\hspace{0.5em}
                   Table \ref{t:cross-match} is too wide to be shown here.
                   It is published in the machine-readable format only
                   as file cross\_match.txt
                  }
    \label{t:cross-match}
\end{deluxetable}

\begin{table}[h]
   \caption{Table of the sources that were observed at 4--8 GHz but 
            have not been detected. 
           }
   \begin{center}
      \begin{tabular}{lll}
         \hline\hline
           R.A.      & decl          & Flux density   \\
                     &               & mJy    \\
         \hline
         00 00 01.53 & $+$68 10 02.4 & 12.0   \\
         00 00 02.87 & $+$09 57 06.6 & 12.0   \\
         00 00 19.40 & $+$55 39 03.0 & 12.0   \\
         00 00 23.80 & $+$62 15 02.0 & 12.0   \\
         00 00 27.09 & $-$33 19 36.7 & 12.0   \\
         \ldots      & \ldots        & \ldots \\
         \hline
      \end{tabular}
   \end{center}
   \tablecomments{The third column shows the upper limit of the expected 
                  correlated flux density in units of mJy.
                  Table~\ref{t:nondet} is published in its entirety in
                  machine-readable format as file nondetections.txt. 
                  A portion is shown here for guidance regarding its 
                  form and content.
                 }
   \label{t:nondet}
\end{table}

\begin{table}[h]
   \caption{The error floor as a function of declination for right ascension
            scaled by $\cos\delta$ and declination. Units are 
            milliarcseconds. The first five lines of the table are shown
            below. 
           }
    \begin{center}
       \begin{tabular}{rrr}
          \hline\hline
          \ntab{c}{$\delta$} & R.A.   & decl   \\
          deg                & mas    & mas    \\
          \hline
          $ -90.0   $ & 0.104  & 0.214  \\
          $ -89.0   $ & 0.104  & 0.214  \\
          $ -88.0   $ & 0.103  & 0.214  \\
          $ -87.0   $ & 0.103  & 0.214  \\
          $ -86.0   $ & 0.103  & 0.214  \\
          \ldots   & \ldots & \ldots \\
          \hline
       \end{tabular}
    \end{center}
    \tablecomments{Table \ref{t:err_floor_mod} is published in its entirety 
                   in the machine-readable format as file
                   error\_floor\_model.txt. A portion is shown here 
                   for guidance regarding its form and content.
                  }
    \label{t:err_floor_mod}
\end{table}

\begin{table}
   \caption{Table of non-canonical RFC designators.
           }
   \begin{center}
      \begin{tabular}{lllr}
         \hline\hline
         \ntab{c}{(1)}  & \ntab{c}{(2)}  & \ntab{c}{(3)}  & \ntab{c}{(4)} \\
         \hline
          RFC J0000$+$030B  & J0000$+$0307     & close  & 262.04 \\
          RFC J0008$-$233A  & J0008$-$2339     & close  & 481.40 \\
          RFC J0023$+$273A  & J0023$+$2734     & pair   &   0.16 \\
          RFC J0031$+$540A  & J0031$+$5401     & pair   &   0.14 \\
          RFC J0116$+$242A  & J0116$+$2422     & close  &  73.63 \\
          \ldots            & \ldots           & \ldots & \ldots \\
         \hline
      \end{tabular}
      \tablecomments{Column descriptions: 
            (1) the non-canonical RFC designator;
            (2) the RFC designator to the closest source with a canonical 
                designator;
            (3) reason of assigning a non-canonical name;
            (4) the distance to the corresponding source with a canonical 
                common name in arcseconds when applicable.
                     Table \ref{t:ncj} is published in its entirety in the 
                     machine-readable format as file noncanonical\_jnames.txt.
                     A portion is shown here for guidance regarding its 
                     form and content.
            }
   \end{center}
   \label{t:ncj}
\end{table}

\begin{table}
   \caption{Table of non-canonical common names.}
   \begin{center}
       \begin{tabular}{llllr}
          \hline\hline
          \ntab{c}{(1)}  & \ntab{c}{(2)}  & \ntab{c}{(3)}  & \ntab{c}{(4)}  & \ntab{c}{(5)} \\
          \hline
          RFC J0009$+$0625 & 0006$+$06A & 0006$+$061 & close    &  244.38 \\
          RFC J0010$+$1058 & IIIZ$W$2   & 0007$+$106 & icrf1    & ...     \\
          RFC J0013$-$3227 & 0011$-$32B & 0011$-$327 & pair     &   53.65 \\
          RFC J0017$+$6750 & 0014$+$67A & 0014$+$675 & pair     &   51.98 \\
          RFC J0022$+$0014 & 4C$+$00.02 & 0019$-$000 & vcs1     & ...     \\
          \ldots           & \ldots     & \ldots     & \ldots   & \ldots  \\
          \hline
       \end{tabular}
       \tablecomments{Column descriptions: 
            (1)~the RFC designator of a source with non-canonical common name;
            (2)~the non-canonical common name;
            (3)~the common canonical name of the closest source;
            (4)~reason of assigning a non-canonical name;
            (5)~the distance to the corresponding source in arcseconds
                with a canonical common name when applicable.
                      Table \ref{t:ncb} is published in its entirety in the 
                      machine-readable format as file noncanonical\_comnams.txt.
                      A portion is shown here for guidance regarding its 
                      form and content.
                     }
   \end{center}
   \label{t:ncb}
\end{table}

\begin{table}
   \caption{Table of pairs of sources within $1'$. }
   \begin{center}
       \begin{tabular}{lllllr}
          \hline\hline
          \ntab{c}{(1)}  & \ntab{c}{(2)}  & \ntab{c}{(3)}   \\
          \hline
              RFC J0031$+$5401 & RFC J0031$+$540A  & 0.1    \\
              RFC J0134$-$093C & RFC J0134$-$093A  & 0.1    \\
              RFC J0904$+$5938 & RFC J0904$+$593A  & 0.1    \\
              RFC J2108$-$210A & RFC J2108$-$2101  & 0.1    \\
              RFC J0023$+$273A & RFC J0023$+$2734  & 0.2    \\
              \ldots           & \ldots            & \ldots \\
          \hline
       \end{tabular}
       \tablecomments{Columns descriptions:
              (1) the RFC designator of the first source in the pair;
              (2) the RFC designator of the second source in the pair;
              (3) the angular distance between sources in the pair in arcseconds.
              Table \ref{t:pairs} is published in its entirety in the 
              machine-readable format as file pairs.txt. 
              A portion is shown here for guidance regarding its 
              form and content.
             }
   \end{center}
   \label{t:pairs}
\end{table}

\begin{table}
   \caption{Table of the session names, experiment codes, and data archive 
            names of astrometric experiments.}
   \begin{center}
          \begin{tabular}{lll}
                 \hline\hline
                 Session name & Exp name & Archive \\
                 \hline
                 19950412\_p  &  br025   & NRAO    \\
                 19950419\_p  &  bb023b  & NRAO    \\
                 19950531\_p  &  rdwps1  & NRAO    \\
                 19950607\_p  &  rdgeo1  & NRAO    \\
                 19950625\_p  &  bb041b  & NRAO    \\
                 \ldots       & \ldots   & \ldots  \\
                 \hline
          \end{tabular}
   \end{center}
   \tablecomments{Table \ref{t:astro} is published in its entirety in the 
                  machine-readable format as file astro\_exp\_names.txt. 
                  A portion is shown here for guidance regarding its 
                  form and content.
                 }
   \label{t:astro}
\end{table}

\begin{table}
   \caption{Table of the session names, experiment codes, and data archive 
            names of 24~hr geodetic experiments.}
   \begin{center}
          \begin{tabular}{lll}
                 \hline\hline
                 Session name  & Exp name & Archive \\
                 \hline
                 19800411\_a  & xus801   & CDDIS   \\
                 19800413\_a  & hdsrvy   & CDDIS   \\
                 19800726\_a  & mert01   & CDDIS   \\
                 19800727\_a  & mert02   & CDDIS   \\
                 19800728\_a  & mert03   & CDDIS   \\
                 \ldots       & \ldots   & \ldots  \\
                 \hline
          \end{tabular}
   \end{center}
   \tablecomments{Table \ref{t:geod24hr} is published in its entirety in the 
                  machine-readable format as file geod24hr\_exp\_names.txt.
                  A portion is shown here for guidance regarding its 
                  form and content.
                 }
   \label{t:geod24hr}
\end{table}

\begin{table}
   \caption{Table of the session names, experiment codes, and data archive 
            names of 1~hr geodetic experiments.}
   \begin{center}
          \begin{tabular}{lll}
                 \hline\hline
                 Session name & Exp name & Archive \\
                 \hline
                 19920103\_i  & i92003   & CDDIS   \\
                 19920104\_i  & i92004   & CDDIS   \\
                 19920105\_i  & i92005   & CDDIS   \\
                 19920111\_i  & i92011   & CDDIS   \\
                 19920112\_i  & i92012   & CDDIS   \\
                 \ldots       & \ldots   & \ldots  \\
                 \hline
          \end{tabular}
   \end{center}
   \tablecomments{Table \ref{t:geod1hr} is published in its entirety in the 
                  machine-readable format as file geod1hr\_exp\_names.txt. 
                  A portion is shown here for guidance regarding its 
                  form and content.
                 }
   \label{t:geod1hr}
\end{table}

\newpage \phantom{a} \newpage \phantom{a} \newpage

\newcommand{\Campcapt}{The list of 72 VLBI absolute astronomy observing campaigns 
                       used for deriving the RFC. \label{t:camp}}
\newcommand{\CampNote}{Principal Investigator name is given for the 
                       observing campaigns that do not have publications.}
%
%
\startlongtable
\begin{deluxetable*}{                  l
                     @{\hspace{0.3em}} l
                     @{\hspace{0.1em}} l
                     @{\hspace{0.3em}} l
                     @{\hspace{-0.8em}}r
                     @{\hspace{0.1em}} r
                     @{\hspace{0.3em}} r
                     @{\hspace{0.3em}} r
                     @{\hspace{0.5em}} l
                     @{\hspace{0.5em}} l
                     @{\hspace{0.3em}} r
                     @{\hspace{0.3em}} r
                     @{\hspace{0.3em}} r
                    }
    \tablecaption{\Campcapt}
    \tablecolumns{13}
    \tabletypesize{\small}
    \tablehead{
               \multicolumn{1}{l}{Campaign}          &
               \colhead{\hspace{-0.7em}Network}      &
               \multicolumn{1}{c}{\hspace{-0.5em}Id} &
               \multicolumn{1}{l}{Reference}         &
               \multicolumn{2}{c}{Frequency}         &
               \colhead{Dur.}                        &
               \colhead{\hspace{-0.8em}Num}          &
               \multicolumn{2}{c}{Dates}             &
               \multicolumn{3}{r}{Number of sources} \vspace{-1ex} \\
               \colhead{}       &
               \colhead{}       &
               \colhead{}       &
               \colhead{}       &
               \colhead{low}    &
               \colhead{\hspace{-0.5em}high}   &
               \colhead{}       &
               \colhead{ses}    &
               \colhead{start}  &
               \colhead{end}    &
               \colhead{obs}    &
               \colhead{det}    &
               \colhead{unique} \vspace{-1ex} \\
               \colhead{}       &
               \colhead{}       &
               \colhead{}       &
               \colhead{}       &
               \colhead{GHz}    &
               \colhead{GHz}    &
               \colhead{hour}   &
               \colhead{}       &
               \colhead{}       &
               \colhead{}       &
               \colhead{}       &
               \colhead{}       &
               \colhead{}       \vspace{-4ex} \\
    }
    \startdata
    & \multicolumn{4}{l}{\sf\large Pathfinder surveys: \hfill}           &       &     &            &            &       &       & \vspace{2ex} \\
    VCS1        & VLBA & bb023    & \citet{r:vcs1}         &  2.3 &  8.4 &   264 &  11 & 1994.08.12 & 1997.08.27 &  1838 &  1823 &     1 \\
    VLBApls     & VLBA & bh019    & \citet{r:vlbapls}      &  2.3 & 22.2 &    16 &   1 & 1996.06.05 & 1996.06.05 &   228 &   214 &     0 \\
                & VLBA & bb041    & PI: T. Beasley, 1995   &  2.3 &  8.4 &    40 &   2 & 1995.06.25 & 1996.02.16 &    57 &    56 &     0 \\
                & VLBA & bu007    & \citet{r:ulv99}        &      &  4.9 &    12 &   1 & 1996.12.19 & 1996.12.19 &   163 &   162 &    67 \\
                & VLBA & bg069    & \citet{r:bolsam}       &      &  5.0 &    60 &   4 & 1997.04.06 & 2005.06.17 &    67 &    61 &     3 \\
                & VLBA & bb119    & \citet{r:bri07}        &      &  5.0 &    72 &   3 & 1999.11.21 & 1999.11.26 &    88 &    87 &     0 \\
                & EVN  & ec013    & \citet{r:cha04}        &  8.4 &  2.3 &    71 &   3 & 2000.05.31 & 2003.10.17 &   161 &   161 &     0 \\
    VCS2        & VLBA & bf071    & \citet{r:vcs2}         &  2.3 &  8.7 &    48 &   2 & 2002.01.31 & 2002.05.14 &   371 &   367 &     2 \\
                & VLBA & bb177    & \citet{r:bol06b}       &      &  5.0 &    12 &   1 & 2004.02.06 & 2004.02.06 &    38 &    35 &     9 \\
    VCS3        & VLBA & bp110    & \citet{r:vcs3}         &  2.3 &  8.7 &    72 &   3 & 2004.04.30 & 2004.05.27 &   533 &   487 &     0 \\
    VCS4        & VLBA & bp118    & \citet{r:vcs4}         &  2.3 &  8.7 &    72 &   3 & 2005.05.12 & 2005.06.30 &   504 &   410 &     0 \\
                & VLBA & bc151    & \citet{r:bol06}        &      &  5.0 &    30 &   4 & 2005.06.16 & 2005.08.04 &    85 &    82 &    23 \\
    VCS5        & VLBA & bk124    & \citet{r:vcs5}         &  2.3 &  8.7 &    72 &   3 & 2005.07.08 & 2005.07.20 &   748 &   701 &     0 \\
    VIPS        & VLBA & bt085    & \citet{r:vips},        &      &  4.9 &   174 &  16 & 2006.01.03 & 2006.08.12 &   858 &   857 &   262 \\
                &      &          & \citet{r:astro_vips}   &      &      &       &     &       &    &            &               &       \\
    NPCS        & VLBA & bk130    & \citet{r:npcs}         &  2.3 &  8.7 &    72 &   3 & 2006.02.14 & 2006.02.23 &   526 &   194 &     5 \\
    VGaPS       & VLBA & bp125    & \citet{r:vgaps}        &      & 24.5 &    72 &   3 & 2006.06.04 & 2006.10.20 &   543 &   388 &    24 \\
                & VLBA & bm252    & \citet{r:majid09}      &      &  8.7 &    20 &   2 & 2006.11.06 & 2006.11.13 &    74 &    53 &    30 \\
    VCS6        & VLBA & bp133    & \citet{r:vcs6}         &  2.3 &  8.7 &    48 &   2 & 2006.12.18 & 2007.01.11 &   347 &   329 &     0 \\
    VEGaPS      & VERA & r07030a  & PI: L. Petrov, 2007    &      & 22.2 &    28 &   2 & 2007.01.30 & 2007.03.21 &   125 &   110 &     0 \\
    LCS-1       & LBA  & v230r    & \citet{r:lcs1}         &      &  8.4 &   108 &   5 & 2007.06.24 & 2009.07.04 &   597 &   574 &   158 \\
    OBRS-1      & EVN  & gc030    & \citet{r:obrs1},       &  2.3 &  8.4 &    48 &   1 & 2008.03.07 & 2008.03.07 &   115 &   115 &     1 \\
                &      &          & \citet{r:bourda11}     &      &      &       &     &       &    &            &               &       \\
    EGaPS       & EVN  & ep066    & \citet{r:egaps}        &      & 22.2 &    48 &   1 & 2009.10.27 & 2009.10.27 &   437 &   183 &    52 \\
    BeSSel-Cal1 & VLBA & br145    & \citet{r:bessel}       &      &  8.4 &   153 &  34 & 2009.11.16 & 2010.08.29 &  1535 &   364 &    95 \\
                & VLBA & bt110    & \citet{r:linford12}    &      &  4.9 &    76 &   7 & 2009.11.22 & 2010.07.30 &   308 &   308 &     1 \\
    LCS-2       & LBA  & v271dr   & \citet{r:lcs2}         &      &  2.3 &   368 &  16 & 2009.12.12 & 2016.06.28 &  1401 &   959 &   480 \\
    BeSSel-Cal2 & VLBA & br149    & PI: M. Reid, 2010      &      &  8.4 &    43 &  14 & 2010.02.06 & 2013.08.04 &   574 &   176 &    37 \\
                & EVN  & gb073    & \citet{r:obrs2},       &  2.3 &  8.4 &   216 &   7 & 2010.03.23 & 2012.05.27 &   378 &   377 &    67 \\
                &      &          & \citet{r:bourda11}     &      &      &       &     &       &    &            &               &       \\
    V2M         & VLBA & bc191    & \citet{r:v2m}          &  4.4 &  7.2 &   637 &  96 & 2010.07.15 & 2013.12.06 &  2701 &  1868 &   458 \\
    1FGL-VLBI   & VLBA & s3111    & PI: Y. Kovalev, 2010   &      &  8.7 &    72 &   3 & 2010.12.05 & 2011.01.09 &   283 &   279 &    84 \\
    VCS7        & VLBA & bp171    & \citet{r:wfcs}         &  4.2 &  7.6 &    73 &  17 & 2013.02.08 & 2013.08.01 &  1626 &   968 &   422 \\
    2FGL-VLBIa  & VLBA & s4195    & PI: Y. Kovalev, 2013   &      &  7.6 &    72 &   3 & 2013.05.07 & 2013.06.22 &   322 &   289 &   136 \\
    2FGL-VLBIc  & VLBA & s5272    & \citet{r:aofus2}       &      &  7.6 &    47 &   4 & 2013.08.06 & 2013.12.05 &   211 &   153 &    47 \\
                & VLBA & bp175    & \citet{r:wfcs}         &  2.3 &  8.1 &    43 &  10 & 2013.10.26 & 2013.12.26 &   405 &   401 &     0 \\
    VCS8        & VLBA & bp177    & \citet{r:wfcs}         &  4.4 &  7.6 &    48 &  10 & 2014.01.07 & 2014.02.23 &  1386 &   927 &   446 \\
    VEPS-1      & CVN  & veps     & \citet{r:veps1}        &      &  8.6 &   425 &  18 & 2015.02.13 & 2017.12.14 &  4571 &   973 &     0 \\
    2FGL-VLBIb  & VLBA & bs241    & \citet{r:aofus2}       &      &  7.6 &    54 &   7 & 2015.02.16 & 2015.07.01 &   451 &   308 &    77 \\
    VCS9        & VLBA & bp192    & \citet{r:wfcs}         &  4.4 &  7.6 &   528 &  99 & 2015.08.07 & 2016.09.07 & 11016 &  5688 &  3945 \\
    3FGL-VLBI   & VLBA & s7104    & \citet{r:aofus3}       &      &  7.6 &    63 &   9 & 2016.06.25 & 2016.07.26 &   607 &   416 &   104 \\
    SOFUS       & LBA  & sofus    & PI: L. Petrov, 2017    &      &  8.5 &    85 &   4 & 2017.04.07 & 2021.05.08 &   324 &   207 &   126 \\
    VOFUS-1     & VLBA & bs262    & \citet{r:aofus4}       &  4.4 &  7.6 &    70 &  21 & 2018.04.08 & 2018.07.24 &   970 &   883 &   319 \\
    VOFUS-2     & VLBA & sb072    & \citet{r:aofus4}       &  4.4 &  7.6 &   110 &  31 & 2018.08.25 & 2019.02.17 &  1467 &  1322 &   551 \\
    AGaPS       & EAVN & ap001a   & PI: L. Petrov, 2018    &      & 22.2 &    24 &   4 & 2018.10.09 & 2019.01.28 &   193 &   121 &     0 \\
    VCS10-CX    & VLBA & bp242    & PI: A. Popkov, 2019    &  4.4 &  7.6 &    94 &  20 & 2019.07.24 & 2020.03.17 &  2779 &  1491 &  1125 \\
    VCS10-SX    & VLBA & bp245u   & PI: A. Popkov, 2020    &  2.3 &  8.7 &    23 &   6 & 2020.03.02 & 2020.03.23 &   638 &   210 &    27 \\
    GC-KVN      & KVN  & n20lp01  & PI: L. Petrov, 2020    & 22.7 & 43.9 &    69 &  14 & 2020.03.05 & 2020.06.16 &   400 &   174 &     0 \\
                & VLBA & bb409    & PI: A. Beasley, 2020   &  4.9 &  6.7 &    24 &   4 & 2020.05.10 & 2020.07.20 &   656 &    28 &    16 \\
    VCS11       & VLBA & br235    & PI: T. Readhead, 2020  &  4.4 &  7.6 &   108 &  18 & 2020.09.11 & 2021.02.16 &  3328 &  2623 &  2150 \\
    VCS12       & VLBA & bp252    & PI: L. Petrov, 2021    &  4.4 &  7.7 &   244 &  53 & 2021.09.21 & 2022.12.18 &  9564 &  3318 &  2284 \\
                &      &          &                        &      &      &       &     &            &            &       &       &       \\
    & \multicolumn{4}{l}{\sf\large Astrometric follow-ups surveys: \hfill}    &         &     &            &             &       &      & \vspace{2ex} \\
    RDV         & VLBA & rv       & \citet{r:rdv}          &  2.3 &  8.4 &  5265 & 219 & 1994.07.08 & 2023.07.04 &  2281 &  2236 &     7 \\
                & VLBA & bf025    & \citet{r:bf025}        &  2.3 &  8.4 &    48 &   2 & 1997.01.10 & 1997.01.11 &   226 &   225 &     0 \\
    VCS-II      & VLBA & bg219    & \citet{r:vcs-ii}       &  2.3 &  8.7 &   196 &   9 & 2014.01.04 & 2015.03.17 &  2597 &  2532 &     0 \\
    VEPS-V1     & VLBA & bs250    & \citet{r:veps1}        &  2.3 &  8.7 &    32 &   4 & 2016.03.22 & 2016.05.19 &   163 &   163 &     0 \\
    VCS-III     & VLBA & uf001    & \citet{r:vcs-eyes}     &  2.3 &  8.7 &   478 &  20 & 2017.01.16 & 2017.10.21 &  3654 &  3647 &     0 \\
    GAIA-L2     & LBA  & v561     & PI: L. Petrov, 2017    &  2.3 &  8.6 &    71 &   2 & 2017.06.16 & 2018.03.14 &   306 &   303 &     0 \\
    SOAP        & LBA  & aua025   & PI: L. Petrov, 2017    &  2.3 &  8.6 &   568 &  24 & 2017.08.22 & 2019.12.04 &   444 &   422 &     6 \\
    VCS-IV      & VLBA & ug002    & \citet{r:vcs-eyes}     &  2.3 &  8.7 &   573 &  24 & 2018.01.18 & 2019.01.21 &  4416 &  4238 &    10 \\
    VEPS-3      & CVN  & epa      & PI: L. Petrov, 2018    &  2.3 &  8.6 &    44 &   2 & 2018.01.24 & 2018.02.10 &   182 &   181 &     0 \\
    VEPS-2      & VLBA & bs264    & PI: F. Shu, 2018       &  2.3 &  8.7 &    48 &   6 & 2018.03.21 & 2018.06.15 &   357 &   357 &     0 \\
    GAIA-V1     & VLBA & bp222    & PI: L. Petrov, 2018    &  2.3 &  8.7 &   304 &  38 & 2018.05.15 & 2020.04.19 &  1367 &  1367 &     0 \\
    VCS-V       & VLBA & ug003    & \citet{r:vcs-eyes}     &  2.3 &  8.7 &   620 &  26 & 2019.01.27 & 2020.08.09 &  4167 &  4162 &     1 \\
    VCS-VI      & VLBA & uh007    & \citet{r:vcs-eyes}     &  2.3 &  8.7 &   667 &  28 & 2020.09.18 & 2022.12.12 &  2545 &  2536 &     0 \\
                &      &          &                        &      &      &       &     &            &            &       &       &       \\
    & \multicolumn{4}{l}{\sf\large High frequency extensions: \hfill}     &         &     &            &             &       &      & \vspace{2ex} \\
    K/Q-Survey  & VLBA & bl115    & \citet{r:kq_astro},    & 24.5 & 43.2 &   336 &  14 & 2002.05.15 & 2011.02.05 &   343 &   334 &     0 \\
                &      &          & \citet{r:kq_image}     &      &      &       &     &       &    &            &               &       \\
    KVNCS       & KVN  & n13jl01  & \citet{r:lee2023};     &      & 23.0 &   196 &   7 & 2013.09.04 & 2014.12.24 &   790 &   752 &     0 \\
                & VLBA & bj083    & \citet{r:witt23}       &      & 24.6 &   105 &   5 & 2015.07.21 & 2016.06.20 &   286 &   286 &     0 \\
                & EVN  & ec076    & \citet{r:gomez23}      &      & 22.3 &    48 &   2 & 2016.06.15 & 2020.10.23 &   172 &   169 &     0 \\
                & VLBA & ud001    & \citet{r:witt23}       &      & 23.6 &   564 &  24 & 2017.01.08 & 2018.07.22 &   738 &   734 &     0 \\
                & VLBA & ud009    & \citet{r:witt23}       &      & 23.6 &   823 &  35 & 2018.09.09 & 2021.06.12 &   821 &   818 &     0 \\
    GAJI        & KVN  & gaji     & PI: L. Petrov, 2018    & 21.7 & 43.8 &    22 &   4 & 2018.09.25 & 2018.12.29 &   151 &    90 &     0 \\
                & EAVN & a20      & PI: S. Xu, 2020        &      & 21.3 &   240 &  10 & 2020.05.07 & 2023.06.08 &   328 &   318 &     0 \\
    GC-VLBA     & VLBA & bp251    & PI: Y. Pihlstrom, 2021 & 24.0 & 43.2 &    34 &   8 & 2021.03.19 & 2022.05.25 &   138 &   138 &     1 \\
                & VLBA & ud015    & PI: A. de Witt, 2021   &      & 23.6 &   426 &  18 & 2021.07.26 & 2023.01.06 &   919 &   914 &     0 \\
                & VLBA & ud018    & PI: A. de Witt, 2023   &      & 23.6 &    16 &   4 & 2023.07.03 & 2023.07.24 &    68 &    68 &     0 \\
    \hline
          Total &      &          &                        &      &     & 16944 & 1140 &            &             & 42469 & 21940 & 13659 \\
    \enddata
    \tablecomments{\CampNote}
    \tablecomments{Two sources, 0528$-$654 and 1144$+$404 were observed only in 
                   geodetic experiments and are not counted here.}
    \tablecomments{The last columns shows the number of detected
                   sources that are unique for that campaign and were 
                   not detected in any other campaign.}
\end{deluxetable*}

\facilities{VLBA,LBA,CVN,EVN,EAVN,IVS,KaVa,VERA,KVN}
\software{SGDASS (L.~Petrov, under review, 2025), AIPS \citep{r:aips}, 
          Difmap \citep{r:difmap}}

\newpage 

\bibliographystyle{aasjournal}
\bibliography{rfc1}

\end{document}